\IfFileExists{venue_preamble.tex}{\def\VenueMode{1}
  \input{venue_preamble}
}{\documentclass[11pt]{article}
  \usepackage[margin=1in]{geometry}
}

\usepackage[T1]{fontenc}
\usepackage{microtype}
\usepackage{amsmath}
\usepackage{amssymb}
\usepackage{amsthm}
\usepackage{aliascnt}
\usepackage{mathtools}
\usepackage{graphicx}
\usepackage{booktabs}
\usepackage{array}
\usepackage{xcolor}
\usepackage{tikz}
\usetikzlibrary{arrows.meta,decorations.markings}
\definecolor{DPBlue}{HTML}{0072B2}
\definecolor{DPOrange}{HTML}{D55E00}
\definecolor{DPPurple}{HTML}{7057A3}
\tikzset{
  dp one/.style={draw=DPBlue,line width=1.1pt},
  dp two/.style={draw=DPOrange,line width=1.1pt},
  dp edge/.style={draw=DPPurple,line width=1.6pt,-{Stealth[length=4pt]}},
  dp arrow/.style={postaction={decorate},decoration={markings,mark=at position 0.58 with {\arrow{Stealth[length=4pt]}}}},
  dp label/.style={inner sep=1.5pt,font=\normalsize}
}
\usepackage{algorithm}
\usepackage{hyperref}
\usepackage[nameinlink,capitalize,noabbrev]{cleveref}

\ifdefined\VenueMode
\else
  \hypersetup{
    colorlinks=true,
    citecolor=blue,
    linkcolor=blue,
    urlcolor=blue
  }
\fi

\graphicspath{{figures/}}
\allowdisplaybreaks

\DeclareMathOperator{\dist}{dist}

\newcommand{\F}{\mathbb{F}}

\DeclareMathOperator{\wt}{wt}
\newcommand{\Ring}{\mathcal{R}}

\theoremstyle{plain}
\newtheorem{theorem}{Theorem}[section]
\newaliascnt{lemma}{theorem}
\newtheorem{lemma}[lemma]{Lemma}
\aliascntresetthe{lemma}
\newaliascnt{proposition}{theorem}
\newtheorem{proposition}[proposition]{Proposition}
\aliascntresetthe{proposition}
\newaliascnt{corollary}{theorem}

\aliascntresetthe{corollary}

\theoremstyle{definition}
\newaliascnt{definition}{theorem}

\aliascntresetthe{definition}

\theoremstyle{remark}
\newaliascnt{remark}{theorem}

\aliascntresetthe{remark}

\theoremstyle{plain}

\newcommand{\PaperTitle}{Faster Algorithms for Disjoint Paths in DAGs and Two Disjoint Shortest Paths}
\newcommand{\AuthorCao}{Yang Cao}
\newcommand{\AuthorQi}{Haoran Qi}
\newcommand{\AuthorWang}{Hanzhi Wang}
\newcommand{\PaperAuthors}{\AuthorCao\thanks{\texttt{yangcao@psu.edu}. Pennsylvania State University.} \and \AuthorQi\thanks{Jilin University.} \and \AuthorWang\thanks{Tsinghua University.}}
\newcommand{\PaperPdfAuthors}{\AuthorCao, \AuthorQi, \AuthorWang}
\newcommand{\PaperDate}{}

\newif\ifPaperPrintReferences
\PaperPrintReferencestrue

\title{\PaperTitle}
\author{\PaperAuthors}
\date{\PaperDate}
\hypersetup{pdftitle={\PaperTitle},pdfauthor={\PaperPdfAuthors}}

\providecommand{\PaperMakeTitle}{\maketitle}
\providecommand{\PaperPrintBibliography}{\ifPaperPrintReferences
    \bibliographystyle{alpha}
    \bibliography{references}
  \fi}
\providecommand{\PaperEndMatter}{}

\begin{document}

\PaperMakeTitle

\begin{abstract}
For $k$ vertex-disjoint paths with prescribed endpoints in a directed acyclic graph (DAG), the $O(mn^{k-1})$-time pebbling algorithm of Fortune, Hopcroft, and Wyllie [TCS 1980] has remained the fastest for every $k\geq3$ for over four decades, even when the paths must be shortest. We break this bound for every $k\geq3$ with randomized algorithms running in expected $O_k(m+n^{k-3+\omega+\varepsilon})$ time for every $\varepsilon>0$, where $\omega<2.372$ is the matrix multiplication exponent; this is $O_k(m+n^{k-0.62})$. For two disjoint shortest paths, we find the paths as fast as we detect them: in expected linear time on DAGs and on undirected graphs with polynomially bounded integer weights, and in expected $O(\min\{mn,n^{\omega+\varepsilon}\})$ time on directed graphs whose cycles have positive weight, which on dense graphs is $o(mn)$ even for detection. All algorithms evaluate a polynomial whose monomials are the solutions: a new first-collision circuit speeds up the evaluation in DAGs, and a correlated extraction step at cuts crossed once by every path finds a solution at the cost of a constant number of evaluations.

\end{abstract}

\clearpage

\section{Introduction}\label{sec:intro}

In the $k$ vertex-disjoint paths problem ($k$-DP), the input is a directed graph with $n$ vertices and $m$ edges and $k$ terminal pairs $(s_i,t_i)$, and the task is to connect every pair by a path so that the paths are pairwise vertex-disjoint. In the $k$ disjoint shortest paths problem ($k$-DSP), every path must in addition be a shortest path between its terminals. Already for two pairs, $k$-DP is NP-hard on general directed graphs \cite[Lemma~3]{fhw80}. On directed acyclic graphs (DAGs) it is polynomial for every fixed $k$ \cite[Theorem~3]{fhw80}, by a pebbling game that extends the two-path algorithm of \cite{ps78}: a configuration records the current vertex of every path, and a move advances the earliest path along an edge. There are $O(n^k)$ configurations and $O_k(mn^{k-1})$ moves, and this bound, stated explicitly in \cite[Lemma~34]{bnr+21} and \cite[Theorems~2 and~4]{epp95}, has remained the fastest for every $k\geq3$, for $k$-DP as well as for weighted $k$-DSP \cite[Section~7]{avw24}.

\cite[Section~7, Open Problem~3]{avw24} asks for an $O(n^{k+1-\delta})$-time algorithm for $k$-DSP in DAGs, for some fixed $k\geq3$ and some constant $\delta>0$. The question includes $k$-DP: giving every edge the difference of the topological positions of its endpoints makes every path between fixed endpoints shortest \cite[Proposition~54]{avw24}.

For two pairs, the shortest-path constraint makes the problem tractable beyond DAGs, and the fastest algorithms are algebraic. The circuits of \cite{avw24} detect two disjoint shortest paths in $O(m+n)$ time on positive-weight DAGs and undirected graphs, and \cite{cks26} detect them in $O(mn\log n)$ time on directed graphs whose cycles have positive weight. Finding the paths has remained a factor of about $n$ slower: the reporting algorithms of \cite{avw24} and of \cite{bnr+21}, for unweighted undirected graphs, take $O(mn)$ time, and for directed graphs \cite{bk17} gives a polynomial-time algorithm and \cite{cks26} report in $O(mn^2\log n)$ time. \cite{cks26} ask whether reporting can match detection and whether directed 2-DSP admits an $o(mn)$-time algorithm \cite[Section~6, Open Problems~1 and~2]{cks26}.

We improve both lines of work; \cref{tab:bounds} summarizes the bounds. All our algorithms evaluate a polynomial over a field of characteristic two whose monomials are the solutions, as the detectors of \cite{avw24} do, and two tools make this approach fast. The first is a circuit that evaluates the polynomial of $k$ disjoint paths in a DAG by subtracting intersecting tuples at their first collision (\cref{thm:dag-k-paths}). The second is a correlated extraction step: at a cut that every path crosses at most once, a nonzero evaluation yields compatible pieces of one solution on both sides of the cut, at the cost of a constant number of evaluations (\cref{lem:extraction-step}). The first tool breaks the pebbling bound in DAGs with superlinearly many edges; the second makes reporting as fast as detection in each of our models.

\begin{table}[t]
\centering
\small
\setlength{\tabcolsep}{4pt}
\renewcommand{\arraystretch}{1.2}
\begin{tabular}{@{}llll@{}}
\toprule
Problem & Previous (detect / report) & This paper & Thm. \\
\midrule
$k$-DP, $k$-DSP in DAGs, $k\geq3$ & $O(mn^{k-1})$ \cite{fhw80} & $O_k(m+n^{k-3+\omega+\varepsilon})$ & \labelcref{thm:dag-k-dsp-informal} \\
\addlinespace
2-DSP, DAGs and undirected & $O(m+n)$ / $O(mn)$ \cite{avw24} & $O(m+n)$ & \labelcref{thm:two-linear-informal} \\
2-DSP, directed & $O(mn\log n)$ / $O(mn^2\log n)$ \cite{cks26} & $O(\min\{mn,n^{\omega+\varepsilon}\})$ & \labelcref{thm:two-directed-informal} \\
\bottomrule
\end{tabular}
\caption{Time bounds for finding the paths. For two pairs, the previous column lists detection and reporting separately; for three or more pairs, one bound covers both. Our bounds are expected times of randomized algorithms with one-sided error, each attained by one algorithm that decides feasibility and reports a solution. The previous two-path bounds for DAGs and undirected graphs assume positive weights; ours allow arbitrary real weights on DAGs and assume positive polynomially bounded integer weights on undirected graphs. Directed graphs have only positive-weight cycles; for negative weights, the directed bound uses the single-source shortest-path algorithm of \cite{hll+26}.}
\label{tab:bounds}
\end{table}

\subsection{Main results}

Throughout, $O_k$ hides factors depending only on $k$, and $\omega$ is the exponent of matrix multiplication over $\F_2$; the computational models are fixed in \cref{sec:prelim}.

\begin{theorem}[Disjoint paths in DAGs, informal version of \cref{thm:dag-k-dsp}]
\label{thm:dag-k-dsp-informal}
For every fixed $k\geq3$ and $\varepsilon>0$, $k$-DSP on DAGs with real edge weights, and hence $k$-DP on DAGs, can be solved by a randomized algorithm with one-sided error in expected $O_k(m+n^{k-3+\omega+\varepsilon})$ time.
\end{theorem}

The pebbling bound $O(mn^{k-1})$ exceeds the bound of \cref{thm:dag-k-dsp-informal} by a factor of $\Omega(n^{3-\omega-\varepsilon})$ for every $m\geq n$ and every fixed $k\geq3$; with $\omega<2.372$, our bound is $O_k(m+n^{k-0.62})$. Keeping one lightest copy of each parallel edge takes $O(m+n)$ time by radix sort and leaves fewer than $n^2$ edges. This answers Open Problem~3 of \cite{avw24}, which asks for $O(n^{k+1-\delta})$ time for some fixed $k\geq3$ and $\delta>0$: the term $m$ only reads the input, and we obtain $\delta=4-\omega-\varepsilon$ for every fixed $k\geq3$. Under the $k$-Clique Hypothesis, $k$-DSP on unweighted DAGs requires $C(n^{1/2},k)^{1-o(1)}$ time, where $C(N,k)$ is the time to detect a $k$-clique in an $N$-vertex graph \cite[Theorem~7 and Corollary~8]{avw24}; for three pairs this lower bound is $n^{\omega/2-o(1)}$.

The algebraic core of \cref{thm:dag-k-dsp-informal} is \cref{thm:dag-k-paths}, which computes the enumerating polynomial of the disjoint path tuples over every commutative ring and for arbitrary commodity subgraphs. Over the integers with every edge variable set to $1$, it counts the vertex-disjoint tuples with prescribed terminal pairs with $O_k(mn^{k-2}+n^k)$ arithmetic operations, compared with $O(mn^{k-1})$ in \cite[Theorem~4]{epp95}.

For two pairs, the extraction step finds the paths within the time of detection, and in directed graphs a faster detector also improves detection.

\begin{theorem}[Linear-time 2-DSP, informal version of \cref{thm:two-linear}]
\label{thm:two-linear-informal}
On DAGs with real edge weights and on undirected graphs with positive polynomially bounded integer weights, two vertex-disjoint shortest paths for two prescribed terminal pairs can be found by a randomized algorithm with one-sided error in expected $O(m+n)$ time.
\end{theorem}

\begin{theorem}[Directed 2-DSP, informal version of \cref{thm:two-directed}]
\label{thm:two-directed-informal}
On directed graphs with real edge weights in which every cycle has positive weight, two vertex-disjoint shortest paths for two prescribed terminal pairs can be found by a randomized algorithm with one-sided error in expected $O(\min\{mn,n^{\omega+\varepsilon}\})$ time, for every fixed $\varepsilon>0$. Deciding whether they exist takes the same time.
\end{theorem}

For comparison, \cref{thm:two-linear-informal} matches the detection time of \cite{avw24} on DAGs and, for integer weights, on undirected graphs, and improves their $O(mn)$ reporting bound by a factor of $n$. Our undirected reporting computes regional distances with the integer-weight single-source shortest-path algorithm of \cite{tho97} (\cref{sec:anchors}), so for positive real weights we claim no linear bound, although the detector of \cite{avw24} is linear there. \Cref{thm:two-directed-informal} answers the reporting question of \cite{cks26} for all real weights, and their $o(mn)$ question in the dense regime: for every fixed $\eta>0$, choosing $\varepsilon<\eta$ makes the bound $o(mn)$ when $m\geq n^{\omega-1+\eta}$, so in this regime both detection and reporting improve on \cite{cks26} by a polynomial factor. The bound depends on the source-distance labelings only through \cref{lem:sssp}. For nonnegative weights they take $O(m+n\log n)$ time \cite{ft87}. For arbitrary real weights, Bellman--Ford \cite{bel58} takes $O(mn)$ time, and the Las Vegas algorithm of \cite{hll+26} takes $m^{1+o(1)}$ time with high probability, the latest in a line of algorithms faster than $O(mn)$ that began with \cite{fin23}; the labelings therefore never dominate the bound. Even for sparse graphs, the $O(mn)$ bound removes the logarithmic factor from the detection bound of \cite{cks26}, whose all-pairs preprocessing \cite[Lemma~2.6]{cks26} takes $O(mn+n^2\log n)$ time with Fibonacci heaps while its remaining detector states cost $O(mn)$ \cite[Section~4.3]{cks26}, and it improves their reporting bound by a factor of $n\log n$. \Cref{thm:two-linear-informal} answers the reporting question for DAGs and, in its integer-weight model, for undirected graphs.

\subsection{Technical overview}\label{sec:overview}

All our algorithms encode the solutions as the monomials of a polynomial $F$ over a field of characteristic two: path tuples are weighted by the product of their edge variables, intersecting tuples cancel, and distinct disjoint tuples leave distinct squarefree monomials (\cref{lem:k-monomials}). One random evaluation therefore decides feasibility. Two costs remain: the size of a circuit for $F$, and the cost of turning a nonzero value into paths. Generic self-reduction fixes one edge at a time and repeats the evaluation, which loses a factor of about $n$. We describe the circuits first, then the extraction step, and then how the reporting recursions stay within the cost of one evaluation.

\paragraph{Evaluation by first collisions.}
For two paths in a DAG, the detector of \cite{avw24} starts from the product of the two source-to-target path sums, which counts every pair of paths, and subtracts the intersecting pairs by their first intersection $v$: such a pair splits into two prefixes that meet only at $v$ and two unconstrained suffixes \cite[Lemma~23]{avw24}. For $k\geq3$ paths, a tuple can collide in many patterns, and tracking all positions is the pebbling game, which examines every edge once for each of the $n^{k-1}$ placements of the other paths. We keep the first-intersection principle and store no placements: we subtract the intersecting tuples, classified by their first collision in a fixed topological order. Let $w$ be this collision and $T$ the set of paths through $w$. Before $w$, the prefixes are disjoint except at $w$ itself. After $w$, nothing is required: the paths through $w$ continue arbitrarily, and every other path continues from its last vertex $d_i<w$ by an edge that jumps over $w$, unless it has already ended.

With more paths, the jumping edges are the new ingredient (\cref{fig:first-event}). A prefix table stores $w$ and the endpoints $d_i$ of the paths outside $T$, so it has at most $k-1$ vertex coordinates instead of $k$. The graph enters only when one coordinate is extended across $w$ by a jumping edge and a path sum, which costs $O_k(mn^{k-2})$ over all choices of $w$ and of the other coordinates. The tables satisfy the same first-collision decomposition, so a topological sweep updates them with $O_k(n^k)$ further operations. Both steps become matrix products, scheduled by divide and conquer over the topological order so that every product reads only final rows. For three paths, the extensions and the updates are batched across rows, which gives $O(m+n^{\omega+\varepsilon})$ operations. For $k\geq4$, each row is extended by matrix products as soon as it is final, and the updates of a pair table become rectangular matrix products once all but one endpoint outside the pair is fixed, which gives $O_k(m+n^{k-3+\omega+\varepsilon})$ operations (\cref{lem:first-event,prop:k-sweep,prop:three-batched,prop:k-batched}).

\begin{figure}[t]
\centering
\begin{tikzpicture}[x=0.88cm,y=0.88cm,font=\normalsize,line cap=round]
\node[anchor=west] at (-0.8,3.9) {(a) Incompatible marginals};
\node at (1.2,3.3) {$Q=z_ez_{f'}+z_{e'}z_f$};
\draw[gray] (0,0.5) rectangle (2,2.5);
\draw[gray] (1,0.5) -- (1,2.5) (0,1.5) -- (2,1.5);
\node at (0.5,2.8) {$f$}; \node at (1.5,2.8) {$f'$};
\node[left] at (0,2) {$e$}; \node[left] at (0,1) {$e'$};
\node[gray] at (0.5,2) {$0$}; \node at (1.5,2) {$1$};
\node at (0.5,1) {$1$}; \node[gray] at (1.5,1) {$0$};
\draw[DPPurple,line width=1.2pt,rounded corners=3pt] (1.1,1.6) rectangle (1.9,2.4);
\node[align=center] at (1.1,-0.15) {Rows and columns can be nonzero\\while their joint coefficient is zero.};
\begin{scope}[shift={(5,0)}]
\node[anchor=west] at (-0.1,3.9) {(b) Compatible cut edges};
\fill[black!3] (0,0.4) rectangle (3,2.7);
\fill[black!7] (3,0.4) rectangle (6,2.7);
\draw[dashed,gray,opacity=0.35] (3,0.25) -- (3,2.85);
\draw[dp one,dp arrow] (0.4,2.2) -- (2.2,2.2);
\draw[dp edge] (2.2,2.2) -- (3.8,2.2) node[midway,above,dp label] {$e$};
\draw[dp one,dp arrow] (3.8,2.2) -- (5.6,2.2);
\draw[dp two,dp arrow] (0.4,0.9) -- (2.2,0.9);
\draw[dp edge] (2.2,0.9) -- (3.8,0.9) node[midway,above,dp label] {$f'$};
\draw[dp two,dp arrow] (3.8,0.9) -- (5.6,0.9);
\foreach \x in {0.4,2.2,3.8,5.6}{\fill (\x,2.2) circle (1.6pt);\fill (\x,0.9) circle (1.6pt);}
\node at (1.5,0) {$F_L^\sigma(\xi)\ne0$};
\node at (4.5,0) {$F_U^\sigma(\xi)\ne0$};
\end{scope}
\end{tikzpicture}
\caption{Why extraction selects a joint coefficient. (a) The coefficient matrix of a quadratic cut polynomial: both $e$ and $f$ occur, but $[z_ez_f]Q=0$. Selecting the highlighted coefficient instead gives the compatible choice $B=\{e,f'\}$. (b) Splitting at these edges, illustrated for a DAG cut. By \cref{lem:cut-factorization}, a nonzero coefficient has an assignment $\sigma$ with two nonzero child evaluations. The two child solutions can then be joined through the selected edges.}
\label{fig:extraction}
\end{figure}
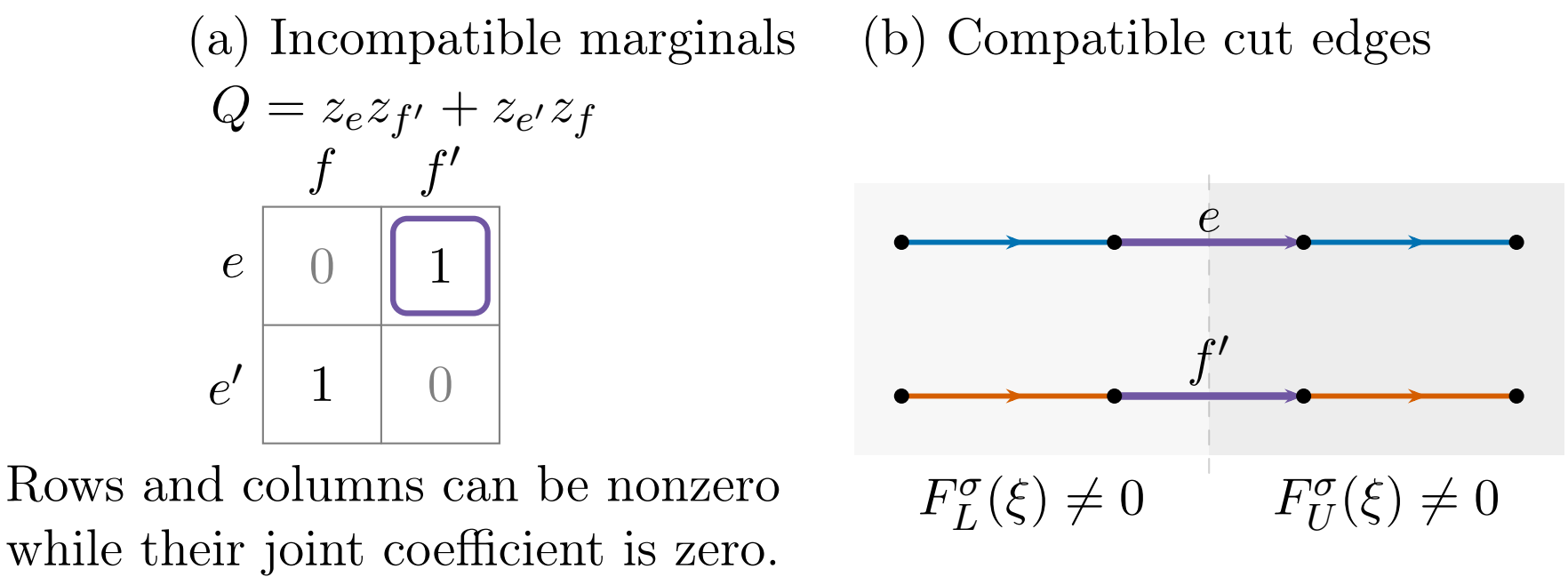

\paragraph{Correlated extraction.}
All first derivatives of a circuit can be evaluated with constant-factor overhead \cite{bs83}, and a nonzero derivative shows that an edge lies in some solution. It does not show which edges lie in a common solution: $z_ez_{f'}+z_{e'}z_f$ contains all four variables but has no monomial $z_ez_f$, so choosing $e$ and $f$ independently can commit to an impossible pair (\cref{fig:extraction}).

We avoid this at a \emph{monotone cut}, which every path crosses at most once: a topological prefix of a DAG, or a threshold of $\Delta\coloneqq d_1-d_2$ for two shortest paths in an undirected or directed graph. Fixing the noncut variables at the inherited evaluation point $\xi$ leaves a polynomial $Q$ in the cut variables that is homogeneous and multilinear of degree $j$, the number of paths whose terminals lie on opposite sides. Its coefficients factor across the cut: for every choice $B$ of $j$ cut edges,
\begin{align*}
[z^B]Q
&=\sum_\sigma F_L^\sigma(\xi)F_U^\sigma(\xi),
\end{align*}
where $\sigma$ ranges over the at most $j!$ assignments of the edges of $B$ to the crossing paths and $F_L^\sigma,F_U^\sigma$ are the polynomials of the two sides (\cref{lem:cut-factorization}). A nonzero coefficient thus yields an assignment whose two child evaluations are nonzero, and any two child solutions join through $B$ into a solution of the parent.

To find such a coefficient without expanding $Q$, we grow a set $A$ of selected variables. Setting every variable of $A$ to a common value $t$ makes the coefficient of the monomials containing $A$ the leading coefficient of a univariate restriction, and interpolating $|A|+1$ gradient evaluations reads it off for all candidate extensions of $A$ at once. After $j$ stages, $A=B$, using an expected $O(j^2)$ gradient evaluations (\cref{lem:interpolated-extraction}). For $j=2$, the coefficients form the Hessian of $Q$, and one gradient exposes a nonzero row; for larger $j$, repeated directional differentiation would introduce a factorial that vanishes in characteristic two, and interpolation avoids it. The temporary samples never replace $\xi$, so every child inherits a nonzero evaluation (\cref{lem:extraction-step}).

\paragraph{Paying for the recursion.}
Reporting cuts every region at a balanced monotone cut and continues on the children. For $k\geq3$ paths in a DAG, we cut at a vertex median and keep both children. The regions at each depth are disjoint and halve in size, so the superlinear circuit sizes sum geometrically over the recursion, and reporting costs a constant factor more than one evaluation (\cref{prop:k-reporting}).

The two-path circuits of \cite{avw24} have linear size, so keeping both children would cost an extra logarithmic factor. Instead we recover one side without recursion. Once two compatible cut edges are selected, compare the routes through them with the routes that exchange them. Shortestness prevents the exchange from shortening both sides. A strict comparison on one side lets a common-sink reduction recover that side by a flow of value two; equality puts the relevant pieces in a common tight DAG, where the algorithm of \cite{tho12} reports them. At most one two-path region remains, and median cuts, with a two-boundary treatment of equal $\Delta$-values, shrink it geometrically (\cref{sec:anchors,sec:linear}).

In a directed graph the exchanged routes need not exist, so we keep both children and speed up the detector instead. On the regional shortest-path supports, the cancellation recurrence of \cite{cks26} enforces its reachability restrictions automatically, through zero path sums, and becomes a triangular matrix computation: finalize the earlier half of the states, multiply their contributions into the later half, and recurse (\cref{lem:causal-recurrence,fig:causal-schedule}). This gives regional circuits of size $O(n_R(m_R+n_R))$ by a row schedule and $O(n_R^{\omega+\varepsilon})$ by fast matrix multiplication (\cref{lem:directed-circuit}).

At a threshold of $\Delta$, the first path crosses in the increasing direction and the second in the decreasing direction, so the assignment of the two selected edges is forced and the selected coefficient is a single product of child evaluations. The circuit sizes of the children contract by a constant factor at balanced cuts, and an equality plateau is resolved after at most two boundary cuts (\cref{lem:directed-cost}). Summing these sizes, together with the two source-distance computations, proves \cref{thm:two-directed-informal}.

\paragraph{Related work.} Generic ways to turn an algebraic detector into a search algorithm either isolate one solution or repeat the decision. The isolation lemma of \cite{mvv87} makes a minimum-weight solution unique under random weights, which their matching algorithm uses to read off its edges. \cite{bkk14} extract a witness of size $k$ with $O(k\log n)$ queries to an oracle that decides whether a given set contains a witness; since our solutions can have $\Theta(n)$ edges, its bound exceeds the detection time by a factor of $O(n)$. Our extraction step needs neither isolation nor repeated decisions: by \cref{lem:cut-factorization}, a nonzero coefficient at a monotone cut already determines compatible child instances, and the recursion then costs a constant factor more than one evaluation.

\section{Preliminaries}\label{sec:prelim}

Graphs have $n$ vertices and $m$ edges. A path contains its endpoints and repeats no vertex; for a vertex $v$, the zero-edge path $(v)$ is a path from $v$ to $v$. For a vertex set $R$, $G[R]$ is the induced subgraph, and $O_k$ hides factors depending only on $k$.

\subsection{Models of computation}

Real edge weights are handled by unit-cost exact additions, subtractions, and comparisons of weight values; the only exception is the undirected case of \cref{thm:two-linear}, where the weights are positive integers bounded by $(m+n)^{O(1)}$ and are stored in words. All other computation takes place on a word RAM with $\Theta(\log(m+n))$-bit words. Polynomial evaluations use a field $\F_{2^q}$ with $q=\Theta(\log(m+n))$, whose elements fit in a constant number of words; field operations and uniform sampling of field elements take unit time. The constant in $q$ depends on the error exponent.

\subsection{Circuits and polynomial identity testing}

An arithmetic circuit, or straight-line program, over a commutative ring $\Ring$ is \emph{division-free} if it uses only additions, subtractions, multiplications, inputs, and constants of $\Ring$. Its size counts gates and input occurrences. A circuit computes a formal polynomial in its inputs, and the identities of this paper are identities of formal polynomials.

The enumerating polynomials of this paper are tested at one random point; the extraction steps evaluate further temporary points, which affect only the running time. By the Schwartz--Zippel lemma, a nonzero polynomial of degree $D$ over a field $\F$ vanishes at a uniform point of $\F^N$ with probability at most $D/|\F|$. The extraction steps also differentiate circuits.

\begin{lemma}[{\cite{bs83}}]
\label{lem:reverse-mode}
Given a division-free circuit of size $S$ for a polynomial in $N$ variables and a point $z$, all $N$ first partial derivatives at $z$ can be computed with $O(S)$ operations and $O(S)$ space.
\end{lemma}

\subsection{Matrix multiplication}

Some of our circuits contain matrix products, and all of them must be division-free, as \cref{lem:reverse-mode} requires. We therefore let $\omega$ be the exponent of square matrix multiplication by bilinear algorithms with coefficients in $\F_2$. The bound $\omega<2.371177$ of \cite{adv+25,dek+26} applies to this definition: it is proved with tensors whose entries are $0$ and $1$, which can be considered over any field \cite[Section~3.1]{adv+25}. A bilinear algorithm with coefficients in $\F_2$ is a division-free straight-line program over every field $\F_{2^q}$. Hence, for every fixed $\varepsilon>0$, applying such an algorithm of exponent below $\omega+\varepsilon$ recursively multiplies two $h\times h$ matrices over $\F_{2^q}$ with $O(h^{\omega+\varepsilon})$ field operations, by a program constructed in time linear in its length. An $h\times p$ by $p\times h$ product, or an $h\times h$ by $h\times p$ product with $h\leq p$, splits into $\lceil p/h\rceil$ square products.

\section{Correlated extraction at monotone cuts}\label{sec:extraction}

A nonzero polynomial evaluation certifies existence; reporting requires compatible choices. We obtain them by selecting a nonzero coefficient at a cut crossed at most once by each path. The cut factorization below turns this coefficient into two child instances with nonzero inherited evaluations. The extraction algorithm finds it without expanding the polynomial and works over every field.

Throughout this section, $\Ring=\F$ is a field with unit-cost operations and uniform sampling, $G$ is a directed or undirected graph with vertex set $V$ and edge set $E$, and for each label $i\in[k]$, $H_i$ is a directed graph on $V$ whose arcs are orientations of edges of $G$; for directed $G$, the only orientation of an edge is its own direction, so $H_i\subseteq G$. Every edge $e$ of $G$ carries one variable $x_e$, shared by all its orientations and all graphs $H_i$. In the setting of \cref{thm:dag-k-paths}, $G$ is the input DAG and $H_1,\ldots,H_k$ are the commodity subgraphs; in \cref{sec:two-linear,sec:directed}, $k=2$ and $H_1,H_2$ are the tight graphs of the two shortest-path systems.

A \emph{regional instance} consists of a set $R\subseteq V$, a label set $I\subseteq[k]$, and terminals $a_i,b_i\in R$ for $i\in I$. Its \emph{solutions} are the tuples $(P_i)_{i\in I}$ of pairwise vertex-disjoint paths such that $P_i$ is an $a_i$--$b_i$ path in $H_i[R]$; a path contains its endpoints, so for $a_i=b_i$ the zero-edge path occupies $a_i$. The \emph{regional polynomial} is
\begin{align*}
F_R
&\coloneqq\sum_{(P_i)_{i\in I}}\prod_{i\in I}\prod_{e\in E(P_i)}x_e,
\end{align*}
where $e$ runs over the edges of $G$ underlying the arcs of $P_i$ and the sum ranges over the solutions. Let $n_R\coloneqq|R|$ and $m_R\coloneqq|E(G[R])|$. For $R=V$, $I=[k]$, and $(a_i,b_i)=(s_i,t_i)$, the regional polynomial is the polynomial $F$ of \cref{thm:dag-k-paths}.

\begin{lemma}[Witness monomials]
\label{lem:k-monomials}
For every regional instance, distinct solutions have distinct squarefree monomials of degree at most $\max\{n_R-1,0\}$, each with coefficient $1$. Hence $F_R$ is a nonzero polynomial exactly when a solution exists, and then $F_R(\xi)\neq0$ with probability at least $1-n_R/|\F|$ for a uniform $\xi\in\F^E$.
\end{lemma}
\begin{proof}
Vertex-disjoint paths share no edge, so the monomial of a solution is the product of the variables of the union of its edge sets, which is squarefree and has at most $\max\{n_R-1,0\}$ edges. If two labels share a terminal, no solution exists. Otherwise the union decomposes into vertex-disjoint paths of $G$. The endpoints of each such path with an edge are the terminals $a_i,b_i$ of a unique label $i$, which also fix its direction from $a_i$ to $b_i$, and every zero-edge path $(a_i)$ is forced. Hence the monomial determines the solution. The last assertion is the Schwartz--Zippel lemma.
\end{proof}

\paragraph{Monotone cuts.} A partition of $R$ into a \emph{low side} $L$ and a \emph{high side} $U$ is a \emph{monotone cut} of a regional instance if, for every $i\in I$, no path of $H_i[R]$ contains two edges between $L$ and $U$. The edges of $G[R]$ between $L$ and $U$ are the \emph{cut edges}; their variables are the \emph{cut variables}, written $z_e$, and all other variables of $F_R$ are \emph{noncut variables}. A label $i\in I$ is \emph{active} if $a_i$ and $b_i$ lie on different sides. Let $J$ be the set of active labels and put $j\coloneqq|J|$. A path whose endpoints lie on one side crosses between the sides an even number of times, and a path whose endpoints lie on different sides an odd number of times. Hence every $a_i$--$b_i$ path in $H_i[R]$ uses exactly one cut edge if $i$ is active and none otherwise.

Two kinds of monotone cuts occur. If $G$ is a DAG containing every $H_i$, every path visits its vertices in increasing topological order, so a topological prefix and its complement form a monotone cut; \cref{sec:k-reporting} cuts at vertex medians. If a vertex coordinate is nondecreasing or nonincreasing along every path of each $H_i$, a path cannot return across a threshold of that coordinate, so a threshold forms a monotone cut; \cref{sec:linear,sec:directed} cut at thresholds of the distance difference of the two shortest-path systems. In the second kind, different labels may cross in opposite directions.

For a set $B$ of $j$ cut edges, a bijection $\sigma\colon J\to B$ is \emph{feasible} if, for every $i\in J$, the edge $\sigma(i)$, oriented from the side of $a_i$ to the side of $b_i$, is an arc $(u_i,v_i)$ of $H_i$ whose underlying edge is $\sigma(i)$. Under a feasible $\sigma$, each active label $i$ has the pair $(a_i,u_i)$ on the side of $a_i$ and the pair $(v_i,b_i)$ on the side of $b_i$. The \emph{low child} is the regional instance on $L$ whose pairs are these pairs inside $L$ for $i\in J$ together with the pairs of the inactive labels with terminals in $L$; the \emph{high child} on $U$ is defined symmetrically. Let $F_L^\sigma$ and $F_U^\sigma$ be their polynomials.

The next lemma records the structure of $F_R$ at a monotone cut: after the noncut variables are fixed, it is a form of degree $j$ in the cut variables, and each coefficient is a sum of products of child evaluations.

\begin{lemma}[Cut factorization]
\label{lem:cut-factorization}
Let $(L,U)$ be a monotone cut of a regional instance $R$. Fix values $\xi_e$ of all noncut variables, and let $Q(z)$ be $F_R$ with the cut variables $z_e$ left free. Then $Q$ is homogeneous multilinear of degree $j$, and every collection $B$ of $j$ cut edges satisfies
\begin{align*}
[z^B]Q
&=\sum_\sigma F_L^\sigma(\xi)F_U^\sigma(\xi),
\end{align*}
where $\sigma$ ranges over the at most $j!$ feasible bijections $J\to B$. In particular, if $[z^B]Q\neq0$, then some feasible $\sigma$ has $F_L^\sigma(\xi)\neq0$ and $F_U^\sigma(\xi)\neq0$.
\end{lemma}
\begin{proof}
A solution of $R$ has exactly one cut edge on each active path and none on the other paths, and its paths share no edge. Its monomial is therefore a product of $j$ distinct cut variables and noncut variables. Since solutions contribute squarefree monomials, fixing the noncut variables and cancellation cannot create other cut monomials, even if a circuit for $F_R$ contains syntactic squares. This proves the first assertion.

A solution whose cut-edge set is $B$ determines $\sigma$ by letting $\sigma(i)$ be the cut edge of $P_i$. Since $P_i$ crosses once, it traverses $\sigma(i)$ from the side of $a_i$ to the side of $b_i$ as an arc $(u_i,v_i)$ of $H_i$, so $\sigma$ is feasible, the part of $P_i$ up to $u_i$ lies on the side of $a_i$, and the part from $v_i$ lies on the side of $b_i$. Every inactive path lies on one side. Hence splitting gives solutions of both children under $\sigma$. Conversely, joining solutions of the two children under one feasible $\sigma$ through the arcs $(u_i,v_i)$ of $H_i$ gives $a_i$--$b_i$ walks in $H_i[R]$ whose two pieces lie on different sides; these walks are paths. They are disjoint inside each side by the child conditions and across the sides because $L\cap U=\emptyset$, and their cut-edge set is $B$. The two maps are inverse. The edge sets of $G[L]$, $G[U]$, and the cut are disjoint, so the weight of the joined solution is $z^B$ times the weights of the two child solutions.

Evaluating the noncut variables at $\xi$ and summing over $\sigma$ gives the identity. A nonzero sum has a nonzero summand, and a nonzero product over a field has two nonzero factors.
\end{proof}

It remains to find a set $B$ with $[z^B]Q\neq0$ without expanding $Q$. Separate first derivatives do not suffice: for $Q=z_ez_{f'}+z_{e'}z_f$, every variable occurs in a monomial, but the coefficient of $z_ez_f$ is zero, so edges indicated by different derivatives need not belong to one solution. For $j=2$, the coefficients of $Q$ form its Hessian matrix, and a nonzero gradient entry at a random point exposes a nonzero row. The next lemma handles every $j$. It finds the variables of a nonzero monomial one at a time: setting the variables already chosen to a common value $t$ turns the coefficient sought into the leading coefficient of a polynomial in $t$, which a few gradient evaluations determine. Unlike repeated differentiation along a line, interpolation does not introduce the factor $r!$, which vanishes in characteristic two for $r\geq2$.

\begin{lemma}[Interpolated coefficient extraction]
\label{lem:interpolated-extraction}
Let $Q$ be a nonzero homogeneous multilinear polynomial of degree $j$ over a field $\F$ with $|\F|\geq2j$, computed by a division-free circuit of size $S$, counting gates and input occurrences. A set $B$ of $j$ variables whose monomial has a nonzero coefficient $c_B$ in $Q$ can be found, together with $c_B$, using expected $O((j^2+1)S)$ field operations and $O(S)$ space.
\end{lemma}
\begin{proof}
Expand $Q$ as $\sum_Mc_Mz^M$ over $j$-sets $M$ of variables, where $z^M\coloneqq\prod_{e\in M}z_e$. For a set $A$ of variables, put $Q_A\coloneqq\sum_{M\supseteq A}c_Mz^{M\setminus A}$. Then $Q_\emptyset=Q$, $Q_A$ is homogeneous multilinear of degree $j-|A|$ in the variables outside $A$, $\partial Q_A/\partial z_f=Q_{A\cup\{f\}}$ for $f\notin A$, and $Q_B=c_B$ when $|B|=j$. We maintain a set $A$ with $Q_A\neq0$, starting from $A=\emptyset$; if $j=0$, one evaluation returns $c_\emptyset=Q$. Consider a stage with $r\coloneqq|A|<j$.

\emph{Recovering candidate coefficients.} Fix distinct $\alpha_0,\ldots,\alpha_r\in\F$. For $f\notin A$, let $p_f(t)$ be $\partial Q/\partial z_f$ evaluated with every variable in $A$ set to $t$ and every other variable $f'$ set to $z_{f'}$. In $\partial Q/\partial z_f=\sum_{M\ni f}c_Mz^{M\setminus\{f\}}$, the term of $M$ has degree $|M\cap A|\leq r$ in $t$, with equality exactly when $A\subseteq M$. Hence $p_f$ has degree at most $r$, and Lagrange interpolation at $\alpha_0,\ldots,\alpha_r$ gives its leading coefficient:
\begin{align*}
Q_{A\cup\{f\}}(z)
&=[t^r]p_f(t)
=\sum_{h=0}^r\frac{p_f(\alpha_h)}{\prod_{g\neq h}(\alpha_h-\alpha_g)}.
\end{align*}
Reverse-mode differentiation evaluates all first partial derivatives of the circuit at one point with $O(S)$ operations and space \cite{bs83}. The point is the same for all $f$, so $r+1$ such evaluations, accumulated with the $O(r^2)$ interpolation weights, give $Q_{A\cup\{f\}}(z)$ for all $f\notin A$ with $O((r+1)S)$ operations. Since a degree-$j$ multilinear monomial uses $j$ distinct inputs, $j\leq S$, so the interpolation-weight computation is included in this bound.

\emph{Extending the selected set.} Sample the coordinates outside $A$ independently and uniformly. The polynomial $Q_A$ is nonzero, multilinear, and of positive degree, so some $Q_{A\cup\{f\}}$ is a nonzero polynomial of degree $j-r-1$. By the Schwartz--Zippel lemma, it vanishes at $z$ with probability at most $(j-r-1)/|\F|\leq1/2$.

The event that all computed values vanish is detected and triggers fresh sampling of $z$. Otherwise, add to $A$ any $f$ with $Q_{A\cup\{f\}}(z)\neq0$, which certifies $Q_{A\cup\{f\}}\neq0$. After $j$ stages, $A=B$ and the last stage computes $c_B\neq0$. Each stage needs an expected constant number of trials, and its gradient evaluations cost $O((r+1)S)$ operations with $O(S)$ space. The total expected cost is $O(\sum_{r<j}(r+1)S)=O(j^2S)$.
\end{proof}

The two lemmas combine into one step of every reporting recursion in this paper. The recursions differ only in how they choose cuts and in which children they keep.

\begin{lemma}[Correlated extraction step]
\label{lem:extraction-step}
Let $(L,U)$ be a monotone cut of a regional instance $R$, let $\xi\in\F^E$ satisfy $F_R(\xi)\neq0$, and suppose that $F_R$ has a division-free circuit of size $S$ and $|\F|\geq2j$. A set $B$ of $j$ cut edges and a feasible bijection $\sigma\colon J\to B$ with $F_L^\sigma(\xi)\neq0$ and $F_U^\sigma(\xi)\neq0$ can be found using expected $O((j^2+1)S)$ field operations and $O(S)$ space, plus the evaluation at $\xi$ of the child polynomials for the at most $j!$ feasible bijections. The computation never changes $\xi$.
\end{lemma}
\begin{proof}
Substituting $\xi$ for the noncut inputs of the circuit of $F_R$ gives a circuit of size at most $S$ for the polynomial $Q$ of \cref{lem:cut-factorization}. Its value at the cut coordinates of $\xi$ is $F_R(\xi)\neq0$, so $Q\neq0$, and \cref{lem:cut-factorization} shows that $Q$ is homogeneous multilinear of degree $j$. \Cref{lem:interpolated-extraction} returns a set $B$ with $[z^B]Q\neq0$; it samples only temporary values of the cut variables. By \cref{lem:cut-factorization}, some feasible $\sigma$ has two nonzero child evaluations at $\xi$, and evaluating the children for every feasible $\sigma$ finds one.
\end{proof}

\section{Disjoint paths in DAGs}\label{sec:dag}

The main algebraic result computes the enumerating polynomial of disjoint path tuples in commodity subgraphs of one DAG. Tight shortest-path subgraphs give $k$-DSP; taking every commodity subgraph to be the input graph gives $k$-DP.

\begin{theorem}
\label{thm:dag-k-paths}
Fix an integer $k\geq3$. Let $G$ be a DAG with $n$ vertices, $m$ edges, and a given topological order, let $H_1,\ldots,H_k$ be spanning edge-subgraphs of $G$, and let $(s_1,t_1),\ldots,(s_k,t_k)$ be vertex pairs, where $s_i=t_i$ is allowed. Assign an element $x_e$ of a commutative ring $\Ring$ to every edge $e$ of $G$, and let
\begin{align*}
F
&\coloneqq\sum_{(P_1,\ldots,P_k)}\prod_{i=1}^k\prod_{e\in E(P_i)}x_e,
\end{align*}
where the sum ranges over all tuples of pairwise vertex-disjoint paths such that $P_i$ is an $s_i$--$t_i$ path in $H_i$. With unit-cost ring operations, $F$ is computed by a division-free straight-line program with $O_k(mn^{k-2}+n^k)$ operations, which is constructed in time linear in its length and executed in $O_k(n^{k-1})$ space. If $h\times h$ matrices over $\Ring$ can be multiplied by a division-free straight-line program that is constructible in time linear in its length and uses $O(h^\alpha)$ operations for some $\alpha>2$, then $O_k(m+n^{k-3+\alpha})$ operations suffice.
\end{theorem}
\begin{proof}
\label{proof:dag-k-paths}
\Cref{lem:first-event} gives the first-collision recurrences. \Cref{prop:k-sweep} evaluates them with the general operation and space bounds, and \cref{prop:three-batched,prop:k-batched} give the matrix multiplication bounds for $k=3$ and for $k\geq4$, respectively.
\end{proof}

The construction has four parts: the first-collision identity, its topological evaluation, and batched evaluations for three pairs and for every fixed $k\geq4$. \Cref{sec:k-reporting} then recovers paths from a nonzero evaluation and proves \cref{thm:dag-k-dsp}.

\subsection{The first-event decomposition}\label{sec:first-event}

We work in the setting of \cref{thm:dag-k-paths}: $G=(V,E)$ is a DAG with an injective topological order $<$, $H_1,\ldots,H_k$ are spanning edge-subgraphs of $G$, and every edge $e$ carries one element $x_e$ of a commutative ring $\Ring$, shared by all subgraphs containing $e$. If two pairs share a terminal, no tuple exists and $F=0$. The weight of a path $P$ is $\wt(P)\coloneqq\prod_{e\in E(P)}x_e$, and the weight of a tuple of paths is the product of their weights. This convention loses no generality for subgraphs with smaller vertex sets: if a prescribed terminal is absent from its commodity subgraph, return $F=0$; otherwise add all missing vertices as isolated vertices.

\paragraph{Private terminals and path sums.} Private terminals make every path start before and end after all real vertices. Then every path avoiding a real vertex $w$ has a last vertex before $w$, and $F$ becomes the collision-free part of a product of path sums. Give commodity $i$ a private source $\sigma_i$ and a private sink $\theta_i$, and let $H_i^+$ consist of $H_i$ and the edges $\sigma_i\to s_i$ and $t_i\to\theta_i$, whose weights are the constant $1$. Extend $<$ so that all private sources precede $V$ and all private sinks follow $V$. Then $P\mapsto\sigma_iP\theta_i$ is a weight-preserving bijection from the $s_i$--$t_i$ paths in $H_i$ to the $\sigma_i$--$\theta_i$ paths in $H_i^+$, and every vertex lying in two of the graphs $H_1^+,\ldots,H_k^+$ is real. Every edge increases the order, so every walk is a path and visits its vertices in increasing order. For vertices $a,b$ of $H_i^+$, let $N_i(a,b)$ be the total weight of the $a$--$b$ paths in $H_i^+$; thus $N_i(a,a)=1$, and $N_i(a,b)=0$ unless $a\leq b$. Put $L_i(b)\coloneqq N_i(\sigma_i,b)$.

\paragraph{Crossing kernels.} The continuation of a path that avoids a collision $w$ is counted by a kernel. For a real vertex $w$, a vertex $c<w$ of $H_i^+$, and a vertex $e$ of $H_i^+$, let the \emph{crossing kernel} $K_i^w(c,e)$ be the total weight of the $c$--$e$ paths in $H_i^+$ whose vertices other than $c$ all follow $w$. If $e<w$, the endpoint $e$ does not follow $w$, so such a path is trivial and $e=c$. If $e=w$, no such path exists, because $e$ neither equals $c<w$ nor follows $w$. Otherwise the path begins with an edge jumping over $w$. Hence
\begin{align*}
K_i^w(c,e)
&=
\begin{cases}
\mathbf 1[c=e], & e<w,\\
0, & e=w,\\
\sum_{\substack{(c,d)\in E(H_i^+)\\d>w}}x_{cd}N_i(d,e), & e>w.
\end{cases}
\end{align*}

\paragraph{Prefix tables.} The collision-free prefixes are counted by the following tables. For $S\subseteq[k]$ with $|S|\geq2$, a real vertex $v$, and a tuple $c=(c_i)_{i\notin S}$ of vertices $c_i<v$ of $H_i^+$, let $\Lambda_S(v;c)$ be the total weight of the tuples $(A_1,\ldots,A_k)$ such that $A_i$ is a $\sigma_i$--$v$ path in $H_i^+$ for $i\in S$, $A_i$ is a $\sigma_i$--$c_i$ path in $H_i^+$ for $i\notin S$, and two paths $A_i,A_j$ intersect exactly in $\{v\}$ when $i,j\in S$ and are disjoint otherwise. The table records only the meeting vertex $v$ and the current endpoint of every other path. It has at most $(n+2k)^{k-|S|+1}$ entries, and $\Lambda_{[k]}(v)$ has no second argument.

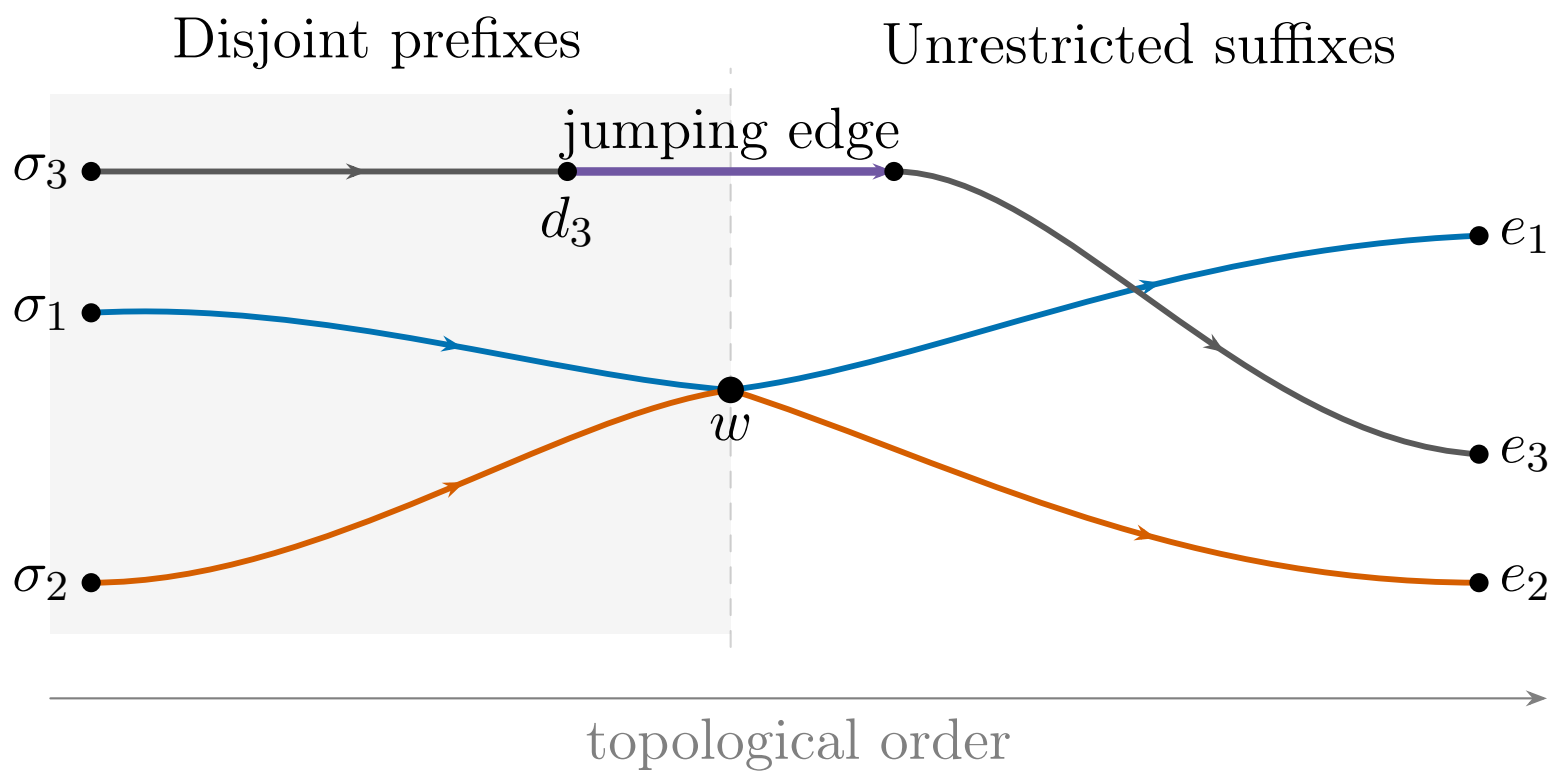
\begin{figure}[t]
\centering
\begin{tikzpicture}[x=0.9cm,y=0.85cm,font=\normalsize,line cap=round]
\fill[black!4] (0,-0.4) rectangle (5,3.8);
\draw[dashed,gray,opacity=0.35] (5,-0.5) -- (5,4);
\node at (2.4,4.2) {Disjoint prefixes};
\node at (8,4.2) {Unrestricted suffixes};
\draw[-{Stealth[length=4pt]},gray] (0,-0.9) -- (11,-0.9) node[midway,below] {topological order};
\draw[dp one,dp arrow] (0.3,2.1) .. controls (2,2.2) and (3.7,1.6) .. (5,1.5);
\draw[dp two,dp arrow] (0.3,0) .. controls (2,0) and (3.7,1.3) .. (5,1.5);
\draw[dp one,dp arrow] (5,1.5) .. controls (6.5,1.7) and (8.2,2.6) .. (10.5,2.7);
\draw[dp two,dp arrow] (5,1.5) .. controls (6.5,1) and (8.2,0) .. (10.5,0);
\draw[draw=black!65,line width=1.1pt,dp arrow] (0.3,3.2) -- (3.8,3.2);
\draw[dp edge] (3.8,3.2) -- (6.2,3.2) node[midway,above,dp label] {jumping edge};
\draw[draw=black!65,line width=1.1pt,dp arrow] (6.2,3.2) .. controls (7.4,3.2) and (8.8,1.1) .. (10.5,1);
\foreach \x/\y/\lab in {0.3/2.1/\sigma_1,0.3/0/\sigma_2,0.3/3.2/\sigma_3}{\fill (\x,\y) circle (1.8pt) node[left] {$\lab$};}
\fill (5,1.5) circle (2.5pt) node[below=3pt,dp label] {$w$};
\fill (3.8,3.2) circle (1.8pt) node[below=3pt,dp label] {$d_3$};
\fill (6.2,3.2) circle (1.8pt);
\foreach \x/\y/\lab in {10.5/2.7/e_1,10.5/0/e_2,10.5/1/e_3}{\fill (\x,\y) circle (1.8pt) node[right] {$\lab$};}
\end{tikzpicture}
\caption{First-collision decomposition for $k=3$ and $T=\{1,2\}$. The prefix table $\Lambda_T(w;d_3)$ stores only the meeting vertex $w$ and the last vertex $d_3<w$ of path $3$. Its continuation must jump over $w$; the three suffixes are counted independently by $N_1(w,e_1)$, $N_2(w,e_2)$, and $K_3^w(d_3,e_3)$. Curves crossing away from a marked vertex do not indicate a graph intersection.}
\label{fig:first-event}
\end{figure}

\paragraph{Extension across a collision.} Applying the kernels to a table row extends the paths that avoid $w$ across $w$, and multiplying by path sums from $w$ extends the paths through $w$ (\cref{fig:first-event}). For a tuple $e=(e_1,\ldots,e_k)$ of vertices of $H_1^+,\ldots,H_k^+$, let $\Omega(e)$ be the set of tuples $A$ of paths $A_1,\ldots,A_k$ such that $A_i$ is a $\sigma_i$--$e_i$ path in $H_i^+$. By distributivity, the total weight of $\Omega(e)$ is $\prod_iL_i(e_i)$. A \emph{collision} of $A$ is a vertex lying on at least two of its paths; collisions are real. For a real vertex $w$ and $T\subseteq[k]$ with $|T|\geq2$, write $e_{-T}\coloneqq(e_i)_{i\notin T}$ and define
\begin{align*}
Y_T^w(e_{-T})
&\coloneqq
\sum_{d}\Lambda_T(w;d)\prod_{i\notin T}K_i^w(d_i,e_i),
\\
B_w(e)
&\coloneqq
\sum_{\substack{T\subseteq[k]\\|T|\geq2}}Y_T^w(e_{-T})\prod_{i\in T}N_i(w,e_i),
\end{align*}
where $d$ ranges over tuples $(d_i)_{i\notin T}$ of vertices $d_i<w$ of $H_i^+$. In particular, $Y_{[k]}^w=\Lambda_{[k]}(w)$. In the terms of the overview, $Y_T^w$ attaches the continuations of the paths avoiding $w$ to a row of prefixes, and $B_w(e)$ also attaches the continuations of the paths through $w$ and sums over the label set $T$.

The next lemma makes the overview precise: $B_w(e)$ is the weight of the tuples whose first collision is $w$. Subtracting these tuples from all tuples gives recurrences in which each table row depends only on rows with earlier meeting vertices.

\begin{lemma}[First-event decomposition]
\label{lem:first-event}
For every endpoint tuple $e$ and real vertex $w$, the tuples $A\in\Omega(e)$ whose earliest collision is $w$ have total weight $B_w(e)$. Consequently, for $S,v,c$ as in the definition of $\Lambda_S$, and for $e_i\coloneqq v$ when $i\in S$ and $e_i\coloneqq c_i$ otherwise,
\begin{align}
\Lambda_S(v;c)
&=\prod_{i=1}^kL_i(e_i)-\sum_{w\in V:w<v}B_w(e).
\label{eq:table-recurrence}
\end{align}
Moreover, with $\theta\coloneqq(\theta_1,\ldots,\theta_k)$,
\begin{align}
F
&=\prod_{i=1}^kL_i(\theta_i)-\sum_{w\in V}B_w(\theta).
\label{eq:F-recurrence}
\end{align}
These identities hold over every commutative ring.
\end{lemma}
\begin{proof}
Let $A\in\Omega(e)$ have earliest collision $w$, and let $T$ be the labels of the paths through $w$. For $i\in T$, split $A_i$ at $w$. For $i\notin T$, split it at its last vertex $d_i<w$, which exists because its private source precedes every real vertex. No two prefixes meet before $w$, so the prefix tuple is counted by $\Lambda_T(w;d)$. The suffix of a label in $T$ is counted by $N_i(w,e_i)$; every vertex of any other suffix after $d_i$ follows $w$, so that suffix is counted by $K_i^w(d_i,e_i)$. The edge weights multiply to the weight of $A$.

Conversely, take a prefix tuple counted by $\Lambda_T(w;d)$ and suffixes counted by these factors. Concatenation gives paths because every edge increases the topological order. All vertices before $w$ belong to prefixes, which are disjoint there. The paths through $w$ have exactly the labels $T$, and for $i\notin T$ the last vertex before $w$ is exactly $d_i$. Thus the earliest collision is $w$, and splitting recovers the chosen pieces. This weight-preserving bijection proves that their total weight is
\begin{align*}
B_w(e)
&=\sum_{\substack{T\subseteq[k]\\|T|\geq2}}\sum_d
\Lambda_T(w;d)\prod_{i\in T}N_i(w,e_i)\prod_{i\notin T}K_i^w(d_i,e_i).
\end{align*}

For the endpoints defining $\Lambda_S(v;c)$, every vertex is at most $v$, and exactly the labels in $S$ end at $v$. Hence a tuple satisfies the table's intersection conditions exactly when it has no collision before $v$. Each excluded tuple has one earliest collision $w<v$; subtracting their weights from $\prod_iL_i(e_i)$ gives \eqref{eq:table-recurrence}. At the private sinks $\theta$, the desired tuples have no collision at all, so subtracting all first-collision classes gives \eqref{eq:F-recurrence}. These are partitions of finite sets of weighted tuples and therefore hold over every commutative ring.
\end{proof}

\subsection{The row sweep}\label{sec:sweep}

\Cref{lem:first-event} computes each row $\Lambda_S(v;\cdot)$ from rows with earlier meeting vertices, so the rows can be finalized in topological order. \Cref{alg:row-sweep} does this. When it reaches $w$, all rows with meeting vertex $w$ are final; it extends them across $w$ to the tensors $Y_T^w$, which costs one path-sum pass per fiber, and then adds $B_w$ to every later table entry. For a table entry with label set $S$, meeting vertex $v$, and endpoints $c$, we write $e(S,v,c)$ for the endpoint tuple with $e_i=v$ for $i\in S$ and $e_i=c_i$ otherwise.

\begin{algorithm}[htb]
\caption{Row sweep}
\label{alg:row-sweep}
\textbf{Input:} the extended graphs $H_1^+,\ldots,H_k^+$ with the order $<$ and the edge values $x_e$. \textbf{Output:} $F$.
\begin{itemize}
\item[\textbf{1.}] Compute the path sums $N_i(a,b)$ for all labels $i$ and vertices $a\leq b$, and put $L_i(b)\coloneqq N_i(\sigma_i,b)$.
\item[\textbf{2.}] Give every table entry $\Lambda_S(v;c)$ an accumulator $A_S(v;c)\coloneqq0$, and put $\Sigma\coloneqq0$.
\item[\textbf{3.}] For every real vertex $w$, in increasing order:
\begin{itemize}
\item[(a)] for every $S$ and $c$, set $\Lambda_S(w;c)\coloneqq\prod_iL_i(e_i)-A_S(w;c)$, where $e=e(S,w,c)$;
\item[(b)] for every $T\neq[k]$, compute $Y_T^w$ by applying the kernel transform of $K_i^w$ to the row $\Lambda_T(w;\cdot)$ along each coordinate $i\notin T$, and put $Y_{[k]}^w\coloneqq\Lambda_{[k]}(w)$;
\item[(c)] for every table entry with meeting vertex $v>w$, add $B_w(e(S,v,c))$ to $A_S(v;c)$, and add $B_w(\theta)$ to $\Sigma$.
\end{itemize}
\item[\textbf{4.}] Return $\prod_iL_i(\theta_i)-\Sigma$.
\end{itemize}
\end{algorithm}

\begin{proposition}[Row sweep]
\label{prop:k-sweep}
In the setting of \cref{thm:dag-k-paths}, a division-free straight-line program with $O_k(mn^{k-2}+n^k)$ ring operations computes $F$. The program is constructed in time linear in its length and can be executed in $O_k(n^{k-1})$ space.
\end{proposition}
\begin{proof}
The program is \cref{alg:row-sweep}. Every operation in it is an addition, subtraction, or multiplication whose pattern depends only on $G$, the subgraphs, the terminals, and $<$, so it is a straight-line program, constructed in time linear in its length. Since $k$ is fixed, the $2k$ private vertices and edges change no bound, and we use $n$ and $m$ also for the sizes of the extended graph. We first show that the algorithm returns $F$, and then bound the cost of each step.

When $w$ is reached, every $w'<w$ has added $B_{w'}(e)$ to all entries with a later meeting vertex in step~3(c). Hence the accumulator of every entry with meeting vertex $w$ equals
\begin{align*}
A_S(w;c)
&=\sum_{w'<w}B_{w'}\bigl(e(S,w,c)\bigr),
\end{align*}
and by \eqref{eq:table-recurrence}, step~3(a) sets $\Lambda_S(w;c)$ to its correct value. Since $B_w$ reads table entries only through the row of $w$, every value is final when it is used, and at the end $\Sigma=\sum_{w\in V}B_w(\theta)$; by \eqref{eq:F-recurrence}, step~4 returns $F$.

Next we bound the path sums of step~1. For fixed $i$ and $a$, the values $N_i(a,b)$ follow from $N_i(a,a)=1$ and
\begin{align*}
N_i(a,b)
&=\sum_{(p,b)\in E(H_i^+)}N_i(a,p)x_{pb}\qquad(b>a),
\end{align*}
evaluated in increasing order of $b$; this classifies the $a$--$b$ paths by their last edge $(p,b)$, whose tail precedes $b$. One pair $(i,a)$ costs $O(m+n)$ operations, so step~1 costs $O_k(n(m+n))$ operations and $O_k(n^2)$ space.

We then show that one kernel transform costs $O(m+n)$ operations. Fix $w\in V$, a label $i$, and a vector $f$ indexed by the vertices $d<w$ of $H_i^+$, and let $g(e)\coloneqq\sum_{d<w}f(d)K_i^w(d,e)$. By the case formula for $K_i^w$, $g(e)=f(e)$ for $e<w$, $g(w)=0$, and for $e>w$
\begin{align*}
g(e)
&=\sum_{d'>w}y(d')N_i(d',e),
\\
y(d')
&\coloneqq\sum_{\substack{(d,d')\in E(H_i^+)\\d<w}}f(d)x_{dd'}.
\end{align*}
One scan of $E(H_i^+)$ computes $y$, and the recurrence
\begin{align*}
g(e)
&=y(e)+\sum_{\substack{(p,e)\in E(H_i^+)\\p>w}}g(p)x_{pe},
\end{align*}
evaluated in increasing order of $e>w$, gives the displayed value: $y(e)$ accounts for the trivial path from $e$, and the term of an edge $(p,e)$ accounts for the nontrivial paths with last edge $(p,e)$, whose first vertices $d'$ satisfy $w<d'\leq p$.

The tensor $Y_T^w$ in step~3(b) has $r\coloneqq k-|T|\leq k-2$ free coordinates. By distributivity, applying the kernel transform of $K_i^w$ along each coordinate $i\notin T$ in turn turns the row $d\mapsto\Lambda_T(w;d)$ into the sum defining $Y_T^w$, and one coordinate pass transforms at most $n^{r-1}$ one-dimensional fibers. Summing over the $n$ vertices $w$ and the $O(2^k)$ sets $T$, step~3(b) costs
\begin{align*}
\sum_{w\in V}\sum_{T\neq[k]}O\bigl(n^{k-|T|-1}(m+n)\bigr)
&=O_k\bigl(mn^{k-2}+n^{k-1}\bigr).
\end{align*}
Given the tensors and the path sums, each value $B_w(e)$ costs $O_k(1)$ operations, so steps~3(a) and~3(c) cost $O_k(n^{k-1})$ per vertex $w$, as there are $O_k(n^{k-1})$ table entries. Altogether, using $k\geq3$,
\begin{align*}
O_k\bigl(n(m+n)\bigr)+O_k\bigl(mn^{k-2}+n^{k-1}\bigr)+O_k\bigl(n\cdot n^{k-1}\bigr)
&=O_k\bigl(mn^{k-2}+n^k\bigr).
\end{align*}
The path sums, tables, and accumulators occupy $O_k(n^{k-1})$ cells, using $n^2\leq n^{k-1}$; the tensors $Y_T^w$ occupy $O_k(n^{k-2})$ cells and are discarded after $w$.
\end{proof}

\subsection{Three pairs in matrix multiplication time}\label{sec:three}

For $k=3$, \cref{prop:k-sweep} gives $O(nm+n^3)$ operations. Its $nm$ term comes from the per-row kernel transforms and its $n^3$ term from the updates; both become matrix products. The updates become products because each row changes a pair table by two outer products, and the kernel transforms become products because each crossing vector is a row of a product of a pair table with adjacency and path-sum matrices. A divide-and-conquer schedule over the order $<$ forms each such product once all rows it reads are final.

\paragraph{Updates as outer products.} Write $\Lambda_{ij}\coloneqq\Lambda_{\{i,j\}}$ and $Y_{ij}^w\coloneqq Y_{\{i,j\}}^w$, and let $\ell$ denote the third label of a pair $\{i,j\}$. For $v>w$ and $c<v$, take $e_i=e_j=v$ and $e_\ell=c$. The four event sets $T\in\{[3],\{i,j\},\{i,\ell\},\{j,\ell\}\}$ contribute the four terms of
\begin{align*}
B_w(e)
&=\Lambda_{[3]}(w)N_i(w,v)N_j(w,v)N_\ell(w,c)+Y_{ij}^w(c)N_i(w,v)N_j(w,v)
\\
&\quad+Y_{i\ell}^w(v)N_i(w,v)N_\ell(w,c)+Y_{j\ell}^w(v)N_j(w,v)N_\ell(w,c).
\end{align*}
The first, third, and fourth sets contain $\ell$ and contribute the factor $N_\ell(w,c)$. Hence
\begin{align*}
B_w(e)
&=\alpha_{ij}^w(v)N_\ell(w,c)+N_i(w,v)N_j(w,v)Y_{ij}^w(c),
\\
\alpha_{ij}^w(v)
&\coloneqq N_i(w,v)N_j(w,v)\Lambda_{[3]}(w)+N_i(w,v)Y_{i\ell}^w(v)+N_j(w,v)Y_{j\ell}^w(v).
\end{align*}
Thus row $w$ changes the table $\Lambda_{ij}$ by the sum of two outer products of vectors indexed by $v$ and by $c$.

\paragraph{Crossing vectors as products.} Index matrices by the vertices of the extended graph in the order $<$. Let $X_i$ be the weighted adjacency matrix of $H_i^+$, whose $(a,b)$ entry sums $x_e$ over the parallel edges $e=(a,b)$ of $H_i^+$, and let $N_i$ be the matrix of path sums. For a pair $T=\{i,j\}$ with third label $\ell$, let $\Lambda_T$ be the matrix with entries $\Lambda_T[w,c]\coloneqq\Lambda_T(w;c)$ for real $w$ and $c<w$, and zero otherwise. Define
\begin{align*}
Z_T[w,d]
&\coloneqq\mathbf 1[d>w]\sum_{c}\Lambda_T[w,c]X_\ell[c,d],\\
\Phi_T[w,e]
&\coloneqq\sum_{d}Z_T[w,d]N_\ell[d,e].
\end{align*}
Comparing with the case formula for $K_\ell^w$, $Y_T^w(e)$ equals $\Lambda_T[w,e]$ for $e<w$, $0$ for $e=w$, and $\Phi_T[w,e]$ for $e>w$, because then
\begin{align*}
Y_T^w(e)
&=\sum_{c<w}\Lambda_T[w,c]\sum_{d>w}X_\ell[c,d]N_\ell[d,e]
=\sum_dZ_T[w,d]N_\ell[d,e].
\end{align*}
Moreover $\Phi_T[w,e]=0$ for $e\leq w$, since $Z_T[w,d]=0$ unless $d>w$ and $N_\ell[d,e]=0$ unless $d\leq e$.

\begin{figure}[t]
\centering
\begin{tikzpicture}[x=0.7cm,y=0.7cm,font=\normalsize]
\fill[black!3] (0,0) rectangle (4,4);
\fill[DPPurple!12] (0,0) rectangle (4,2);
\fill[DPBlue!30] (0,4) -- (0,2) -- (2,2) -- cycle;
\fill[DPPurple!40] (0,2) -- (0,0) -- (4,0) -- (2,2) -- cycle;
\draw (0,0) rectangle (4,4);
\draw[gray] (0,2) -- (4,2);
\draw[densely dashed,gray] (0,4) -- (4,0);
\node[left] at (0,3) {$I_1$};
\node[left] at (0,1) {$I_2$};
\node[above] at (2,4) {column $c$};
\node[rotate=90] at (-1.2,2) {row $v$};
\node at (2,5.2) {(a) Pair table $\Lambda_{ij}$};
\fill[DPPurple!25] (6,0) rectangle (10,2);
\draw (6,0) rectangle (10,2);
\draw[gray] (8,0) -- (8,2);
\node at (7,1) {$P_1$};
\node at (9,1) {$P_2$};
\node[left] at (6,1) {$I_2$};
\node[above] at (7,2) {$I_1$};
\node[above] at (9,2) {$I_1$};
\node at (10.8,1) {$\times$};
\fill[DPBlue!18] (11.6,-1) rectangle (15.6,3);
\draw (11.6,-1) rectangle (15.6,3);
\draw[gray] (11.6,1) -- (15.6,1);
\node at (13.6,2) {$Q_1$};
\node at (13.6,0) {$Q_2$};
\node[left] at (11.6,2) {$I_1$};
\node[left] at (11.6,0) {$I_1$};
\node[above] at (13.6,3) {column $c$};
\node at (10.8,5.2) {(b) Batched update};
\end{tikzpicture}
\caption{The cross update of \cref{alg:three-pairs} for a pair table $\Lambda_{ij}$ at a call on $I=I_1\sqcup I_2$. (a) Rows $v$ and columns $c$ of $\Lambda_{ij}$ in the order $<$; only entries below the dashed diagonal, with $c<v$, are read. After the call on $I_1$, the rows in $I_1$ are final (blue). Each final row $w$ changes every later row by two outer products, so all updates from $I_1$ to the rows in $I_2$ (purple) are known at once; the entries with $c\geq v$ (light) are filled but never read. (b) The sum of these updates is one product of an $|I_2|\times2|I_1|$ and a $2|I_1|\times n$ matrix. Column $w$ of $P_1$ and $P_2$ holds the row factors $\alpha_{ij}^w$ and $N_i(w,\cdot)N_j(w,\cdot)$, and row $w$ of $Q_1$ and $Q_2$ holds the column factors $N_\ell(w,\cdot)$ and $Y_{ij}^w$. The updates within $I_1$ and within $I_2$ are made by the two recursive calls.}
\label{fig:three-schedule}
\end{figure}
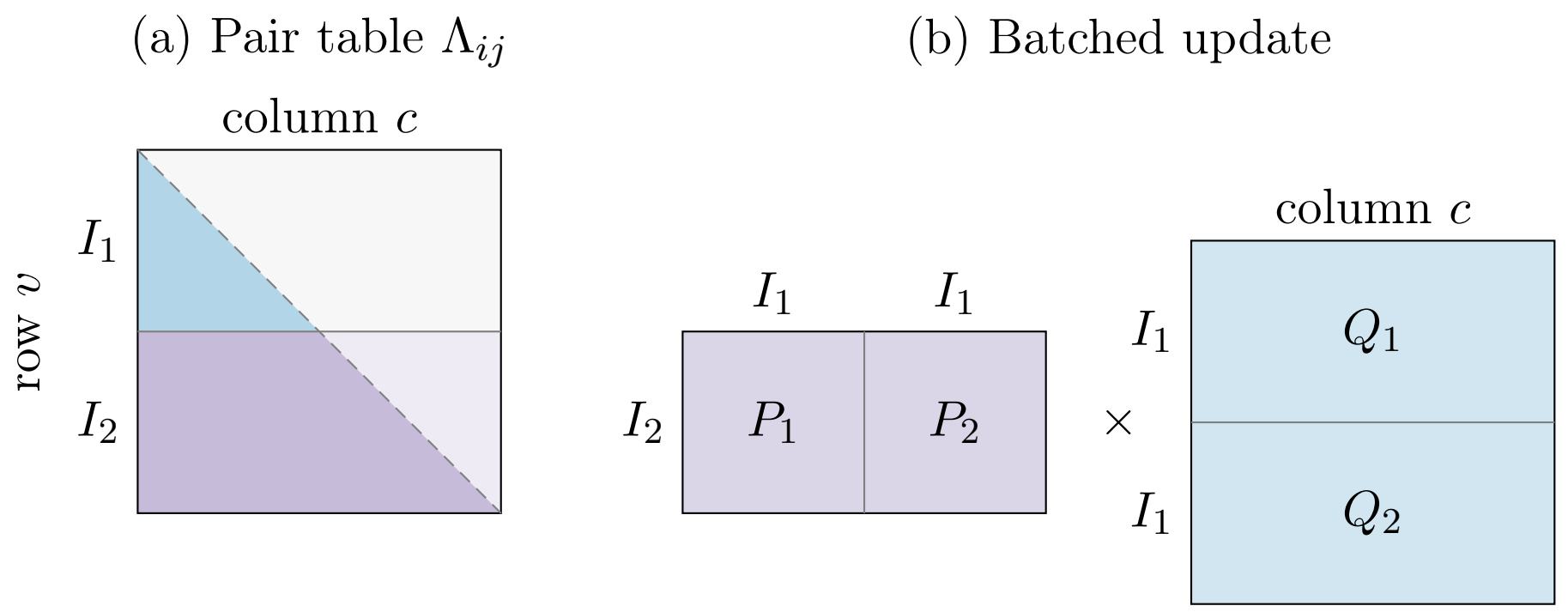

\paragraph{Batching the updates.} These products cannot be formed all at once, because the rows depend on each other: row $w$ of a pair table is final only after every earlier row has sent its update, and the update sent by row $w$ uses the crossing vectors of row $w$. The row sweep respects this dependence by processing one row at a time, which leaves no room for fast matrix multiplication. \Cref{alg:three-pairs} batches the updates by divide and conquer over the order $<$ (\cref{fig:three-schedule}). Once the rows of the first half $I_1$ of an interval are final, the updates that they send to the rows of the second half $I_2$ are known together, and by the outer-product form their sum is a single matrix product. The crossing vectors that these updates read are again products of final blocks. Every ordered pair $w<v$ is separated by exactly one call, so every update is sent exactly once, and since $\alpha>2$ the cost is dominated by the top level of the recursion.

The schedule \cref{alg:three-pairs} sweeps the real vertices $v_1<\cdots<v_n$ by divide and conquer. It keeps accumulators for all entries $\Lambda_{[3]}(v)$ and $\Lambda_{ij}(v;c)$, and the matrices $Z_T$ and $\Phi_T$ restricted to the real vertices. Its cross updates use, for a target pair $\{i,j\}$ with third label $\ell$, the factor matrices
\begin{align*}
P_1[v,w]
&\coloneqq\alpha_{ij}^w(v),
\\
Q_1[w,c]
&\coloneqq N_\ell(w,c),
\\
P_2[v,w]
&\coloneqq N_i(w,v)N_j(w,v),
\\
Q_2[w,c]
&\coloneqq Y_{ij}^w(c).
\end{align*}
A call on an interval $I=[v_l,v_r)$ of real vertices assumes that the accumulators of every $v\in I$ contain the contributions $B_w$ of all real $w<v_l$. It finalizes all rows $w\in I$ and computes $Z_T[w,d]$ and $\Phi_T[w,d]$ for all $w<d$ in $I$.

\begin{algorithm}[htb]
\caption{\textsc{Solve}$(I)$ for three pairs, on an interval $I=[v_l,v_r)$ of real vertices}
\label{alg:three-pairs}
\begin{itemize}
\item[\textbf{1.}] If $I=\{w\}$, finalize the row of $w$ by \eqref{eq:table-recurrence} and return.
\item[\textbf{2.}] Split $I$ into a first half $I_1=[v_l,v_b)$ and a second half $I_2=[v_b,v_r)$, and call \textsc{Solve}$(I_1)$.
\item[\textbf{3.}] For each pair $T=\{i,j\}$ with third label $\ell$, compute the crossing blocks
\begin{align*}
Z_T[I_1,I_2]
&=\Lambda_T[I_1,\{c:c<v_b\}]X_\ell[\{c:c<v_b\},I_2],
\\
\Phi_T[I_1,I_2]
&=Z_T[I_1,I_1]N_\ell[I_1,I_2]+Z_T[I_1,I_2]N_\ell[I_2,I_2].
\end{align*}
\item[\textbf{4.}] For each target pair $\{i,j\}$, add $P_1[I_2,I_1]Q_1[I_1,\cdot]+P_2[I_2,I_1]Q_2[I_1,\cdot]$ to the accumulators of the entries $\Lambda_{ij}(v;c)$ with $v\in I_2$ and $c<v_r$, and add $B_w(v,v,v)$ to the accumulator of $\Lambda_{[3]}(v)$ for all $w\in I_1$ and $v\in I_2$.
\item[\textbf{5.}] Call \textsc{Solve}$(I_2)$.
\end{itemize}
\end{algorithm}

\begin{proposition}[Three pairs]
\label{prop:three-batched}
Suppose that $k=3$ and that $h\times h$ matrices over $\Ring$ can be multiplied by a division-free straight-line program that is constructible in time linear in its length and uses $O(h^\alpha)$ operations for some $\alpha>2$. Then a division-free straight-line program with $O(m+n^\alpha)$ operations computes $F$. It is constructed in time linear in its length and stores $O(n^2)$ ring elements in addition to the workspace of the multiplication program.
\end{proposition}
\begin{proof}
The program forms $X_1,X_2,X_3$, computes the path-sum matrices, runs \cref{alg:three-pairs} on all real vertices, and evaluates \eqref{eq:F-recurrence}. Rectangular products are tiled into square blocks: a product of an $h\times p$ and a $p\times q$ matrix with $h\leq p,q\leq n+6$ costs $O(pqh^{\alpha-2})$ operations. We first compute the path sums, then show that \cref{alg:three-pairs} meets its specification, and finally count its operations.

Split the ordered index set into a first half $A$ and a second half $B$. A path from $a\in A$ to $b\in B$ has a unique edge from $A$ to $B$, because its vertices increase, and no path leads from $B$ to $A$. Hence
\begin{align*}
N_i[A,B]
&=N_i[A,A]X_i[A,B]N_i[B,B],
\end{align*}
and $N_i[A,A],N_i[B,B]$ are the path-sum matrices of the subgraphs induced by the two halves, since a path between two vertices of one half stays in that half. Recursing on both halves, with base case $N_i[a,a]=1$, the products at depth $t$ cost $O(2^t(2n/2^t)^\alpha)$, which decreases geometrically in $t$ because $\alpha>2$; the path sums cost $O(n^\alpha)$ operations.

Next we verify the two computations of a call on $I$ with $|I|\geq2$. The first crossing identity holds because $\Lambda_T[w,c]=0$ for $c\geq w$ and every $d\in I_2$ follows every $w\in I_1$. For the second, $Z_T[w,d]=0$ unless $d>w\geq v_l$ and $N_\ell[d,e]=0$ unless $d\leq e<v_r$, so the sum defining $\Phi_T[w,e]$ runs over $d\in I$; the block $Z_T[I_1,I_1]$ was computed by the call on $I_1$, and the rows $\Lambda_T[I_1,\cdot]$ are final. For the cross updates, every $w\in I_1$ precedes every $v\in I_2$, so the outer-product form of $B_w(e)$ applies, and for $c<v$
\begin{align*}
\bigl(P_1Q_1+P_2Q_2\bigr)[v,c]
&=\sum_{w\in I_1}B_w(e),
\end{align*}
where $e_i=e_j=v$ and $e_\ell=c$.
Every factor is available when it is formed: $\alpha_{ij}^w(v)$ uses $\Lambda_{[3]}(w)$ and $\Phi_{i\ell}[w,v],\Phi_{j\ell}[w,v]$, and $Y_{ij}^w(c)$ equals $\Lambda_{ij}[w,c]$ for $c<w$, zero for $c=w$, and $\Phi_{ij}[w,c]$ for $w<c<v_r$, all of which are final or were computed for $w<c$ in $I$. The value $B_w(v,v,v)$ uses $\Lambda_{[3]}(w)$, the path sums, and $Y_T^w(v)=\Phi_T[w,v]$ for the three pairs $T$. Entries with $c\geq v$ are never read.

We then show by induction on $|I|$ that a call meets its specification. If $I=\{w\}$, the assumption says that the accumulators of $w$ contain the contributions of all real $w'<w$, so \eqref{eq:table-recurrence} finalizes the row of $w$. Otherwise the call on $I_1$ meets its specification by induction. After step~4, the accumulators of every $v\in I_2$ contain the contributions of all $w<v_l$ and of all $w\in I_1$, which is the assumption of the call on $I_2$. Every ordered pair $w<v$ of real vertices is added exactly once, at the unique call with $w\in I_1$ and $v\in I_2$. The entries of $Z_T$ and $\Phi_T$ for $w<d$ in $I$ lie in $I_1\times I_1$, $I_1\times I_2$, or $I_2\times I_2$, and they are computed by the call on $I_1$, step~3, and the call on $I_2$, respectively.

The accumulators start at zero, so the call on all real vertices meets its assumption, and afterwards every table entry is final. The values $B_w(\theta)$ need $Y_T^w(\theta_\ell)=\Phi_T[w,\theta_\ell]$, since $\theta_\ell>w$. For each pair $T$,
\begin{align*}
Z_T[\cdot,\theta_\ell]
&=\Lambda_TX_\ell[\cdot,\theta_\ell],
\\
\Phi_T[\cdot,\theta_\ell]
&=Z_T[\cdot,V]N_\ell[V,\theta_\ell]+Z_T[\cdot,\theta_\ell],
\end{align*}
which costs $O(n^2)$, and \eqref{eq:F-recurrence} gives $F$ with $O(n)$ further operations.

Finally we count the operations of \cref{alg:three-pairs}. A call on $h\geq2$ vertices performs constantly many products of shapes $(h\times n)(n\times h)$, $(h\times h)(h\times h)$, and $(h\times h)(h\times n)$, costing $O(nh^{\alpha-1})$, and forms their factors with $O(nh)$ operations. The calls at depth $t$ number at most $2^t$ and have at most $\lceil n/2^t\rceil\leq2n/2^t$ vertices each, so depth $t$ costs
\begin{align*}
O\bigl(2^t\cdot n(2n/2^t)^{\alpha-1}\bigr)
&=O\bigl(n^\alpha2^{-(\alpha-2)t}\bigr)
\end{align*}
operations for the products and $O(n^2)$ for the factors. Since $\alpha>2$, the products sum to $O(n^\alpha)$ and the factors to $O(n^2\log n)=O(n^\alpha)$. The leaves, the direct updates of $\Lambda_{[3]}$, and the final step cost $O(n^2)$. With the $O(m+n^2)$ operations for $X_1,X_2,X_3$ and the $O(n^\alpha)$ operations for the path sums, this gives $O(m+n^\alpha)$. The matrices $X_i,N_i,\Lambda_T,Z_T,\Phi_T$ and the accumulators occupy $O(n^2)$ cells, and the factors of a call can be released before its second recursive call.
\end{proof}

\subsection{Fixed \texorpdfstring{$k\geq4$}{k>=4} in matrix multiplication time}\label{sec:k-batched}

For $k\geq4$, the row extensions of \cref{prop:k-sweep} can be computed in full when their rows are finalized; unlike for three pairs, this costs no more than the target bound. The remaining bottleneck is updating the pair tables. Fixing all but one outsider endpoint makes each update a matrix product, even when the extension tensor depends on the destination vertex.

Use $X_i$ for the weighted adjacency matrix of $H_i^+$, as above, and include all private vertices in endpoint indices. For each real $w$, form the matrices $K_i^w$ from their defining case formula: their columns $e<w$ are the identity on rows $c<w$, their column $w$ is zero, and
\begin{align*}
K_i^w[\{c:c<w\},\{e:e>w\}]
&=X_i[\{c:c<w\},\{d:d>w\}]N_i[\{d:d>w\},\{e:e>w\}].
\end{align*}
Apply these matrices along the outsider coordinates of a finalized row $\Lambda_T(w;\cdot)$ to obtain the entire tensor $Y_T^w$. Invalid prefix indices are zero; the extension tensors include endpoints on either side of $w$ and at the private sinks.

For a target pair $S$, a source event $T$, and endpoints $e=e(S,v,c)$, the contribution of $T$ to $B_w(e)$ is
\begin{align*}
Y_T^w(e_{-T})
\prod_{i\in T\cap S}N_i(w,v)
\prod_{i\in T\setminus S}N_i(w,c_i),
\end{align*}
where $e_{-T}$ has coordinate $v$ for $i\in S\setminus T$ and coordinate $c_i$ for $i\notin S\cup T$. At a split into $I_1<I_2$, take $w\in I_1$ and $v\in I_2$. If $T=S$, define factors with the entire outsider tuple as a column index:
\begin{align*}
P[v,w]&\coloneqq\prod_{i\in S}N_i(w,v),\\
Q[w,c]&\coloneqq Y_S^w(c).
\end{align*}
If $T\neq S$, choose $j\in T\setminus S$, which exists since $|T|\geq2=|S|$. Fix $c_i$ for all $i\notin S\cup\{j\}$ and define
\begin{align*}
P[v,w]&\coloneqq Y_T^w(e_{-T})
\prod_{i\in T\cap S}N_i(w,v)
\prod_{i\in T\setminus(S\cup\{j\})}N_i(w,c_i),\\
Q[w,c_j]&\coloneqq N_j(w,c_j).
\end{align*}
Since $j\in T$, the tensor $Y_T^w$ has no $c_j$ coordinate, so each product $PQ$ sums exactly the contribution of $T$ over $w\in I_1$. Destination masks $c_i<v$ are imposed when the products are added to the accumulators.

\begin{algorithm}[htb]
\caption{\textsc{Solve}$(I)$ for fixed $k\geq4$, on an interval of real vertices}
\label{alg:k-batched}
\begin{itemize}
\item[\textbf{1.}] If $I=\{w\}$, finalize every row $\Lambda_S(w;\cdot)$ by \eqref{eq:table-recurrence}, form the matrices $K_i^w$, compute and retain every tensor $Y_T^w$ by coordinate transforms, and return.
\item[\textbf{2.}] Split $I$ into a first half $I_1$ and a second half $I_2$, and call \textsc{Solve}$(I_1)$.
\item[\textbf{3.}] For each target pair $S$ and event $T$, add the product $PQ$ defined above to the corresponding accumulators in $I_2$, retaining only entries with $c_i<v$ for every $i\notin S$; when $T\neq S$, do this for every fixed tuple $(c_i)_{i\notin S\cup\{j\}}$.
\item[\textbf{4.}] For every $|S|\geq3$, $w\in I_1$, and table entry $\Lambda_S(v;c)$ with $v\in I_2$, add $B_w(e(S,v,c))$ directly to its accumulator.
\item[\textbf{5.}] Call \textsc{Solve}$(I_2)$.
\end{itemize}
\end{algorithm}

\begin{proposition}[Fixed-$k$ batching]
\label{prop:k-batched}
Fix $k\geq4$. Under the matrix multiplication hypothesis of \cref{prop:three-batched}, a division-free straight-line program with $O_k(m+n^{k-3+\alpha})$ ring operations computes $F$. It is constructed in time linear in its length and stores $O_k(n^{k-1})$ ring elements in addition to the workspace of the multiplication program.
\end{proposition}
\begin{proof}
The program forms the matrices $X_i$, computes the path sums by the block recursion in \cref{prop:three-batched}, initializes all accumulators to zero, runs \cref{alg:k-batched} on the real vertices, and evaluates \eqref{eq:F-recurrence}. As in \cref{prop:k-sweep}, the $2k$ private vertices change no bound, and $n$ below also bounds the extended index sets up to a constant factor.

The correctness invariant is the one used for \cref{alg:three-pairs}: on entry to a call on $I$, every accumulator in $I$ contains the contributions of all real vertices preceding $I$. A singleton therefore finalizes its rows by \cref{lem:first-event} and \eqref{eq:table-recurrence}. The kernel formula and distributivity show that its coordinate transforms compute exactly $Y_T^w$. At a nontrivial call, induction makes all rows and all extension tensors with $w\in I_1$ final before any cross update. The displayed factorizations give exactly the contributions of these rows to the pair tables; step~4 gives exactly their contributions to the other tables. This establishes the invariant for $I_2$. Every ordered pair $w<v$ is used at its unique split with $w\in I_1$ and $v\in I_2$, so no contribution is omitted or repeated. Every factor is available: $N_i$ is precomputed, and $Y_T^w$ was computed in full at the leaf of $w$, including all future endpoints. Only legal destination entries are read in finalizing rows. After the recursion, all terms $B_w(\theta)$ are available, and \eqref{eq:F-recurrence} returns $F$.

Forming the adjacency matrices and path sums costs $O_k(m+n^\alpha)$ operations. Each leaf forms its kernels with $O_k(n^\alpha)$ operations. If $r\coloneqq k-|T|\geq2$, one coordinate transform is a product of an $n^{r-1}\times n$ matrix and an $n\times n$ matrix, after zero padding. Splitting its rows into blocks of $n$ costs $O(n^{r-2+\alpha})$ operations. For $r=1$ use a direct vector--matrix product costing $O(n^2)$; for $r=0$ the tensor is the scalar $\Lambda_{[k]}(w)$. Over all leaves and event sets the kernels and transforms therefore cost
\begin{align*}
O_k\bigl(n^{\alpha+1}+n^{k-3+\alpha}+n^3\bigr)
&=O_k(n^{k-3+\alpha}),
\end{align*}
since $k\geq4$ and $\alpha>2$.

Consider a call on $h\geq2$ vertices, padding its halves to size $O(h)$. For $T=S$ the product has shape $(h\times h)(h\times n^{k-2})$, costing $O(n^{k-2}h^{\alpha-1})$ by square tiling. For $T\neq S$, each of the $O_k(n^{k-3})$ fixed tuples gives a product of shape $(h\times h)(h\times n)$, costing $O(nh^{\alpha-1})$. These cases exhaust all event sets: $T=S$, $T\supsetneq S$, $|T\cap S|=1$, and $T\cap S=\emptyset$; in the last three, the chosen $j\in T\setminus S$ removes the only free outsider coordinate from $Y_T^w$, whatever its dependence on $v$. There are $O_k(1)$ choices of $S$ and $T$, so the products at depth $t$ cost
\begin{align*}
O_k\bigl(2^tn^{k-2}(2n/2^t)^{\alpha-1}\bigr)
&=O_k\bigl(n^{k-3+\alpha}2^{-(\alpha-2)t}\bigr).
\end{align*}
Their sum is $O_k(n^{k-3+\alpha})$. Forming factors, accumulating the tiled outputs, and applying the masks cost $O_k(n^{k-2}h)$ per call and hence $O_k(n^{k-1}\log n)$ overall, which is absorbed because $\alpha>2$. For fixed $|S|\geq3$, the direct updates have $O(n^{k-|S|+1})$ destination entries and $n$ possible earlier rows, so their total cost is $O_k(n^{k-|S|+2})\leq O_k(n^{k-1})$. Row finalization costs $O_k(n^{k-1})$, and the root sum costs $O_k(n)$.

All tables, accumulators, and retained tensors together use $O_k(n^{k-1})$ cells. Kernels and coordinate-transform buffers are reused between leaves. Products are tiled and accumulated one block at a time, and the factors of a call are released before its second recursive call, so temporary storage satisfies the same bound, apart from the workspace of the multiplication program. The schedule, masks, and indices depend only on the input graphs and the order, and every arithmetic step is an addition, subtraction, or multiplication, so emitting the program takes time linear in its length.
\end{proof}

\subsection{Reporting a solution}\label{sec:k-reporting}

We now use the regional instances and monotone cuts of \cref{sec:extraction} in the setting of \cref{thm:dag-k-paths}, with a field $\F$ in place of $\Ring$. Adding, for each $i\in[k]\setminus I$, a new isolated vertex carrying a zero-edge pair for commodity $i$ changes neither the solutions nor $F_R$ of a regional instance. With this padding, \cref{prop:k-sweep,prop:three-batched,prop:k-batched} give a division-free circuit for $F_R$, constructed in time linear in its size, of size $\Gamma(R)$ with
\begin{align*}
\Gamma(R)
&\coloneqq
\begin{cases}
O_k(m_R+n_R^{k-3+\alpha}), & \text{under the hypothesis of \cref{prop:three-batched}},\\
O_k(m_Rn_R^{k-2}+n_R^k), & \text{otherwise},
\end{cases}
\end{align*}
for $n_R\geq1$. When the matrix multiplication bound is used, the extraction below differentiates this circuit, so we assume that the matrix multiplication program of \cref{prop:three-batched} computes the product as an identity of formal polynomials and is constructible in time linear in its length; the bilinear algorithms of \cref{sec:prelim} used in \cref{thm:dag-k-dsp} have both properties.

Let $L$ be the first $\lceil n_R/2\rceil$ vertices of $R$ in the order $<$, and $U\coloneqq R\setminus L$. This is a monotone cut, and every path in $H_i[R]$ visits its vertices in increasing order, so no path exists for a label with $a_i\in U$ and $b_i\in L$. Hence, if $F_R\neq0$, every active label has $a_i\in L$ and $b_i\in U$, the cut edges used by solutions go from $L$ to $U$, and under a feasible $\sigma$ the low child has the pairs $(a_i,u_i)$ and the high child the pairs $(v_i,b_i)$ for $i\in J$.

The reporting algorithm keeps both children of every cut. The regional circuits have superlinear size and the regions halve, so the costs decrease geometrically with the recursion depth.

\begin{algorithm}[htb]
\caption{\textsc{ReportDAG}$(R,\xi)$, for a regional instance $R$ with $F_R(\xi)\neq0$}
\label{alg:report-dag}
\begin{itemize}
\item[\textbf{1.}] If $I=\emptyset$, return the empty tuple. If $n_R=1$, return the zero-edge path of the unique label in $I$.
\item[\textbf{2.}] Split $R$ into the vertex-median cut $(L,U)$. By \cref{lem:extraction-step}, applied to the circuit of $F_R$, find a set $B$ of cut edges and a feasible bijection $\sigma$ with $F_L^\sigma(\xi)F_U^\sigma(\xi)\neq0$.
\item[\textbf{3.}] Call \textsc{ReportDAG} on the low and the high child under $\sigma$, and join the returned tuples through the edges $\sigma(i)$.
\end{itemize}
\end{algorithm}

\begin{proposition}[Reporting]
\label{prop:k-reporting}
In the setting of \cref{thm:dag-k-paths} over a field $\F$ with $|\F|\geq2k$, there is a randomized algorithm that returns either \textnormal{\textsc{no}} or a solution. It never returns an invalid tuple, always returns \textnormal{\textsc{no}} when no solution exists, and returns a solution with probability at least $1-n/|\F|$ otherwise. Its expected running time is $O_k(mn^{k-2}+n^k)$; under the hypothesis of \cref{prop:three-batched}, it is $O_k(m+n^{k-3+\alpha})$.
\end{proposition}
\begin{proof}
The algorithm reduces parallel edges, tests $F(\xi)\neq0$ at a uniform point $\xi\in\F^E$, returns \textnormal{\textsc{no}} if the value is zero, and otherwise runs \cref{alg:report-dag} on the input instance. We first describe the reduction and the root test, then show that \cref{alg:report-dag} returns a solution, and finally bound its running time.

For each ordered vertex pair $(u,v)$ and each label $i$, keep one edge from $u$ to $v$ that belongs to $H_i$, if any, and delete the other edges; every kept edge retains its memberships in all subgraphs. With an endpoint-indexed table this takes $O_k(m+n^2)$ time and leaves at most $kn^2$ edges. A solution uses at most one edge from $u$ to $v$, which can be replaced by the edge kept for its label, and every solution of the reduced instance is a solution of the original one, so the existence of solutions is preserved, although the polynomial changes. All polynomials below refer to the reduced instance, and $m$ denotes its edge count. The root test evaluates $F(\xi)$ with the program of \cref{prop:k-sweep}, \cref{prop:three-batched}, or \cref{prop:k-batched}; by \cref{lem:k-monomials}, it returns \textnormal{\textsc{no}} always when no solution exists and with probability at most $n/|\F|$ otherwise.

Next we show by induction on $n_R$ that a call on a regional instance $R$ with $F_R(\xi)\neq0$ returns a solution of $R$, where $\xi$ is restricted to the variables of $G[R]$. If $I=\emptyset$, the empty tuple is the solution. If $n_R=1$, the invariant forces $I=\{i\}$ with $a_i=b_i$, and the zero-edge path is a solution. Otherwise $(L,U)$ is a monotone cut with $j\leq k\leq|\F|/2$, and the circuit of $F_R$ has size $\Gamma(R)$, so \cref{lem:extraction-step} applies. Both children satisfy the invariant and have fewer vertices than $R$, so by induction they return solutions, and by \cref{lem:cut-factorization} joining them through $\sigma$ gives a solution of $R$. Together with the root test, this shows that the algorithm never returns an invalid tuple, returns \textnormal{\textsc{no}} when no solution exists, and returns a solution with probability at least $1-n/|\F|$ otherwise. The samples of \cref{lem:interpolated-extraction} never replace $\xi$, so after the root test randomness affects only the running time.

Finally we bound the running time. Conditional on the history, the expected cost of a call on $R$, including the circuit, the extraction, the at most $k!$ child evaluations, and the partition, is $O_k(\Gamma(R)+m_R+n_R)$, since $m_L+m_U\leq m_R$ and $n_L,n_U\leq n_R$ give child circuits of total size $O(\Gamma(R))$. The regions at recursion depth $t$ are vertex-disjoint, their graphs are edge-disjoint, and each has at most $\lceil n/2^t\rceil\leq2n/2^t$ vertices. Hence, using $\sum_Rm_R\leq m$, $\sum_Rn_R\leq n$, and $k-2\geq1$,
\begin{align*}
\sum_{R\text{ at depth }t}\bigl(m_Rn_R^{k-2}+n_R^k\bigr)
&\leq(2n/2^t)^{k-2}\sum_Rm_R+(2n/2^t)^{k-1}\sum_Rn_R
\leq2^{k-1}2^{-t}\bigl(mn^{k-2}+n^k\bigr).
\end{align*}
Summing over $t$ gives $O_k(mn^{k-2}+n^k)$, which also dominates the $O((m+n)\log n)$ partition cost and the $O_k(m+n^2)$ reduction. Under the hypothesis of \cref{prop:three-batched}, put $\beta\coloneqq k-3+\alpha>2$; the same argument gives
\begin{align*}
\sum_{R\text{ at depth }t}n_R^\beta
&\leq2^{\beta-1}2^{-(\beta-1)t}n^\beta,
\end{align*}
and the terms $m_R$ sum to $O(m\log n)=O_k(n^2\log n)=O_k(n^\beta)$ over the $O(\log n)$ depths, because the reduced instance has at most $kn^2$ edges; with the $O_k(m+n^2)$ reduction this gives $O_k(m+n^{k-3+\alpha})$. These bounds hold for every recursion tree, so linearity of expectation gives the claimed expected running time.
\end{proof}

We now prove the formal version of \cref{thm:dag-k-dsp-informal}.

\begin{theorem}[Disjoint paths in DAGs, formal version of \cref{thm:dag-k-dsp-informal}]
\label{thm:dag-k-dsp}
Fix $k\geq3$ and $c>0$. On a DAG with $n$ vertices, $m$ edges, and real edge weights, $k$-DSP can be solved by a randomized algorithm in expected
\begin{align*}
O_k(m+n^{k-3+\omega+\varepsilon})
\end{align*}
time, for every fixed $\varepsilon>0$. The algorithm never returns an invalid tuple, returns \textnormal{\textsc{no}} on every infeasible instance, and reports a solution with probability at least $1-n^{-c}$ on every feasible instance. Decision with the same one-sided error has the same worst-case time bound. These guarantees also hold for $k$-DP in DAGs.
\end{theorem}
\begin{proof}
\label{proof:dag-k-dsp}
We reduce both problems to the setting of \cref{thm:dag-k-paths} and apply \cref{prop:k-reporting} over a field of polynomial size. First keep one lightest copy of every set of parallel edges, which takes $O(m+n)$ time by radix sorting the edges by their endpoints. A shortest path through a parallel class can use a lightest copy, so feasibility is unchanged, and afterwards $m<n^2$. For $k$-DSP, compute $d_i(v)\coloneqq\dist_G(s_i,v)$ by topological dynamic programming with $O(k(m+n))$ exact operations, and let $H_i$ consist of the edges $(u,v)$ with $d_i(u)<+\infty$ and $d_i(v)=d_i(u)+w(u,v)$. Along a path of $H_i$ from $s_i$ to $t_i$ the labels telescope to length $d_i(t_i)$, and every edge of a shortest $s_i$--$t_i$ path is tight because its prefixes are shortest; hence the $s_i$--$t_i$ paths in $H_i$ are exactly the shortest ones. For $k$-DP, put $H_i\coloneqq G$.

Use $\F\coloneqq\F_{2^q}$ with $q\coloneqq\lceil\log_2(2k)\rceil+\lceil(c+1)\log_2(n+1)\rceil$, so that $|\F|\geq2k$, $n/|\F|\leq n^{-c}$, and field elements fit in $O(\log n)$-bit words for fixed $k$ and $c$. By \cref{sec:prelim}, a bilinear algorithm with coefficients in $\F_2$ multiplies $h\times h$ matrices over $\F$ as an identity of formal polynomials with $O(h^{\omega+\varepsilon})$ operations, so the hypothesis of \cref{prop:three-batched} holds with $\alpha\coloneqq\omega+\varepsilon$. \Cref{prop:k-reporting} gives the reporting algorithm. For decision, evaluate $F(\xi)$ at a uniform $\xi$ with the program of \cref{prop:three-batched} or \cref{prop:k-batched} and answer according to whether it is nonzero; \cref{lem:k-monomials} bounds the error.
\end{proof}

\section{Two disjoint shortest paths in linear time}
\label{sec:two-linear}

To obtain linear reporting time from a linear-size circuit, the recursion must leave at most one unresolved two-path region. The key is to recover one side of a two-active cut directly, using shortest-path comparisons to enforce the prescribed pairing. We first establish the regional circuits and recursion invariant, then prove this one-sided recovery and its linear total cost.

\subsection{Regional circuits and the recursive schedule}
\label{sec:regional}

This subsection specializes the regional polynomial of \cref{sec:extraction} to two shortest paths, gives it circuits of linear size on DAGs and undirected graphs (\cref{lem:regional-circuits}), and states the recursive schedule shared by the three two-path reporting algorithms; \cref{sec:directed} gives the circuits for directed graphs. Every region keeps the tight graphs of the whole instance, so its shortest paths are tight for the global labels (\cref{lem:regional-closure}) and the cancellation identities of \cite{avw24} remain valid with regional terminals. All reporting algorithms evaluate these circuits at one point sampled at the root \eqref{eq:inherited-nonzero}.

We use the characteristic-two field and arithmetic model of \cref{sec:prelim}. The field-size choice for the error guarantee is given in \cref{lem:probability}.

We begin with the shortest-path subgraphs used by both algorithms. For $i\in\{1,2\}$, let $H_i$ be the graph of edges tight with respect to the source-distance labels
\begin{align*}
d_i(v)
&\coloneqq \dist_G(s_i,v).
\end{align*}
Thus a directed edge $(u,v)$ belongs to $H_i$ when $d_i(u)<+\infty$ and $d_i(v)=d_i(u)+w(u,v)$. In an undirected graph, each physical edge is oriented in every tight direction satisfying the same finite-distance condition. Each $H_i$ is acyclic: it is a subgraph of the input DAG in the DAG case, positive weights orient undirected tight edges by increasing distance, and a directed tight cycle would have total weight zero.

A region $R\subseteq V$ in the recursion has terminal pairs $(a_1,b_1)$ and $(a_2,b_2)$. The $i$th regional paths are the $a_i$--$b_i$ paths in $H_i[R]$. We attach one variable $x_e$ to every physical edge, so that the regional polynomial of \cref{sec:extraction} with $k=2$ reads
\begin{align*}
F_R(x)
&\coloneqq \sum_{(P_1,P_2)}\prod_{e\in E(P_1)\cup E(P_2)}x_e,
\end{align*}
where the sum ranges over vertex-disjoint regional path pairs, the \emph{solutions} of $R$. By \cref{lem:k-monomials}, distinct pairs contribute distinct squarefree monomials, and $F_R\neq0$ exactly when $R$ has a solution. The circuits below start from products of path sums, which also count intersecting pairs. The identities of \cite{avw24} remove these terms by cancellation: an involution swaps subpaths between intersection vertices, which keeps the multiset of edges, so two matched terms carry the same monomial and sum to zero in characteristic two.

The algorithms of \cref{sec:anchors} compute shortest paths inside a region. The next lemma shows that such paths are tight for the global labels, and that on monotone intervals regional and global shortest paths agree.

\begin{lemma}[Regional shortest-path closure]
\label{lem:regional-closure}
Suppose $x$ precedes $y$ on a path in $H_i[R]$. Every shortest $x$--$y$ path contained in $G[R]$ is tight for $d_i$. If $R$ is an interval of a coordinate monotone on $H_i$, every globally shortest $x$--$y$ path stays in $R$, and the global and regional shortest-path polynomials between $x$ and $y$ agree.
\end{lemma}
\begin{proof}
We compare every path with the inherited $x$--$y$ part of the given $H_i$-path, whose length telescopes to $d_i(y)-d_i(x)$ because its edges are tight. The triangle inequality gives
\begin{align*}
d_i(y)
&\leq d_i(x)+\dist_G(x,y),
\end{align*}
so the inherited path is globally shortest. It lies in $G[R]$, hence $\dist_{G[R]}(x,y)=\dist_G(x,y)=d_i(y)-d_i(x)$, and every shortest $x$--$y$ path in $G[R]$ is globally shortest. Prefixing any globally shortest $x$--$y$ path by an $s_i$--$x$ shortest path gives a walk of length $d_i(y)$, that is, a shortest walk to $y$. In a DAG this walk cannot repeat a vertex; in the other models, deleting a repeated positive-weight cycle would shorten it. Hence the walk is a path and each of its edges is tight for $d_i$, which proves the first assertion. If the region is an interval of a monotone coordinate, such a path lies in $H_i$, so the coordinate is monotone along it, and the path cannot leave the interval and later return. Thus the globally shortest $x$--$y$ paths are exactly the shortest $x$--$y$ paths in $G[R]$, and the two shortest-path polynomials agree.
\end{proof}

The next lemma provides the circuits evaluated by the DAG and undirected algorithms. Its proof checks that the identities of \cite{avw24} survive the passage to regional terminals.

\begin{lemma}[Regional cancellation circuits]
\label{lem:regional-circuits}
For a DAG or an undirected graph, $F_R$ has a division-free arithmetic circuit over a characteristic-two field of size $O(m_R+n_R)$, where $n_R\coloneqq |R|$ and $m_R\coloneqq |E(G[R])|$. The statement holds for all regional terminals produced by the algorithms below, including zero-edge terminal pairs.
\end{lemma}
\begin{proof}
If $a_i=b_i=v$, the $i$th path has zero edges and $F_R$ is the path-sum polynomial for the other pair in $H_{3-i}[R-v]$. If both paths have zero edges, $F_R$ is $1$ when their vertices are distinct and $0$ otherwise. A terminal shared by the two pairs also makes $F_R=0$. Each case has a circuit of size $O(m_R+n_R)$, since a path sum in the acyclic graph $H_{3-i}[R-v]$ takes one pass in topological order.

Hence assume that both paths contain an edge and the boundary vertices required to be disjoint are distinct.

For the remaining regions we use the identities of \cite{avw24} that compute the polynomial of two disjoint shortest paths: the source-linkage and first-intersection identities \cite[Lemmas~22 and~23]{avw24} for DAGs, and in addition the agreeing- and disagreeing-pair identities \cite[Lemmas~30, 36, 38, and~39]{avw24} for undirected graphs. \Cref{lem:regional-transfer} in \cref{app:regional} shows that they hold with $s_i$, $t_i$, and the shortest-path graphs replaced by $a_i$, $b_i$, and $H_i[R]$: regional paths are shortest paths of $G$, and every exchange of subpaths in the proofs of \cite{avw24} stays in $H_i[R]$. The same lemma evaluates the identities with $O(m_R+n_R)$ additions and multiplications, which proves the claimed circuit size.
\end{proof}

We also use the linear-time algorithm of \cite{tho12} that, given a DAG and two prescribed terminal pairs, reports two vertex-disjoint paths or determines that none exist.

We use one random assignment $\xi\in\F^{E}$ sampled at the root and maintain the invariant
\begin{align}
F_R(\xi|_{E(R)})
&\neq 0
\label{eq:inherited-nonzero}
\end{align}
for the unique active region in the linear algorithms and for every active region in the directed recursion. A nonzero value certifies that $R$ has a solution. The root is the only place where polynomial identity testing may reject a feasible instance. Every later recursive choice proceeds only to subinstances whose evaluated factors are nonzero.

\Cref{alg:recursive-report} is the recursive schedule shared by the two-path reporting algorithms. It applies \cref{lem:extraction-step} at monotone cuts, and the three algorithms differ in how they choose cuts and in which children of a two-active cut they keep; a cut is \emph{zero-}, \emph{one-}, or \emph{two-active} according to its number of active pairs. In a DAG the cuts are prefixes of a topological order. In undirected and directed graphs they are thresholds of a difference $\Delta$ of the labelings $d_1$ and $d_2$ (\cref{sec:linear,sec:directed}), which is nondecreasing along the edges of $H_1$ and nonincreasing along those of $H_2$. Every region is then a union of consecutive $\Delta$-levels, and the $i$th regional paths take their $\Delta$-values in the \emph{terminal range} $J_i$ between the values at $a_i$ and $b_i$. A \emph{plateau} is the set of vertices of a region with one common $\Delta$-value, and it is \emph{heavy} if it contains more than half of the region. A median threshold leaves at most three quarters of the vertices on each side unless the region has a heavy plateau, which is then split off by cuts at its two boundaries. If the two terminal ranges are disjoint, the pairs cannot meet. If they meet in a single value, the paths can meet only on the plateau of that value, and the \emph{plateau base case} of \cref{lem:plateau-base} resolves the region in linear time by a reduction to two prescribed paths in one DAG.

\begin{algorithm}[htb]
\caption{\textsc{RecursiveReport}$(R,\xi)$, for a regional instance $R$ with two terminal pairs and $F_R(\xi)\neq0$}
\label{alg:recursive-report}
\begin{itemize}
\item[\textbf{1.}] If a path has zero edges or two terminals coincide, resolve $R$ directly. In a distance-difference region, report the two pairs independently if their terminal ranges are disjoint, and apply the plateau base case if the ranges meet in one value.
\item[\textbf{2.}] Choose a median monotone cut of $R$.
\item[\textbf{3.}] If no pair is active, recurse on the side that contains both pairs, or report the two pairs separately.
\item[\textbf{4.}] If one pair is active, select its crossing edge by \cref{lem:extraction-step}, report the path of the single-path child, and recurse on the two-path child.
\item[\textbf{5.}] If both pairs are active, select by \cref{lem:extraction-step} two cut edges and an assignment whose two child evaluations at $\xi$ are nonzero. In a DAG or an undirected graph, report one side, or both, as in \cref{sec:anchors}, and recurse on any remaining child; in a directed graph, recurse on both children.
\item[\textbf{6.}] If a retained child is a heavy plateau, cut its other boundary next; a finite two-path plateau is resolved by the plateau base case, and only exterior children continue.
\item[\textbf{7.}] Join the returned paths through the selected cut edges.
\end{itemize}
\end{algorithm}

\subsection{Recovering one side without recursion}
\label{sec:anchors}

This subsection proves \cref{prop:one-sided-reporting} for DAG cuts and its undirected counterpart \cref{lem:undirected-anchor}. Once \cref{lem:extraction-step} has selected an assignment with two nonzero child evaluations at a two-active cut, the path pieces on one side of the cut can be reported in linear time without recursion, or the entire pair can be reported. \Cref{sec:linear} uses this to keep at most one two-path child per call. Reporting a side directly is a prescribed-pair problem: the two pieces lie in the different shortest-path DAGs $H_1$ and $H_2$, and each must end at its own endpoint of a selected cut edge. The key idea is to compare the selected solution with the two cross routes that exchange the selected cut edges (\cref{fig:cross-route}). Shortestness prevents the exchange from shortening both sides. On a side where it strictly lengthens the routes, a common sink with suitable attachment weights steers each source to its own endpoint, so a flow of value $2$ that ignores the pairing reports that side.

Fix an assignment selected by \cref{lem:extraction-step}, so both child evaluations are nonzero.

Handle the zero-edge and repeated-terminal cases of \cref{lem:regional-circuits} before applying the common-sink construction. Hence every side considered below has two nonempty prescribed pairs with distinct boundary vertices.

Throughout this subsection, a difference of extended distances is used only when the two operands are not both $+\infty$, with $r-(+\infty)=-\infty$ and $(+\infty)-r=+\infty$ for finite $r$.

\subsubsection{DAG cuts}

In this subsection, $e=(u,v)$ belongs to path $1$, $f=(x,y)$ to path $2$, and $u,x$ lie on the low side of a DAG topological cut while $v,y$ lie on its high side. For the current terminals $(a_1,b_1),(a_2,b_2)$, define
\begin{align*}
\delta(z)
&\coloneqq \dist_R(a_1,z)-\dist_R(a_2,z),
\\
\gamma(z)
&\coloneqq \dist_R(z,b_1)-\dist_R(z,b_2).
\end{align*}
Thus $\delta(z)$ is the extra cost of reaching $z$ from $a_1$ rather than from $a_2$, and $\gamma(z)$ is the extra cost of leaving $z$ toward $b_1$ rather than toward $b_2$. The two nonzero child factors provide the corresponding path pieces incident to $u,x,v,y$, so $\dist_R(a_1,u)$, $\dist_R(a_2,x)$, $\dist_R(v,b_1)$, and $\dist_R(y,b_2)$ are finite and every difference below is well defined. Hence $\delta(u),\gamma(v)\in\mathbb R\cup\{-\infty\}$ and $\delta(x),\gamma(y)\in\mathbb R\cup\{+\infty\}$, so whenever one of the comparisons below is an equality, all four distances in that comparison are finite.

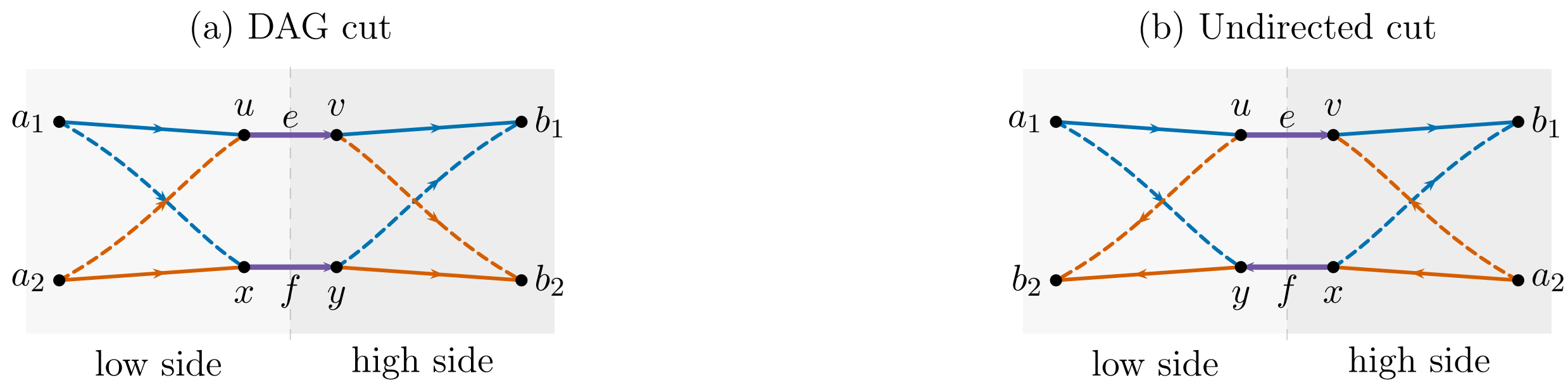
\begin{figure}[t]
\centering
\begin{tikzpicture}[x=0.69cm,y=0.69cm,font=\normalsize,line cap=round]
\fill[black!3] (0,0) rectangle (4,4);
\fill[black!7] (4,0) rectangle (8,4);
\draw[dashed,gray,opacity=0.35] (4,-0.1) -- (4,4.1);
\node at (4,4.6) {(a) DAG cut};
\node at (2,-0.45) {low side}; \node at (6,-0.45) {high side};
\draw[dp one,dp arrow] (0.5,3.2) -- (3.3,3);
\draw[dp one,dp arrow] (4.7,3) -- (7.5,3.2);
\draw[dp two,dp arrow] (0.5,0.8) -- (3.3,1);
\draw[dp two,dp arrow] (4.7,1) -- (7.5,0.8);
\draw[dp one,densely dashed,dp arrow] (0.5,3.2) .. controls (1.8,2.5) and (2.5,1.5) .. (3.3,1);
\draw[dp one,densely dashed,dp arrow] (4.7,1) .. controls (5.5,1.5) and (6.2,2.5) .. (7.5,3.2);
\draw[dp two,densely dashed,dp arrow] (0.5,0.8) .. controls (1.8,1.5) and (2.5,2.5) .. (3.3,3);
\draw[dp two,densely dashed,dp arrow] (4.7,3) .. controls (5.5,2.5) and (6.2,1.5) .. (7.5,0.8);
\draw[dp edge] (3.3,3) -- (4.7,3) node[midway,above,dp label] {$e$};
\draw[dp edge] (3.3,1) -- (4.7,1) node[midway,below,dp label] {$f$};
\fill (0.5,3.2) circle (1.8pt) node[left] {$a_1$};
\fill (0.5,0.8) circle (1.8pt) node[left] {$a_2$};
\fill (7.5,3.2) circle (1.8pt) node[right] {$b_1$};
\fill (7.5,0.8) circle (1.8pt) node[right] {$b_2$};
\fill (3.3,3) circle (1.8pt) node[above=2pt] {$u$};
\fill (4.7,3) circle (1.8pt) node[above=2pt] {$v$};
\fill (3.3,1) circle (1.8pt) node[below=2pt] {$x$};
\fill (4.7,1) circle (1.8pt) node[below=2pt] {$y$};
\end{tikzpicture}
\hfill
\begin{tikzpicture}[x=0.69cm,y=0.69cm,font=\normalsize,line cap=round]
\fill[black!3] (0,0) rectangle (4,4);
\fill[black!7] (4,0) rectangle (8,4);
\draw[dashed,gray,opacity=0.35] (4,-0.1) -- (4,4.1);
\node at (4,4.6) {(b) Undirected cut};
\node at (2,-0.45) {low side}; \node at (6,-0.45) {high side};
\draw[dp one,dp arrow] (0.5,3.2) -- (3.3,3);
\draw[dp one,dp arrow] (4.7,3) -- (7.5,3.2);
\draw[dp two,dp arrow] (3.3,1) -- (0.5,0.8);
\draw[dp two,dp arrow] (7.5,0.8) -- (4.7,1);
\draw[dp one,densely dashed,dp arrow] (0.5,3.2) .. controls (1.8,2.5) and (2.5,1.5) .. (3.3,1);
\draw[dp one,densely dashed,dp arrow] (4.7,1) .. controls (5.5,1.5) and (6.2,2.5) .. (7.5,3.2);
\draw[dp two,densely dashed,dp arrow] (3.3,3) .. controls (2.5,2.5) and (1.8,1.5) .. (0.5,0.8);
\draw[dp two,densely dashed,dp arrow] (7.5,0.8) .. controls (6.2,1.5) and (5.5,2.5) .. (4.7,3);
\draw[dp edge] (3.3,3) -- (4.7,3) node[midway,above,dp label] {$e$};
\draw[dp edge] (4.7,1) -- (3.3,1) node[midway,below,dp label] {$f$};
\fill (0.5,3.2) circle (1.8pt) node[left] {$a_1$};
\fill (0.5,0.8) circle (1.8pt) node[left] {$b_2$};
\fill (7.5,3.2) circle (1.8pt) node[right] {$b_1$};
\fill (7.5,0.8) circle (1.8pt) node[right] {$a_2$};
\fill (3.3,3) circle (1.8pt) node[above=2pt] {$u$};
\fill (4.7,3) circle (1.8pt) node[above=2pt] {$v$};
\fill (3.3,1) circle (1.8pt) node[below=2pt] {$y$};
\fill (4.7,1) circle (1.8pt) node[below=2pt] {$x$};
\end{tikzpicture}
\caption{Selected path pieces (solid) and cross routes (dashed). Blue denotes pair $1$, orange pair $2$; arrows indicate traversal direction. In (a), the cross routes exchange $e$ and $f$ in their forward directions. In (b), pair $2$ crosses from high to low, and each cross route uses the other selected edge backwards. The low- and high-side length changes are respectively $\delta(x)-\delta(u)$ and $\gamma(y)-\gamma(v)$ in (a), and $\ell(y)-\ell(u)$ and $h(x)-h(v)$ in (b). The sum cannot be negative, since each cross route connects a prescribed pair. Dashed routes need not be disjoint; crossings without marked vertices are schematic.}
\label{fig:cross-route}
\end{figure}

\paragraph{Two routes through the selected edges.} The selected solution sends path $1$ through $e$ and path $2$ through $f$. The \emph{cross routes} use the other edge: the $a_1$--$b_1$ route reaches $x$, crosses $f$, and continues from $y$, while the $a_2$--$b_2$ route reaches $u$, crosses $e$, and continues from $v$ (\cref{fig:cross-route}(a)). Passing to the cross routes exchanges the endpoints of the pieces on each side of the cut. On the low side, the length $\dist_R(a_1,u)+\dist_R(a_2,x)$ becomes $\dist_R(a_1,x)+\dist_R(a_2,u)$, a change of $\delta(x)-\delta(u)$; on the high side, the change is $\gamma(y)-\gamma(v)$. Each cross route connects a prescribed pair, so the total change is not negative, and the exchange cannot shorten both sides. The next lemma records this dichotomy and its equality case. A side on which the exchange strictly lengthens the routes, such as the low side when $\delta(u)<\delta(x)$, is reported by \cref{lem:common-sink}. If neither side is strict, the exchange is free on both sides, both selected edges are tight for both labelings, and the prescribed-pair algorithm of \cite{tho12} finishes.

\begin{lemma}[Cross-route dichotomy]
\label{lem:cross-route}
At least one of
\begin{align*}
\delta(u)
&\leq \delta(x),\\
\gamma(v)
&\leq \gamma(y)
\end{align*}
holds. If neither comparison is strict in the displayed direction, both are equalities and both $e$ and $f$ are tight in $H_1$ and $H_2$.
\end{lemma}
\begin{proof}
Put $D_i\coloneqq\dist_R(a_i,b_i)$. The selected edges and their feasible child paths are tight, so
\begin{align*}
D_1&=\dist_R(a_1,u)+w(e)+\dist_R(v,b_1),\\
D_2&=\dist_R(a_2,x)+w(f)+\dist_R(y,b_2).
\end{align*}
We first compare both sums of differences with $D_1-D_2$. Expanding $\delta$ and $\gamma$ with these two identities, the edge weights cancel, and
\begin{align*}
\delta(u)+\gamma(v)
&=D_1-\bigl(\dist_R(a_2,u)+w(e)+\dist_R(v,b_2)\bigr)\\
&\leq D_1-D_2\\
&\leq\bigl(\dist_R(a_1,x)+w(f)+\dist_R(y,b_1)\bigr)-D_2\\
&=\delta(x)+\gamma(y).
\end{align*}
The two inequalities are the triangle inequalities for the cross routes: the route through $e$ connects $a_2$ to $b_2$, and the route through $f$ connects $a_1$ to $b_1$. The left side lies in $\mathbb R\cup\{-\infty\}$ and the right side in $\mathbb R\cup\{+\infty\}$, so the chain is well defined.

If $\delta(u)>\delta(x)$ and $\gamma(v)>\gamma(y)$, the allowed extended values force all four values to be finite, and adding the two inequalities contradicts the chain. This proves the dichotomy. If neither comparison is strict in the displayed direction, then $\delta(u)\geq\delta(x)$ and $\gamma(v)\geq\gamma(y)$, the four values are again finite, and the chain forces $\delta(u)+\gamma(v)=\delta(x)+\gamma(y)$. Hence both comparisons are equalities, and both inequalities in the chain are equalities, so both cross routes have their prescribed shortest length. By \cref{lem:regional-closure}, the route for pair $2$ makes $e$ tight in $H_2$, and the route for pair $1$ makes $f$ tight in $H_1$. Each edge was already tight for its selected pair, as required.
\end{proof}

The next lemma reports a side on which the exchange strictly lengthens the routes.

\begin{lemma}[Common-sink reduction]
\label{lem:common-sink}
If $\delta(u)<\delta(x)$, the two disjoint shortest low-side path pieces ending at $u,x$ can be reported in $O(m_R+n_R)$ time. If $\gamma(v)<\gamma(y)$, the two high-side pieces can be reported within the same bound.
\end{lemma}
\begin{proof}
We prove the low-side statement by making the endpoint pairing follow from shortestness. Attach both boundary vertices to one sink, choosing the two attachment weights so that each source strictly prefers its designated boundary vertex. Since the low side is a topological prefix, every path ending at $u$ or $x$ lies in it, so the regional distances below are also low-side distances. Double all edge weights in the low side, add a new sink $q$, and add edges $u\to q$ and $x\to q$ of respective weights $\alpha,\beta>0$. Put
\begin{align*}
A
&\coloneqq \dist_R(a_1,u)-\dist_R(a_1,x),\\
B
&\coloneqq \dist_R(a_2,u)-\dist_R(a_2,x).
\end{align*}
Rearranging $\delta(u)<\delta(x)$, the strict potential inequality is $A<B$ in the extended order. Doubling ensures that, for integer inputs, the open interval $(2A,2B)$ contains an integer.

To force the pairing, when both endpoints are finite, choose $z\coloneqq A+B$; if only the lower endpoint is finite, choose $z\coloneqq2A+1$; if only the upper endpoint is finite, choose $z\coloneqq2B-1$; if $A=-\infty$ and $B=+\infty$, choose $z\coloneqq0$. Set
\begin{align*}
\alpha
&\coloneqq 1+\max\{0,-z\},\\
\beta
&\coloneqq \alpha+z.
\end{align*}
Then $\alpha\geq1$ and $\beta=1+\max\{z,0\}\geq1$, and both are integers when the input weights are integers. Moreover, $\beta-\alpha=z$, and in each case $z$ lies strictly between $2A$ and $2B$ because $A<B$:
\begin{align*}
2A
&<\beta-\alpha<2B.
\end{align*}
After doubling, the shortest $a_i$--$q$ route through $u$ has length $2\dist_R(a_i,u)+\alpha$, and the one through $x$ has length $2\dist_R(a_i,x)+\beta$. Hence
\begin{align*}
\bigl(2\dist_R(a_1,x)+\beta\bigr)-\bigl(2\dist_R(a_1,u)+\alpha\bigr)
&=(\beta-\alpha)-2A>0,
\\
\bigl(2\dist_R(a_2,u)+\alpha\bigr)-\bigl(2\dist_R(a_2,x)+\beta\bigr)
&=2B-(\beta-\alpha)>0,
\end{align*}
so every shortest $a_1$--$q$ path enters $q$ through $u$, and every shortest $a_2$--$q$ path through $x$.
Compute distance-to-$q$ labels in linear time, by topological dynamic programming for a DAG or by undirected integer-weight single-source shortest paths \cite{tho97}. Form the DAG of tight edges, split every vertex except $q$ into a unit-capacity in/out pair, and connect a super-source to $a_1,a_2$ with unit capacities. Feasibility of the low-side child supplies a flow of value $2$, which two augmentations find in linear time: by the strict preferences just proved, the child pieces followed by $u\to q$ and $x\to q$ are shortest routes to $q$, so all their edges are tight. Conversely, in a flow of value $2$, the unit path through $a_1$ is a tight, hence shortest, $a_1$--$q$ path, so it enters $q$ through $u$; the path through $a_2$ enters through $x$. Deleting $q$ yields disjoint shortest $a_1$--$u$ and $a_2$--$x$ paths. By \cref{lem:regional-closure}, they lie in $H_1,H_2$, respectively.
For undirected positive-integer inputs, let $\lambda$ bound the finite regional distances. Then the finite values of $A$ and $B$ lie in $[-\lambda,\lambda]$, so $|z|\leq2\lambda+1$ and $\max\{\alpha,\beta\}\leq2\lambda+2$, and the attachment weights fit in one RAM word. In a DAG, the real attachment weights take a constant number of exact operations. The high-side statement follows by reversing the DAG on the high side and applying the same construction to $b_1,b_2$ and boundary vertices $v,y$: after the reversal, distances from $b_i$ are the original distances to $b_i$, so $\gamma(v)<\gamma(y)$ becomes the low-side hypothesis.
\end{proof}

\paragraph{Double equality.} Suppose both comparisons of \cref{lem:cross-route} are equalities. Then the condition $2A<\beta-\alpha<2B$ of \cref{lem:common-sink} is empty: no attachment weights make $a_1$ prefer $u$ while $a_2$ prefers $x$, and a flow could pair $a_1$ with $x$. Instead, we build one DAG in which every $a_1$--$u$ path and every $a_2$--$x$ path is shortest, and call the prescribed-pair algorithm of \cite{tho12}. By \cref{lem:cross-route}, both selected edges are tight in $H_1$ and $H_2$. On the low side $L$, topological closure keeps all four source-to-boundary distances inside $L$, and
\begin{align*}
A
&\coloneqq \dist_L(a_1,u)-\dist_L(a_1,x)
=\dist_L(a_2,u)-\dist_L(a_2,x).
\end{align*}
Double the weights and attach $u,x$ to a common sink $q$ with positive weights $\alpha,\beta$ satisfying $\beta-\alpha=2A$. The distance from a low-side vertex $z$ to $q$ is then
\begin{align*}
D_q(z)
&=\min\{2\dist_L(z,u)+\alpha,2\dist_L(z,x)+\beta\}.
\end{align*}
The triangle inequality gives $\dist_L(u,x)\geq\dist_L(a_1,x)-\dist_L(a_1,u)=-A$ and $\dist_L(x,u)\geq\dist_L(a_2,u)-\dist_L(a_2,x)=A$, so neither port reaches $q$ more cheaply through the other port. For $a_1$ and $a_2$, the two terms of the minimum coincide because $\beta-\alpha=2A$. Therefore
\begin{align*}
D_q(u)&=\alpha,\\
D_q(x)&=\beta,
\\
D_q(a_1)&=2\dist_L(a_1,u)+\alpha,\\
D_q(a_2)&=2\dist_L(a_2,x)+\beta.
\end{align*}
Thus the child paths lie in the common $D_q$-tight DAG, since each realizes the drop of $D_q$ between its endpoints. Any tight $a_1$--$u$ path $P$ satisfies
\begin{align*}
2w(P)
&=D_q(a_1)-D_q(u)
=2\dist_L(a_1,u),
\end{align*}
and the same calculation holds for $(a_2,x)$. Hence the prescribed-pair DAG algorithm of \cite{tho12}, run on this DAG, returns disjoint shortest $a_1$--$u$ and $a_2$--$x$ paths, which lie in $H_1$ and $H_2$ by \cref{lem:regional-closure}. Repeating on the reversed high side, the algorithm of \cite{tho12} returns both sides, which join through $e,f$.

Combining the three cases gives the main result of this subsection.

\begin{proposition}[One-sided reporting]
\label{prop:one-sided-reporting}
For every assignment selected at a two-active DAG cut with two nonzero child evaluations, one complete side can be reported in $O(m_R+n_R)$ time, or the entire pair can be reported within the same bound. At most one two-path child remains unresolved.
\end{proposition}
\begin{proof}
The four distance labels defining $\delta$ and $\gamma$ take linear time by topological dynamic programming. By \cref{lem:cross-route}, a strict displayed inequality invokes \cref{lem:common-sink} on the corresponding side, and only the opposite child remains. If neither is strict, the common-tight equality construction reports both sides. Concatenation through the selected crossing edges is valid by \cref{lem:cut-factorization}.
\end{proof}

\subsubsection{Undirected cuts}

At an undirected distance-difference cut, path $1$ crosses from the low side to the high side and path $2$ from the high side to the low side (\cref{sec:linear}), so each side contains the beginning of one path and the end of the other. Orient a selected cut pair by
\begin{align*}
e&:u\text{ (low)}\longrightarrow v\text{ (high)},\\
f&:x\text{ (high)}\longrightarrow y\text{ (low)}.
\end{align*}
Reverse the low suffix of path $2$ and the high suffix of path $1$. The low-side paths run from $a_1$ to $u$ and from $b_2$ to $y$; after the reversal, the high-side paths run from $b_1$ to $v$ and from $a_2$ to $x$. Define
\begin{align*}
\ell(z)
&\coloneqq \dist_R(a_1,z)-\dist_R(b_2,z),
\\
h(z)
&\coloneqq \dist_R(b_1,z)-\dist_R(a_2,z).
\end{align*}
Each side now carries two pieces that run from terminals to the cut, with sources $a_1,b_2$ on the low side and $b_1,a_2$ on the high side, and $\ell,h$ play the roles of $\delta,\gamma$. The cross routes traverse the other selected edge backwards (\cref{fig:cross-route}(b)), which is legal because the graph is undirected. One new point arises: a shortest route in $R$ between two low-side vertices may pass through the high side, so restricting to one side can change distances. The lemma therefore also checks that the order needed by the common-sink construction survives this restriction.

\begin{lemma}[Undirected one-sided reporting]
\label{lem:undirected-anchor}
At least one of $\ell(u)\leq\ell(y)$ and $h(v)\leq h(x)$ holds. On a side satisfying its weak inequality, restriction to that side preserves the order needed by the common-sink construction. A strict local order reports the side by a flow of value $2$; local equality reduces to two prescribed vertex-disjoint paths in one common DAG. Hence \cref{prop:one-sided-reporting} holds for an undirected distance-difference cut.
\end{lemma}
\begin{proof}
\emph{Cross-route comparison.} Suppose both inequalities fail strictly. Since $\dist_R(a_1,u)$ and $\dist_R(b_2,y)$ are finite, $\ell(u)>\ell(y)$ forces both values to be finite, as in the proof of \cref{lem:cross-route}; the same holds for $h(v)>h(x)$. Traverse $f$ backwards in the cross route from $a_1$ to $b_1$, and traverse $e$ backwards in the cross route from $a_2$ to $b_2$:
\begin{align*}
a_1&\leadsto y\mathbin{\circ}f\mathbin{\circ}x\leadsto b_1,\\
a_2&\leadsto v\mathbin{\circ}e\mathbin{\circ}u\leadsto b_2.
\end{align*}
As in \cref{lem:cross-route}, the low-side exchange changes the total length by $\ell(y)-\ell(u)$ and the high-side exchange by $h(x)-h(v)$. The sum of the cross-route lengths minus the two prescribed distances is therefore
\begin{align*}
-\bigl(\ell(u)-\ell(y)\bigr)-\bigl(h(v)-h(x)\bigr)
&<0,
\end{align*}
a contradiction.

Suppose $\ell(u)\leq\ell(y)$, and let $L$ be the low-side induced region. For $X\in\{R,L\}$, put
\begin{align*}
A_X
&\coloneqq\dist_X(a_1,u)-\dist_X(a_1,y),\\
B_X
&\coloneqq\dist_X(b_2,u)-\dist_X(b_2,y).
\end{align*}

We claim that $A_L\leq A_R\leq B_R\leq B_L$. The child paths lie in $L$ and are shortest in $R$, so they preserve the two source-to-boundary distances $\dist_X(a_1,u)$ and $\dist_X(b_2,y)$ in $L$, while the cross distances $\dist_X(a_1,y)$ and $\dist_X(b_2,u)$ can only increase from $R$ to $L\subseteq R$. This gives the outer inequalities. The middle one is $\ell(u)\leq\ell(y)$ rearranged.

For positive attachment weights $\alpha,\beta$, put
\begin{align*}
D_q(z)
&\coloneqq
\min\{2\dist_L(z,u)+\alpha,2\dist_L(z,y)+\beta\},
\end{align*}
the distance from $z$ to a new sink $q$ attached to $u$ and $y$ after doubling the weights of $L$. Orient $\{r,s\}$ from $r$ to $s$ only if $D_q(r)<+\infty$ and $D_q(r)=2w(r,s)+D_q(s)$; positive weights make this a DAG.

\emph{Strict local order.} If $A_L<B_L$, choose $\alpha,\beta$ by \cref{lem:common-sink}, with $A_L,B_L$ in place of $A,B$. As there, every shortest $a_1$--$q$ path enters $q$ through $u$ and every shortest $b_2$--$q$ path through $y$. The child paths then supply a flow of value $2$, whose pieces have the required endpoint distances.

\emph{Local equality.} Suppose $A_L=B_L$. Since $A_L\in\mathbb R\cup\{-\infty\}$ and $B_L\in\mathbb R\cup\{+\infty\}$, this value is finite. Choose positive integer weights $\alpha,\beta$ with $\beta-\alpha=2A_L$. The triangle inequalities $\dist_L(u,y)\geq-A_L$ and $\dist_L(y,u)\geq B_L=A_L$, together with the coincidence of the two terms of the minimum at $a_1$ and at $b_2$, give
\begin{align*}
D_q(u)&=\alpha,\\
D_q(y)&=\beta,
\\
D_q(a_1)&=2\dist_L(a_1,u)+\alpha,\\
D_q(b_2)&=2\dist_L(b_2,y)+\beta.
\end{align*}
The child paths belong to this common DAG, and any tight $a_1$--$u$ path $P$ satisfies
\begin{align*}
2w(P)
&=D_q(a_1)-D_q(u)
=2\dist_L(a_1,u),
\end{align*}
with the same identity for $(b_2,y)$. Hence the prescribed-pair DAG algorithm of \cite{tho12} returns shortest $a_1$--$u$ and $b_2$--$y$ paths.

Two undirected single-source shortest-path computations give all values $D_q$ in linear time \cite{tho97}; the high-side case is symmetric. Reversing a returned path piece restores its original orientation. One side satisfies its weak inequality, and the corresponding construction reports that side in linear time. This proves \cref{prop:one-sided-reporting} for an undirected distance-difference cut.
\end{proof}

\subsection{The recursion is linear}
\label{sec:linear}

This subsection completes the proof of \cref{thm:two-linear}. By \cref{prop:one-sided-reporting} and its undirected counterpart \cref{lem:undirected-anchor}, every cut of the reporting recursion leaves at most one two-path child, so it suffices that this child shrinks geometrically and that each region is handled in time linear in its size. We first make the number of edges of every region linear in its number of vertices by subdividing edges, and then treat DAGs and undirected graphs in turn.

Each region stores compact vertex and internal-adjacency lists obtained while partitioning its parent. Timestamped arrays avoid reinitializing $O(m+n)$ entries at every recursive call, and data used by the common-sink construction are discarded before recursion. Returned subpaths and selected cut edges are stored in concatenation trees and flattened once at the end, so each path prefix is stored once.

\begin{lemma}[Sparsity after edge subdivision]
\label{lem:subdivision}
Replace every edge $e=(p,q)$ of weight $w_e$ by a private vertex $z_e$ and the two-edge path $p$--$z_e$--$q$, with each new edge of weight $w_e$. Expansion and contraction give a bijection between original path pairs and expanded path pairs, preserve vertex-disjointness, and multiply every path length by two. Moreover, every induced region $R$ of the expanded graph satisfies $m_R\leq2n_R$.
\end{lemma}
\begin{proof}
Every original path expands uniquely, and every path between original vertices in the expanded graph contracts uniquely because each $z_e$ has degree two and is incident only to the two halves of $e$. Lengths double, so shortest paths correspond. Private subdivision vertices cannot create or remove an intersection between two expanded original-terminal paths.

Every expanded edge is incident to a subdivision vertex. Assign each regional edge to such an endpoint. Every subdivision vertex receives at most two edges, whence
\begin{align*}
m_R
&\leq2|R\cap\{z_e:e\in E\}|
\leq2n_R.
\end{align*}
\end{proof}

\subsubsection{DAGs}

Fix a topological order of the expanded DAG. At an unresolved region, cut at a median vertex of the full regional order. \Cref{alg:recursive-report}, together with \cref{prop:one-sided-reporting}, leaves at most one two-path child.

\begin{lemma}[DAG recurrence]
\label{lem:dag-recurrence}
The DAG algorithm has expected running time $O(m+n)$ and uses $O(m+n)$ storage cells.
\end{lemma}
\begin{proof}
Every unresolved child contains at most $\lceil n_R/2\rceil$ vertices. By \cref{lem:subdivision}, its induced graph and cancellation circuit have size $O(n_R)$. Circuit construction, coefficient extraction, the at most two assignment tests, common-sink construction, flow, and path reporting require a constant expected number of linear passes. Therefore
\begin{align*}
T(N)
&\leq cN+T(\lceil N/2\rceil)
=O(N)
=O(m+n).
\end{align*}
Only one child survives. Releasing the parent circuit and temporary flow arrays before recursion gives $O(N)$ peak space.
\end{proof}

\subsubsection{Undirected graphs}

If some $t_i$ is disconnected from $s_i$, the instance is infeasible. If the two terminal pairs lie in different connected components, report their shortest paths independently. Otherwise restrict the instance to their common component; all values $d_1(v),d_2(v)$ below are then finite.

For the expanded undirected graph define
\begin{align*}
\Delta(v)
&\coloneqq d_1(v)-d_2(v).
\end{align*}
If an edge $(u,v)$ is oriented tightly in $H_1$, the triangle inequality for $d_2$ gives $\Delta(u)\leq\Delta(v)$; on $H_2$ the inequality is reversed. A nonvertex threshold therefore is a monotone cut in the sense of \cref{sec:extraction}. If an $H_1$- or $H_2$-edge has $\Delta(u)=\Delta(v)$, it is tight in both systems with the same orientation. Hence the two tight graphs induce the same DAG on every equality plateau.

Choose a median $\mu$ of the multiset of $\Delta$-values in $R$ and put
\begin{align*}
A&\coloneqq |\{v\in R:\Delta(v)<\mu\}|,\\
P&\coloneqq |\{v\in R:\Delta(v)=\mu\}|,\\
B&\coloneqq |\{v\in R:\Delta(v)>\mu\}|.
\end{align*}
If $A\geq n_R/4$, cut immediately below $\mu$. If $B\geq n_R/4$, cut immediately above $\mu$. In either case both sides contain at most $3n_R/4$ vertices. Otherwise $P>n_R/2$; cut at one boundary of this heavy plateau. If the plateau is retained, the next call cuts at its other boundary. After at most two such cuts the remaining region is an exterior with fewer than $n_R/2$ vertices or the plateau itself.

The single-value plateau base case occurs when the two terminal ranges have only one common $\Delta$-value; the corresponding plateau may contain many vertices. The following lemma gives the exact reduction and reconstruction bound.

\begin{lemma}[Single-value plateau base case]
\label{lem:plateau-base}
Let
\begin{align*}
J_1
&\coloneqq[\Delta(a_1),\Delta(b_1)],\\
J_2
&\coloneqq[\Delta(b_2),\Delta(a_2)],
\end{align*}
and suppose $J_1\cap J_2=\{\delta\}$. Put $Z\coloneqq\{v\in R:\Delta(v)=\delta\}$, and suppose $K\coloneqq H_1[Z]=H_2[Z]$ is a DAG. In $O(m_R+n_R)$ time, one can either report a path pair in $R$ when one path avoids $Z$, or construct an auxiliary instance on $K$ with two prescribed terminal pairs and size $O(m_R+n_R)$. Every solution of the auxiliary instance lifts to a solution in $R$ in $O(m_R+n_R)$ time. Over all plateau leaves of the separator recursion, the total reduction and lifting cost is $O(m+n)$.
\end{lemma}
\begin{proof}
A common vertex of two paths in $R$ has a $\Delta$-value in $J_1\cap J_2$, so every possible intersection lies in $Z$. Search $H_i[R-Z]$ for an $a_i$--$b_i$ path for each $i$. If one exists, combine it with an arbitrary path for the other pair in $R$. The resulting pair is disjoint because the first path contains no vertex of $Z$.

\emph{Entry and exit attachments.} Assume now that every path in $R$ visits $Z$. For $i\in\{1,2\}$, define the possible first and last plateau vertices by
\begin{align*}
S_i^Z
&\coloneqq
\begin{cases}
\{a_i\}, & a_i\in Z,\\
\{z\in Z:\text{an $a_i$--$z$ path $P$ in $H_i[R]$ has $V(P)\cap Z=\{z\}$}\}, & a_i\notin Z,
\end{cases}
\\
T_i^Z
&\coloneqq
\begin{cases}
\{b_i\}, & b_i\in Z,\\
\{z\in Z:\text{a $z$--$b_i$ path $P$ in $H_i[R]$ has $V(P)\cap Z=\{z\}$}\}, & b_i\notin Z.
\end{cases}
\end{align*}
Add private vertices $\alpha_i,\beta_i$, edges $\alpha_i\to z$ for $z\in S_i^Z$, and edges $z\to\beta_i$ for $z\in T_i^Z$. Denote the resulting DAG by $K^+$.

Any path in $R$ meets $Z$ in one contiguous segment by $\Delta$-monotonicity. Its first and last plateau vertices therefore define an $\alpha_i$--$\beta_i$ path in $K^+$. Conversely, an $\alpha_i$--$\beta_i$ path selects an entry $z_i\in S_i^Z$, an exit $w_i\in T_i^Z$, and a $z_i$--$w_i$ path in $K$. Join this plateau segment to witnessing exterior paths for the two selected attachments.

This lifting preserves disjointness. The auxiliary paths are disjoint on $Z$, and every exterior witness avoids $Z$ internally. Moreover, at most one path has a nonempty exterior below $\delta$: otherwise $\Delta(a_1)<\delta$ and $\Delta(b_2)<\delta$, contradicting $J_1\cap J_2=\{\delta\}$. The symmetric argument applies above $\delta$. Hence exterior pieces of the two paths cannot intersect. Within each path, the prefix, plateau segment, and suffix occupy strictly ordered $\Delta$-ranges. All lifted edges belong to $H_i$, so both lifted paths remain shortest.

\emph{Construction and cost.} When $a_i\notin Z$, the set $S_i^Z$ is found by one forward reachability scan from $a_i$ through $R-Z$, followed by a scan of edges entering $Z$; when $a_i\in Z$, its unique first plateau vertex is $a_i$. The reverse rule from $b_i$ gives $T_i^Z$. Store one predecessor or successor per reached exterior vertex. These four forests recover the exterior witnesses only for the entries and exits selected by the auxiliary solution.

The prescribed-pair DAG algorithm of \cite{tho12} reports the paths in $K^+$ in $O(|E(K^+)|+|V(K^+)|)$ time. Each plateau vertex creates at most four attachment edges, so constructing $K^+$, solving it, and following the selected forest paths costs $O(m_R+n_R)$. Plateau leaves are disjoint regions of the separator partition; summing their vertex lists, edge lists, attachment edges, and four forests gives $O(m+n)$.
\end{proof}

\begin{lemma}[Undirected recurrence]
\label{lem:undirected-recurrence}
The undirected algorithm has expected running time $O(m+n)$ and uses $O(m+n)$ words of space.
\end{lemma}
\begin{proof}
The active-count cases are handled exactly as in the DAG algorithm, using \cref{lem:undirected-anchor} at a two-active cut and \cref{lem:plateau-base} at a single-value plateau base case. Deterministic linear selection finds $\mu$, and each cut uses a constant expected number of linear circuit evaluations, shortest-path computations, differentiation sweeps, assignment tests, and common-sink constructions. The single-source shortest-path calls take linear time \cite{tho97}.

Grouping a balanced cut or a two-boundary heavy phase gives one of the recurrences
\begin{align*}
T(N)
&\leq cN+T(3N/4),
\\
T(N)
&\leq2cN+T(N/2).
\end{align*}
Both solve to $O(N)$. By the final assertion of \cref{lem:plateau-base}, the total cost of all single-value plateau base cases is linear. Depth-first execution releases all temporary data of the parent before entering the unique child, proving the space bound.
\end{proof}

\begin{lemma}[Randomization and output verification]
\label{lem:probability}
Let $N$ be the number of vertices of the expanded graph in the linear algorithms and the input size $N\coloneqq m_0+n_0$ in the directed algorithm of \cref{sec:directed}. For every constant $c>0$, a field of size at least $2N^{c+2}$ makes the feasible-instance false-negative probability at most $N^{-(c+1)}$. Internal cut sampling changes only the expected running time. Every reported pair is valid.
\end{lemma}
\begin{proof}
Two vertex-disjoint paths use fewer edges than the graph has vertices, so the root polynomial has degree at most $2(N-1)$: the expanded graph has $N$ vertices, and in the directed algorithm the degree is at most $n_0-2\leq2(N-1)$. Schwartz--Zippel bounds the probability that the root evaluation vanishes by $2(N-1)/|\F|\leq N^{-(c+1)}$. All internal sampling is the detected retry of \cref{lem:interpolated-extraction} and affects only the expected running time.

Finally, contract every subdivision vertex in the linear algorithms, and verify the endpoints, edge membership, total lengths, and vertex-disjointness of the two output paths in linear time. Thus randomization can cause only a false \textsc{no}, never an invalid witness.
\end{proof}

We now prove the formal version of \cref{thm:two-linear-informal}.

\begin{theorem}[Linear-time 2-DSP, formal version of \cref{thm:two-linear-informal}]
\label{thm:two-linear}
Let $G$ have $n$ vertices, $m$ edges, and four distinct terminals. Suppose $G$ is either a DAG with arbitrary real edge weights, or an undirected graph with positive integer edge weights bounded by $(m+n)^{O(1)}$. In the respective computational models of \cref{sec:prelim}, for every constant $c>0$, there is a randomized algorithm using expected $O(m+n)$ time and $O(m+n)$ storage cells that never returns an invalid pair, returns \textnormal{\textsc{no}} on every infeasible instance, and reports two vertex-disjoint shortest paths with probability at least $1-(m+n)^{-c}$ on every feasible instance. Under a feasibility promise, root resampling gives a Las Vegas algorithm with the same expected running time.
\end{theorem}
\begin{proof}
\label{proof:two-linear}
\Cref{lem:subdivision,lem:dag-recurrence,lem:undirected-recurrence,lem:probability} give the running time, space, and error guarantees. Under a feasibility promise, root resampling succeeds with probability $1-N^{-\Omega(1)}$ and therefore takes a constant expected number of trials.
\end{proof}

\section{Two disjoint shortest paths in directed graphs}
\label{sec:directed}

This section proves \cref{thm:two-directed}. The undirected one-sided argument used cross routes traversing the selected edges in reverse. A directed graph need not contain these routes, so here we retain both children of a cut. This is affordable because the detector has superlinear regional cost: splitting a region decreases the sum of its children's costs by a constant factor. The remaining task is to build the regional detector cheaply.

We do this using only the two source-distance labelings. On the regional shortest-path supports, the cancellation recurrence of \cite{cks26} becomes a triangular matrix computation: each column uses only earlier columns. Evaluating one column at a time gives $O(n_R(m_R+n_R))$ gates; batching earlier-to-later contributions gives $O(n_R^{\omega+\varepsilon})$ gates. We first justify the recurrence, then give both evaluation schedules, and finally sum their costs over the reporting recursion.

The algorithm needs the two source-distance labelings. The next lemma computes them within the claimed time, also for negative weights.

\begin{lemma}[Source distances]
\label{lem:sssp}
Let $G$ be a directed graph with $n$ vertices, $m$ edges, no parallel edges, and real edge weights in which every directed cycle has positive total weight, and let $s$ be a vertex. The labeling $\dist_G(s,\cdot)$, together with a shortest-path tree, can be computed in expected $O(m+n)+m^{1+o(1)}$ time, and the output is always correct.
\end{lemma}
\begin{proof}
A labeling $d$ equals $\dist_G(s,\cdot)$ if $d(s)=0$, $d$ is finite exactly on the vertices reachable from $s$, $d(v)\leq d(u)+w(u,v)$ for every edge $(u,v)$ with $d(u)$ finite, and every reachable vertex $v\neq s$ has an incoming edge $(u,v)$ with $d(v)=d(u)+w(u,v)$. Indeed, the edge inequalities, applied along a shortest $s$--$v$ path, give $d(v)\leq\dist_G(s,v)$. Conversely, fix one such tight incoming edge for every reachable vertex other than $s$, and follow the fixed edges backwards from $v$. The walk never repeats a vertex, since a repeated vertex would close a cycle of tight edges, whose weight telescopes to zero, whereas every cycle has positive weight. It can stop only at $s$, and reversed it is an $s$--$v$ path of weight $d(v)-d(s)=d(v)$; hence $\dist_G(s,v)\leq d(v)$. These conditions, and the tree formed by the tight edges, are checked in $O(m+n)$ time.

First restrict $G$ to the vertices reachable from $s$ in $O(m+n)$ time, and give every other vertex the label $+\infty$. The restricted graph has $n'$ vertices and $m'\geq n'-1$ edges, since every vertex other than $s$ has an incoming edge from the search tree. On it, run the Bellman--Ford algorithm \cite{bel58} and the Las Vegas algorithm of \cite{hll+26} alternately, one step at a time, verify every labeling that either returns, and stop at the first verified labeling. Bellman--Ford returns the distances within $O(m'n')$ steps. With high probability, in particular with probability at least $1-1/n'$, the algorithm of \cite{hll+26} returns the distances within $m'^{1+o(1)}$ steps. The expected number of steps is therefore at most $m'^{1+o(1)}+O(m'n')/n'=m'^{1+o(1)}$, and the total expected time is $O(m+n)+m^{1+o(1)}$.
\end{proof}

Let $n_0,m_0$ be the input sizes. The preprocessing has three phases.
\begin{itemize}
\item Delete self-loops, which are positive-weight cycles and never lie on a shortest path, and keep one minimum-weight copy of every set of parallel edges; only such copies can be tight, and vertex-disjointness does not distinguish parallel copies.
\item Compute the source-distance labels $d_i(v)=\dist_G(s_i,v)$ for $i\in\{1,2\}$, together with their shortest-path trees, by \cref{lem:sssp}, in expected $O(m_0+n_0)+m_0^{1+o(1)}$ time; for nonnegative weights, Dijkstra's algorithm with Fibonacci heaps takes $O(m_0+n_0\log n_0)$ time \cite{ft87}. These two labelings are the only distances used by the algorithm.
\item If $d_i(t_i)=+\infty$ for some $i$, the instance is infeasible. If the two terminal pairs lie in different weak components, the two shortest-path trees report their paths. Otherwise retain their common weak component and write $n,m$ for its sizes; then $n\leq m+1$ because the component is weakly connected, $m\leq m_0$, and $m\leq n^2$ because no self-loops or parallel edges remain.
\end{itemize}

Every tight shortest-path graph is acyclic because every directed cycle has positive total weight, whereas the weight of a directed cycle of edges tight for $d_i$ telescopes to zero.

Delete every vertex with $d_1(v)=d_2(v)=+\infty$, since no prescribed path can use it. On the remaining vertices, extend the distance-difference coordinate by
\begin{align*}
\Delta(v)
&\coloneqq
\begin{cases}
d_1(v)-d_2(v), & d_1(v),d_2(v)<+\infty,\\
-\infty, & d_1(v)<+\infty=d_2(v),\\
+\infty, & d_2(v)<+\infty=d_1(v).
\end{cases}
\end{align*}
Thus vertices reached only from $s_1$ sit at the bottom of the extended order, and vertices reached only from $s_2$ at the top.

The next lemma is the directed counterpart of the monotonicity used for undirected graphs in \cref{sec:linear}. It makes thresholds of $\Delta$ monotone cuts in the sense of \cref{sec:extraction} and forces the crossing-edge assignment at each of them.

\begin{lemma}[Directed distance-difference separator]
\label{lem:directed-delta}
The coordinate $\Delta$ is nondecreasing on $H_1$ and nonincreasing on $H_2$, including in the extended order. Every finite equality plateau is a common DAG. At a strict nonvertex threshold, the first path crosses on a strictly increasing edge and the second on a strictly decreasing edge, so the two cut-edge sets are disjoint and their assignment to the terminal pairs is forced.
\end{lemma}
\begin{proof}
If $(u,v)\in H_1$ and $d_2(u),d_2(v)$ are finite, then
\begin{align*}
d_1(v)-d_1(u)
&=w(u,v)
\geq d_2(v)-d_2(u),
\end{align*}
where the equality is the tightness of $(u,v)$ for $d_1$ and the inequality is the directed triangle inequality along $(u,v)$. Thus $\Delta(u)\leq\Delta(v)$. For $(u,v)\in H_2$ with $d_1(u),d_1(v)$ finite, exchanging the roles of $d_1$ and $d_2$ gives $d_2(v)-d_2(u)=w(u,v)\geq d_1(v)-d_1(u)$, hence $\Delta(u)\geq\Delta(v)$. It remains to treat infinite values. Both endpoints of an $H_1$-edge have finite $d_1$, so neither has $\Delta=+\infty$. Along an $H_1$-edge, reachability from $s_2$ can change only from false to true, so if one of the two values is $-\infty$, then $\Delta(u)=-\infty\leq\Delta(v)$. On an $H_2$-edge the value $-\infty$ does not occur, reachability from $s_1$ can change only from false to true, and a value $+\infty$ forces $\Delta(u)=+\infty\geq\Delta(v)$.

Let $(u,v)\in H_1$ with $\Delta(u)=\Delta(v)=\delta$. Then $d_2(u),d_2(v)$ are finite, and equality at a finite value in the displayed inequality gives $d_2(v)=d_2(u)+w(u,v)$, so the edge is tight in both $H_1$ and $H_2$. The same computation with $d_1$ and $d_2$ exchanged shows that every $H_2$-edge inside the plateau lies in $H_1$. A directed cycle of finite equality edges would be tight for $d_1$ and would have total weight zero, so each finite plateau is acyclic.

Finally, consider a strict nonvertex threshold. An $H_1$-path crossing it uses an edge whose tail lies below and whose head lies above the threshold, so $\Delta$ strictly increases along this edge; by the first step, an $H_2$-path crosses on an edge along which $\Delta$ strictly decreases. No edge has both properties, so the two cut-edge sets are disjoint, and the direction of a cut edge determines its terminal pair.
\end{proof}

\subsection{Aligned regions and their supports}

Call an instance $(R;a_1,b_1,a_2,b_2)$ \emph{aligned} when $R=\{v:\Delta(v)\in I\}$ for an interval $I$ of the extended line and, for each $i$, the endpoints $a_i,b_i$ occur in this order on an $H_i[R]$-path. Its \emph{supports} are the subgraphs $H_i^\star\subseteq H_i[R]$ formed by the vertices and edges of $a_i$--$b_i$ paths in $H_i[R]$. Thus the $i$th paths of the instance are exactly the $a_i$--$b_i$ paths of $H_i^\star$, and the supports are the only parts of the tight graphs that the regional polynomial $F_R$ sees. Both supports are acyclic as subgraphs of $H_i$. Every vertex of $H_i^\star$ lies on an $a_i$--$b_i$ path, so $a_i$ is the unique source of $H_i^\star$, and $b_i$ is its unique sink. The edges of $H_i[R]$ are recognized from $d_i$, and one forward search from $a_i$ and one backward search from $b_i$ construct $H_i^\star$ in $O(m_R+n_R)$ time.

The detector of \cite{cks26} exchanges subpaths between the two terminal pairs. In \cref{lem:first-intersection} below, a piece of a first path and a piece of a second path both run from a vertex $r$ to a vertex $t$, and the cancellation swaps their $r$--$t$ segments; the segment taken from $H_1^\star$ must then be a path of $H_2^\star$, and conversely. The following lemma shows that the two supports agree on every segment they share, so these exchanges stay inside $H_1^\star$ and $H_2^\star$ without any further distance information.

\begin{lemma}[Shared segments]
\label{lem:directed-exchange}
Let $R\subseteq V$, let $H_i^\star$ be formed by the vertices and edges of $a_i$--$b_i$ paths in $H_i[R]$, and suppose that $r$ reaches $t$ in both $H_1^\star$ and $H_2^\star$. Then $H_1^\star$ and $H_2^\star$ have the same $r$--$t$ paths, namely the shortest $r$--$t$ paths of $G$ contained in $G[R]$.
\end{lemma}
\begin{proof}
Fix $i\in\{1,2\}$. An $r$--$t$ path exists in $H_i^\star$, and its tight edges give weight $d_i(t)-d_i(r)$. The triangle inequality gives the reverse bound
$\dist_G(r,t)\geq d_i(t)-d_i(r)$, so this path, and every other $r$--$t$ path of $H_i^\star$, is shortest.

Conversely, let $P\subseteq G[R]$ be any globally shortest $r$--$t$ path. Prefix it by a shortest $s_i$--$r$ path. The resulting walk has length $d_i(r)+\dist_G(r,t)=d_i(t)$; a repeated vertex would create a positive-weight cycle that could be deleted to shorten it. Thus every edge of $P$ is tight for $d_i$. Since $r,t\in H_i^\star$, join $P$ to an $a_i$--$r$ path and a $t$--$b_i$ path of $H_i^\star$. This is a walk in the acyclic graph $H_i[R]$, hence a path, so $P\subseteq H_i^\star$.
\end{proof}

\subsection{A causal matrix form of the cancellation recurrence}
\label{sec:directed-recurrence}

This subsection expresses $F_R$ for an aligned instance through auxiliary states collected in a matrix $\Xi$. The main identity (\cref{lem:first-intersection}) writes every state through $\Xi$ and the path matrices $M_1,M_2$ of the supports, and \cref{lem:causal-recurrence} shows that each column of $\Xi$ depends only on earlier columns in a topological order of $H_1^\star$.

Boundary instances need no recurrence. If $a_i=b_i=v$, the $i$th path has zero edges and $F_R$ is the path sum of the other pair in $H_{3-i}[R-v]$; if both paths have zero edges, $F_R$ is $1$ when their vertices differ and $0$ otherwise; and a terminal shared by the two pairs gives $F_R=0$. Each case costs one path-sum pass. Fix now an aligned instance with four distinct terminals and two nonempty paths. Index rows and columns by $V_R^\star\coloneqq V(H_1^\star)\cup V(H_2^\star)$, let $W_i$ be the matrix with entry $x_{pq}$ at every edge $(p,q)$ of $H_i^\star$ and zero elsewhere, and let
\begin{align*}
M_i
&\coloneqq(I-W_i)^{-1}.
\end{align*}
Since $H_i^\star$ is acyclic, $W_i$ is nilpotent, and $M_i=\sum_{\ell\geq0}W_i^\ell$. The entry $W_i^\ell[x,y]$ sums the monomials of the $x$--$y$ walks of $H_i^\star$ with $\ell$ edges, which are paths by acyclicity. Hence $M_i[x,y]$ is the sum of the monomials $\pi(P)\coloneqq\prod_{e\in E(P)}x_e$ over the $x$--$y$ paths $P$ of $H_i^\star$, with $M_i[x,x]=1$. Put $U_R\coloneqq V(H_1^\star)\cap V(H_2^\star)$ and $C_R\coloneqq E(H_1^\star)\cap E(H_2^\star)$; a path of $H_1^\star$ and a path of $H_2^\star$ can share only vertices of $U_R$ and edges of $C_R$.

The auxiliary states of the recurrence fix the first path's source $a_1$ and the second path's sink $b_2$ and vary the two remaining endpoints. For $y\in V(H_1^\star)$ and $x\in V(H_2^\star)$, define
\begin{align*}
\widehat F(y,x)
&\coloneqq\sum_{(P_1,P_2)}\pi(P_1)\pi(P_2),
\end{align*}
where $P_1$ ranges over the $a_1$--$y$ paths of $H_1^\star$, $P_2$ over the $x$--$b_2$ paths of $H_2^\star$, and the pair must satisfy $V(P_1)\cap V(P_2)\subseteq\{y\}\cap\{x\}$; an empty path at a vertex $v$ has vertex set $\{v\}$. Thus the two paths are disjoint, except that they may share the free endpoint when $y=x$. The output state has disjoint vertex sets, because $b_1\neq a_2$, and therefore no shared edge, so
\begin{align*}
F_R
&=\widehat F(b_1,a_2).
\end{align*}
The recurrence uses the \emph{diagonal states} $\widehat F(v,v)$ for $v\in U_R$ and the \emph{edge states} $\widehat F(a,v)$ for $(a,v)\in C_R$. Collect them in the matrix $\Xi$ indexed by $V_R^\star$, with
\begin{align*}
\Xi[v,v]
&\coloneqq\widehat F(v,v)
\qquad(v\in U_R),
\\
\Xi[a,v]
&\coloneqq x_{av}^2\widehat F(a,v)
\qquad\bigl((a,v)\in C_R\bigr),
\end{align*}
and all other entries zero. Column $v$ of $\Xi$ thus consists of the states whose second endpoint is $v$. A diagonal state counts two paths that meet exactly at $v$, the first ending and the second starting there. An edge state arises from configurations in which both paths traverse the common edge $(a,v)$, and the factor $x_{av}^2$ records the two traversals.

\begin{figure}[t]
\centering
\begin{tikzpicture}[x=0.66cm,y=0.82cm,font=\normalsize,line cap=round]
\node at (4,3.2) {(a) Exchange the two $r$--$t$ segments};
\draw[dp one,dp arrow] (0,2.4) .. controls (1.1,2.4) and (1.5,1.4) .. (2.3,1.2);
\draw[dp two,dp arrow] (0,0) .. controls (1.1,0) and (1.5,1) .. (2.3,1.2);
\draw[dp one,dp arrow] (2.3,1.2) .. controls (3.2,2.5) and (4.8,2.5) .. (5.7,1.2);
\draw[dp two,dp arrow] (2.3,1.2) .. controls (3.2,-0.1) and (4.8,-0.1) .. (5.7,1.2);
\draw[dp one,dp arrow] (5.7,1.2) .. controls (6.5,1.4) and (6.9,2.4) .. (8,2.4);
\draw[dp two,dp arrow] (5.7,1.2) .. controls (6.5,1) and (6.9,0) .. (8,0);
\node[above,DPBlue] at (4,2.25) {$\beta[r,t]$};
\node[below,DPOrange] at (4,0.15) {$\alpha[r,t]$};
\node[below=3pt,dp label] at (2.3,1.2) {$r$};
\node[above=3pt,dp label] at (5.7,1.2) {$t$};
\foreach \x/\y/\lab/\pos in {0/2.4/a_1/left,0/0/x/left,8/2.4/y/right,8/0/b_2/right}{\fill (\x,\y) circle (1.8pt) node[\pos] {$\lab$};}
\fill (2.3,1.2) circle (1.8pt);\fill (5.7,1.2) circle (1.8pt);
\begin{scope}[shift={(10.2,0)}]
\node at (4,3.2) {(b) Fixed point: one shared edge};
\draw[dp one,dp arrow] (0,2.4) .. controls (1.2,2.4) and (2,1.4) .. (3,1.2);
\draw[dp two,dp arrow] (0,0) .. controls (1.2,0) and (2,1) .. (3,1.2);
\draw[dp edge] (3,1.2) -- (5,1.2);
\draw[dp one,dp arrow] (5,1.2) .. controls (6,1.4) and (6.8,2.4) .. (8,2.4);
\draw[dp two,dp arrow] (5,1.2) .. controls (6,1) and (6.8,0) .. (8,0);
\node[below=3pt,dp label] at (4,1.2) {weight $x_{rt}^2$};
\node[above] at (3,1.3) {$r$};\node[above] at (5,1.3) {$t$};
\foreach \x/\y/\lab/\pos in {0/2.4/a_1/left,0/0/x/left,8/2.4/y/right,8/0/b_2/right}{\fill (\x,\y) circle (1.8pt) node[\pos] {$\lab$};}
\fill (3,1.2) circle (1.8pt);\fill (5,1.2) circle (1.8pt);
\end{scope}
\end{tikzpicture}
\caption{The involution in \cref{lem:first-intersection}. The first path runs from $a_1$ to $y$ (blue), and the second from $x$ to $b_2$ (orange). In (a), $r$ is the last common vertex before $t$ on the blue prefix; the two $r$--$t$ segments lie in both supports by \cref{lem:directed-exchange}. Exchanging them preserves the monomial, so distinct paired configurations cancel in characteristic two. In (b), both segments are the same edge. These fixed points survive with the squared edge weight, giving the term $M_2[x,r]\Xi[r,t]M_1[t,y]$.}
\label{fig:directed-swap}
\end{figure}
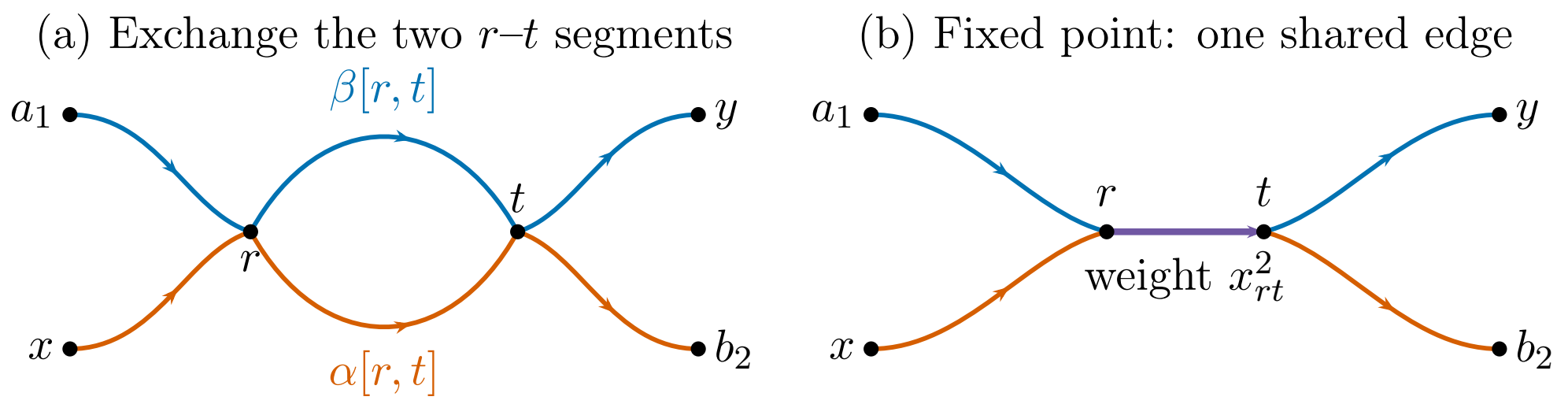

\Cref{fig:directed-swap} illustrates the identity behind the recurrence. The product $M_1[a_1,y]M_2[x,b_2]$ counts all pairs of an $a_1$--$y$ path of $H_1^\star$ and an $x$--$b_2$ path of $H_2^\star$, so in characteristic two $\widehat F(y,x)$ is this product plus the contribution of the pairs that meet. Cut a meeting pair at the first vertex $t$ of its first path that lies on its second path, not counting a shared endpoint. This gives four pieces: $\beta$ from $a_1$ to $t$ and $\delta$ from $t$ to $y$ on the first path, and $\alpha$ from $x$ to $t$ and $\gamma$ from $t$ to $b_2$ on the second. The pieces $\alpha$ and $\delta$ are counted by $M_2[x,t]$ and $M_1[t,y]$, and the pair $(\beta,\gamma)$, which meets only at $t$, by the diagonal state $\Xi[t,t]$. This product ignores one constraint, namely that $\beta$ avoids $\alpha$ before $t$. Swapping the last common segment of $\alpha$ and $\beta$ cancels the violating quadruples in pairs, and the swap stays in the supports by \cref{lem:directed-exchange}. The only unpaired violations are those in which $\alpha$ and $\beta$ enter $t$ through one common edge $(r,t)$, and they are counted by the edge state $\Xi[r,t]$. Altogether, the meeting pairs with first intersection $t$ contribute $(M_2\Xi)[x,t]M_1[t,y]$.

The next identity is the first-intersection cancellation of \cite[Section~4]{cks26}, specialized to the two supports. Its proof needs only acyclicity and \cref{lem:directed-exchange}; in particular, every sum ranges over all of $U_R$ and $C_R$, and the restrictions to admissible intersections are carried by the path matrices themselves. For instance, a vertex $t$ that does not reach $y$ in $H_1^\star$ contributes nothing, because $M_1[t,y]=0$. This uniformity over all states is what turns the recurrence into a matrix product.

\begin{lemma}[First-intersection identity]
\label{lem:first-intersection}
For every $y\in V(H_1^\star)$ and $x\in V(H_2^\star)$,
\begin{align*}
\widehat F(y,x)
&=M_1[a_1,y]M_2[x,b_2]
+\sum_{t\in U_R\setminus(\{y\}\cap\{x\})}(M_2\Xi)[x,t]M_1[t,y].
\end{align*}
\end{lemma}
\begin{proof}
Put $S\coloneqq\{y\}\cap\{x\}$, and let $\pi(\mathcal X)$ denote the sum of the path-product monomials of a family $\mathcal X$. The product $M_1[a_1,y]M_2[x,b_2]$ counts all path pairs. The desired pairs may meet only in $S$; in characteristic two, adding the contribution of all other pairs removes them.

\emph{First-intersection classes.} Let $t\in U_R\setminus S$ be the first vertex of the first path that lies on the second path outside $S$. Split the paths into
\begin{align*}
\alpha&=P_2[x,t],\\
\beta&=P_1[a_1,t],\\
\gamma&=P_2[t,b_2],\\
\delta&=P_1[t,y].
\end{align*}
For fixed $t$, let $\mathcal Q_t$ contain all such quadruples with the indicated endpoints and supports, subject only to $V(\beta)\cap V(\gamma)=\{t\}$. Then
\begin{align*}
\pi(\mathcal Q_t)&=M_2[x,t]\Xi[t,t]M_1[t,y].
\end{align*}
Acyclicity makes $\beta\delta$ and $\alpha\gamma$ paths. Call a quadruple \emph{good} when $V(\alpha)\cap V(\beta)=\{t\}$, and \emph{bad} otherwise, with families $\mathcal G_t,\mathcal B_t$.
Splitting at the first intersection gives a monomial-preserving bijection onto $\mathcal G_t$: vertices of $\beta$ before $t$ avoid both $\alpha$ and $\gamma$, since $S$ is contained in the last vertex $y$ of the first path. Conversely these two avoidance conditions make $t$ the first forbidden intersection of the joined pair.

\emph{Cancellation.} For a bad quadruple, let $r$ be the last vertex of $\beta$ before $t$ lying on $\alpha$. Both supports contain an $r$--$t$ path, so \cref{lem:directed-exchange} allows us to exchange the two segments:
\begin{align*}
\phi(\alpha,\beta,\gamma,\delta)
&=\bigl(\alpha[x,r]\beta[r,t],\ \beta[a_1,r]\alpha[r,t],\ \gamma,\ \delta\bigr).
\end{align*}
Each concatenation stays in its support and is a path by acyclicity. The new second path avoids $\gamma$ before $t$: its $\beta$-prefix already did, and its $\alpha$-suffix does because $\alpha\gamma$ is a path. The image is bad with the same last shared vertex $r$. Indeed, the interior of $\alpha[r,t]$ avoids $\alpha[x,r]$ by simplicity and avoids $\beta[r,t]$ by the choice of $r$. Applying $\phi$ again therefore exchanges the same segments back. The map preserves the multiset of edges, so nonfixed quadruples cancel in pairs.

\emph{Fixed points.} A fixed point has $\alpha[r,t]=\beta[r,t]$. This segment has no interior shared vertex by the choice of $r$, hence is one common edge $(r,t)\in C_R$. Removing its two traversals leaves an $x$--$r$ path of $H_2^\star$, a pair counted by $\widehat F(r,t)$, and a $t$--$y$ path of $H_1^\star$. Conversely, these pieces reconstruct a fixed point: inserting the common edge preserves disjointness of the middle pair before $t$, and acyclicity makes both concatenations paths. Thus
\begin{align*}
\pi(\mathcal B_t)
&=\sum_{(r,t)\in C_R}M_2[x,r]x_{rt}^2\widehat F(r,t)M_1[t,y]\\
&=\sum_{(r,t)\in C_R}M_2[x,r]\Xi[r,t]M_1[t,y].
\end{align*}
The first equality uses cancellation and the fixed-point bijection; the second is the definition of $\Xi[r,t]$. Since $\mathcal Q_t=\mathcal G_t\sqcup\mathcal B_t$, characteristic two gives
\begin{align*}
\pi(\mathcal G_t)
&=\pi(\mathcal Q_t)+\pi(\mathcal B_t)\\
&=(M_2\Xi)[x,t]M_1[t,y],
\end{align*}
using the diagonal term and the common-edge terms that constitute column $t$ of $\Xi$. Summing over $t\in U_R\setminus S$ counts every unwanted pair once and proves the identity.
\end{proof}

Only some states are needed: the diagonal states, the edge states, and the output state. For a state in column $v$ of $\Xi$, the first argument $y$ in \cref{lem:first-intersection} is $v$ itself or the tail $a$ of an edge $(a,v)$ of $H_1^\star$. Since $M_1[t,y]=0$ unless $t$ reaches $y$ in $H_1^\star$, only first intersections $t$ before $v$ in a topological order of $H_1^\star$ contribute. Thus column $v$ of $\Xi$ depends only on earlier columns, and $\Xi$ can be filled column by column, as in forward substitution for a triangular system. The following accumulated sums collect the contributions of the earlier columns: row $v$ of $\Pi$ is row $v$ of $M_2\Xi M_1$ with the middle index restricted to vertices before $v$.

Order $V(H_1^\star)$ by a topological order $\prec$ of $H_1^\star$. For $v\in U_R$ and $a\in V_R^\star$, define the accumulated sums
\begin{align*}
\Pi[v,a]
&\coloneqq\sum_{\substack{u\in U_R\\u\prec v}}(M_2\Xi)[v,u]M_1[u,a].
\end{align*}

\begin{lemma}[Causal recurrence]
\label{lem:causal-recurrence}
For every $v\in U_R$ and every $(a,v)\in C_R$,
\begin{align*}
\widehat F(v,v)
&=M_1[a_1,v]M_2[v,b_2]+\Pi[v,v],
\\
\widehat F(a,v)
&=M_1[a_1,a]M_2[v,b_2]+\Pi[v,a],
\end{align*}
and
\begin{align*}
F_R
&=M_1[a_1,b_1]M_2[a_2,b_2]+(M_2\Xi M_1)[a_2,b_1].
\end{align*}
In particular, column $v$ of $\Xi$ is determined by $M_1$, $M_2$, and the columns $u\prec v$ of $\Xi$.
\end{lemma}
\begin{proof}
Apply \cref{lem:first-intersection}; in each case, the structural zeros of $M_1$ cut the sum over $t$ down to the stated range. For the diagonal state, $S=\{v\}$, and $M_1[t,v]=0$ unless $t$ precedes or equals $v$ in $H_1^\star$; hence the sum ranges exactly over $t\prec v$. For the edge state, $S=\{a\}\cap\{v\}=\varnothing$, and $M_1[t,a]=0$ unless $t\preceq a$, while $a\prec v$ because $(a,v)$ is an edge of $H_1^\star$; hence again only $t\prec v$ contributes. For the output state, $S=\varnothing$ because the terminals are distinct, and the sum over all $t\in U_R$ is the entry $(M_2\Xi M_1)[a_2,b_1]$, since $\Xi$ has no nonzero entry outside the columns $U_R$. The final assertion holds because $\Pi[v,\cdot]$ involves only the columns $u\prec v$ of $\Xi$.
\end{proof}

\subsection{Two evaluation schedules}
\label{sec:directed-schedules}

Throughout this subsection, fix $\varepsilon>0$ and put $\tau\coloneqq\omega+\varepsilon>2$. By \cref{sec:prelim}, two $h\times h$ matrices over $\F_{2^q}$ are multiplied with $O(h^\tau)$ field operations, so an $h\times p$ by $p\times h$ product, or an $h\times h$ by $h\times p$ product with $h\leq p$, splits into $\lceil p/h\rceil$ square products and costs $O(ph^{\tau-1})$ operations.

\Cref{lem:causal-recurrence} reduces the regional detector to computing the entries $\Pi[v,v]$ and $\Pi[v,a]$ in the order $\prec$. Writing $u_1\prec\cdots\prec u_\nu$ for the vertices of $U_R$, the definition of $\Pi$ reads
\begin{align*}
\Pi[u_h,\cdot]
&=\sum_{g<h}(M_2\Xi)[u_h,u_g]M_1[u_g,\cdot],
\end{align*}
and the states in column $u_h$ of $\Xi$ are computed from the row $\Pi[u_h,\cdot]$. Thus column $u_g$ feeds every later row, but only after it is final. This is the dependence pattern of the three-pair schedule of \cref{sec:three}, with the columns of $\Xi$ in the role of the finished table rows and the rows of $\Pi$ in the role of the accumulators. The first schedule is forward substitution: it computes one row of $\Pi$ at a time by a path-sum pass over $H_1^\star$, which costs $O(m_R+n_R)$ per row; the second batches rows by divide and conquer and fast matrix multiplication. It splits the indices into a lower half $L$ and an upper half $H$ (\cref{fig:causal-schedule}). Once all columns in $L$ are final, their entire contribution to the rows in $H$ is one product of three rectangular matrices, and the contributions within $L$ and within $H$ are handled recursively. Every pair $g<h$ is separated at exactly one node of this recursion. A node whose halves have $\ell$ indices multiplies matrices with dimensions $\ell$ and $|V_R^\star|$, and since $\tau>2$, the cost of the recursion is dominated by its top level.

\begin{figure}[t]
\centering
\begin{tikzpicture}[x=0.9cm,y=0.9cm,font=\normalsize]
\fill[black!3] (0,0) rectangle (4,4);
\fill[DPPurple!45] (0,0) rectangle (2,2);
\fill[DPPurple!24] (0,2) rectangle (1,3);
\fill[DPPurple!24] (2,0) rectangle (3,1);
\foreach \x/\y in {0/3,1/2,2/1,3/0}{\fill[DPPurple!12] (\x,\y) rectangle (\x+0.5,\y+0.5);}
\draw (0,0) rectangle (4,4);
\draw[gray] (2,0) -- (2,4) (0,2) -- (4,2);
\draw[densely dashed,gray] (0,4) -- (4,0);
\node[align=center] at (1,1) {$L\to H$};
\node[above] at (1,4) {$L$}; \node[above] at (3,4) {$H$};
\node[left] at (0,3) {$L$}; \node[left] at (0,1) {$H$};
\node at (2,4.85) {source column $g$ of $\Xi$};
\draw[-{Stealth[length=4pt]},gray] (0,4.6) -- (4,4.6);
\node[rotate=90] at (-1,2) {target row $h$ of $\Pi$};
\draw[-{Stealth[length=4pt]},gray] (-0.65,4) -- (-0.65,0);
\node[anchor=west] at (5,3.2) {1. \textsc{SolveColumns}$(L)$: its columns become final.};
\node[anchor=west] at (5,2.1) {2. Add their contribution to $H$:};
\node[anchor=west] at (5.25,1.45) {$M_2[u_H,\cdot]\Xi[\cdot,u_L]M_1[u_L,\cdot]$};
\node[anchor=west] at (5,0.4) {3. \textsc{SolveColumns}$(H)$ with these contributions.};
\end{tikzpicture}
\caption{The causal matrix schedule. A source column $g$ contributes to target row $h$ only for $g<h$, below the dashed diagonal. The dark rectangle batches the contributions from the first half $L$ into the second half $H$; lighter rectangles are batches at deeper recursion levels. Each ordered pair $g<h$ is processed exactly once, after its source column is final and before its target row is used. The rectangles depict dependency pairs, rather than submatrices of the final output.}
\label{fig:causal-schedule}
\end{figure}

\begin{lemma}[Directed circuit for a region]
\label{lem:directed-circuit}
Given $d_1$ and $d_2$, the polynomial $F_R$ of an aligned instance has division-free straight-line circuits of size $O(n_R(m_R+n_R))$ and, for every fixed $\varepsilon>0$, of size $O(n_R^{\omega+\varepsilon})$. Each circuit is constructed in time linear in its size, and its structure is determined by $R$, the terminals, and $d_1,d_2$.
\end{lemma}
\begin{proof}
We evaluate the quantities of \cref{lem:causal-recurrence} in the order $\prec$, once row by row and once by divide and conquer, and count the gates of each schedule.

Construct $H_1^\star$ and $H_2^\star$ in $O(m_R+n_R)$ time. By \cref{lem:causal-recurrence}, it suffices to compute the needed entries of $M_1$ and $M_2$, the entries $\Pi[v,v]$ and $\Pi[v,a]$ for $(a,v)\in C_R$ in the order $\prec$, filling each column of $\Xi$ as soon as its states are known, and finally the root expression. Every index set and every order used below is determined by $H_1^\star$ and $H_2^\star$, and no step tests a field value.

The first schedule evaluates rows by path-sum passes. \begin{itemize}
\item For each $v\in U_R$, one pass over $H_2^\star$ in topological order computes the row $M_2[v,\cdot]$ by
\begin{align*}
M_2[v,v]
&=1,\\
M_2[v,q]
&=\sum_{(p,q)\in E(H_2^\star)}M_2[v,p]x_{pq}
\qquad(q\neq v).
\end{align*}
Analogous passes compute the row $M_1[a_1,\cdot]$, the row $M_2[a_2,\cdot]$, and the columns $M_1[\cdot,b_1]$ and $M_2[\cdot,b_2]$. These passes use $O(n_R(m_R+n_R))$ gates.
\item Process $v\in U_R$ in the order $\prec$. At that time every column $u\prec v$ of $\Xi$ is known, and a scan of $U_R$ and $C_R$ gives
\begin{align*}
X_v[u]
&\coloneqq(M_2\Xi)[v,u]
=M_2[v,u]\Xi[u,u]+\sum_{(p,u)\in C_R}M_2[v,p]\Xi[p,u]
\end{align*}
for all $u\prec v$ with $O(m_R+n_R)$ gates. Put $X_v[u]\coloneqq0$ for every other $u\in V_R^\star$. One pass over $H_1^\star$ in topological order then computes
\begin{align*}
\Pi[v,a]
&=X_v[a]+\sum_{(p,a)\in E(H_1^\star)}\Pi[v,p]x_{pa}
\end{align*}
for all $a\in V(H_1^\star)$. \Cref{lem:causal-recurrence} now gives the states in column $v$ of $\Xi$ with $O(1)$ gates each.
\item Evaluate the root expression $\sum_u(M_2\Xi)[a_2,u]M_1[u,b_1]$ from the row $M_2[a_2,\cdot]$ and the column $M_1[\cdot,b_1]$.
\end{itemize}

The last-edge recurrence verifies the row pass. Induct along $\prec$: every edge $(p,a)$ of $H_1^\star$ has $p\prec a$, so the induction hypothesis applies to $p$, and the computed entry $\widetilde \Pi[v,a]$ satisfies
\begin{align*}
\widetilde \Pi[v,a]
&=X_v[a]+\sum_{(p,a)\in E(H_1^\star)}\sum_uX_v[u]M_1[u,p]x_{pa}\\
&=\sum_uX_v[u]\Bigl([u=a]+\sum_{(p,a)\in E(H_1^\star)}M_1[u,p]x_{pa}\Bigr)\\
&=\sum_uX_v[u]M_1[u,a].
\end{align*}
Since $X_v[u]=(M_2\Xi)[v,u]$ for $u\in U_R$ with $u\prec v$ and $X_v[u]=0$ otherwise, the right-hand side is $\Pi[v,a]$.

Each $v$ costs $O(m_R+n_R)$ gates, and the root expression $\sum_u(M_2\Xi)[a_2,u]M_1[u,b_1]$ costs $O(m_R+n_R)$ more. With the initial passes, the row schedule has $O(n_R(m_R+n_R))$ gates.

The second schedule batches the earlier-to-later contributions. Let $n^\star\coloneqq|V_R^\star|\leq n_R$; simple graphs have $m_R\leq n_R^2$, so the support construction is within the bound.

\emph{Path matrices.}
For each $i$, order $V_R^\star$ by a topological order of $H_i^\star$ followed by the vertices outside $H_i^\star$, so that $W_i$ is strictly upper triangular and $I-W_i$ is unit upper triangular; the reordering is a relabeling of rows and columns. In characteristic two, a block upper triangular matrix with unit-triangular diagonal blocks satisfies
\begin{align*}
\begin{pmatrix}A&B\\0&D\end{pmatrix}^{-1}
&=\begin{pmatrix}A^{-1}&A^{-1}BD^{-1}\\0&D^{-1}\end{pmatrix},
\end{align*}
where the sign of the upper right block disappears in characteristic two. The diagonal blocks of a unit upper triangular matrix are again unit upper triangular. Recursing on the two diagonal blocks computes $M_i=(I-W_i)^{-1}$ with $T(n^\star)\leq2T(n^\star/2)+O((n^\star)^\tau)=O((n^\star)^\tau)$ operations and no division.

Let $u_1\prec\cdots\prec u_\nu$ enumerate $U_R$. For an index interval $J$, the procedure $\textsc{SolveColumns}(J)$ does the following.
\begin{itemize}
\item If $J=\{h\}$, it computes the states in column $u_h$ of $\Xi$ from the current row $\Pi[u_h,\cdot]$ by \cref{lem:causal-recurrence}.
\item Otherwise it splits $J$ into a lower half $L$ and an upper half $H$, calls $\textsc{SolveColumns}(L)$, adds
\begin{align*}
M_2[u_H,\cdot]\Xi[\cdot,u_L]M_1[u_L,\cdot]
\end{align*}
to the rows $\Pi[u_H,\cdot]$, and calls $\textsc{SolveColumns}(H)$. Here $u_L$ and $u_H$ denote the vertex sets indexed by $L$ and $H$.
\end{itemize}
Starting from $\Pi=0$, call $\textsc{SolveColumns}(\{1,\ldots,\nu\})$.

\emph{Causality of \textnormal{\textsc{SolveColumns}}.}
For indices $g<h$, exactly one call has $g\in L$ and $h\in H$, namely the call on the smallest interval of the recursion that contains both; the product at that call adds exactly the terms with $g\in L$ and $h\in H$. At that call the column $u_g$ of $\Xi$ is final because $\textsc{SolveColumns}(L)$ has returned, and $u_h$ has not yet been processed. Hence each term of $\Pi[u_h,\cdot]$ is added exactly once before $u_h$ is processed, as \cref{lem:causal-recurrence} requires.

\emph{Cost of \textnormal{\textsc{SolveColumns}}.}
A call on an interval of size $2\ell$ multiplies an $\ell\times n^\star$ by an $n^\star\times\ell$ matrix and then an $\ell\times\ell$ by an $\ell\times n^\star$ matrix, using $O(n^\star\ell^{\tau-1})$ operations. There are $O(n^\star/\ell)$ such calls, so the intervals of size $2\ell$ cost $O((n^\star)^2\ell^{\tau-2})$ in total. Summing over $\ell=1,2,4,\ldots,n^\star$ gives $O((n^\star)^\tau)$, since $\tau>2$ makes the sum geometric.

The base calls cost $O(n^\star+|C_R|)$, and the root expression $(M_2[a_2,\cdot]\Xi)M_1[\cdot,b_1]$ costs $O((n^\star)^2)$. Together with the path-matrix construction and $m_R\leq n_R^2$, the circuit therefore has $O(n_R^{\omega+\varepsilon})$ gates.
\end{proof}

\subsection{The reporting recursion}\label{sec:directed-recursion}

We now check that \cref{alg:recursive-report} applies to aligned instances: thresholds of $\Delta$ have forced crossing-edge assignments, and every cut produces aligned children.

For an aligned instance, put $J_1\coloneqq[\Delta(a_1),\Delta(b_1)]$, $J_2\coloneqq[\Delta(b_2),\Delta(a_2)]$, and $I_R\coloneqq J_1\cap J_2$. By \cref{lem:directed-delta}, every vertex of an $i$th path has its $\Delta$-value in $J_i$. A feasible $H_2$-path excludes $-\infty$ and a feasible $H_1$-path excludes $+\infty$, so $I_R$ contains neither infinite value. If $I_R=\varnothing$, the two path ranges are disjoint, and arbitrary shortest paths for the two terminal pairs are vertex-disjoint. If $I_R$ is a singleton, \cref{lem:plateau-base,lem:directed-delta} apply: its value is finite, so \cref{lem:directed-delta} makes its plateau a common DAG, as \cref{lem:plateau-base} requires. That lemma uses the full graphs $H_i[R]$, which are available from $d_1$ and $d_2$.

Consider a threshold between consecutive extended $\Delta$-values. Since $\Delta$ is monotone along every $i$th path, the $i$th pair is active exactly when the threshold lies in the interior of its interval $J_i$. With two active pairs, \cref{lem:cut-factorization} gives a squarefree quadratic cut polynomial. For $e=(e^-,e^+)\in H_1$ and $f=(f^+,f^-)\in H_2$, the child terminal pairs are $(a_1,e^-),(f^-,b_2)$ below the cut and $(e^+,b_1),(a_2,f^+)$ above it. The edge directions in \cref{lem:directed-delta} determine the edge-to-pair assignment, so the sum in \cref{lem:cut-factorization} has at most one term, and exactly one for the edge pair $\{e,f\}$ returned by \cref{lem:extraction-step}, since its coefficient is nonzero:
\begin{align*}
[z_ez_f]Q
&=F_L^\sigma(\xi)F_U^\sigma(\xi),
\end{align*}
where $L$ and $U$ are the sides below and above the threshold and $\sigma$ is the forced assignment. Thus \cref{lem:extraction-step} selects two nonzero child evaluations directly.

The remaining active-count cases follow \cref{alg:recursive-report}. Hence every strict cut produces at most two aligned two-path children and preserves the nonzero-evaluation invariant.

The next lemma bounds the total cost: the circuit size of a region contracts by a constant factor from the region to its recursive children, so the root circuit dominates.

\begin{lemma}[Directed recursion cost]
\label{lem:directed-cost}
After the two source-distance computations, the directed reporting recursion takes expected $O(mn)$ time and $O(mn)$ storage cells with the row schedule of \cref{lem:directed-circuit}, and, for every fixed $\varepsilon>0$, expected $O(n^{\omega+\varepsilon})$ time and $O(n^{\omega+\varepsilon})$ storage cells with its matrix schedule.
\end{lemma}
\begin{proof}
\emph{Balancing the regions.} Treat a heavy plateau as one phase. Sort the vertices once by extended $\Delta$. Every region is a union of consecutive $\Delta$-levels. For a region $R$, choose a median $\mu$ of its vertex $\Delta$-values and put
\begin{align*}
A
&\coloneqq|\{v\in R:\Delta(v)<\mu\}|,\\
P
&\coloneqq|\{v\in R:\Delta(v)=\mu\}|,\\
B
&\coloneqq|\{v\in R:\Delta(v)>\mu\}|,
\end{align*}
and let $Z\coloneqq\{v\in R:\Delta(v)=\mu\}$ be the plateau of $\mu$. The cut is chosen as follows.
\begin{itemize}
\item {\bf Balanced cut.} If $A\geq n_R/4$, cut immediately below $\mu$; if $B\geq n_R/4$, cut immediately above it. In either case every child has at most $3n_R/4$ vertices, since $A,B\leq n_R/2$ for a median.
\item {\bf Heavy finite plateau.} Otherwise $P>n_R/2$. For finite $\mu$, process at most the two boundary cuts of the plateau $Z$. A surviving two-path child on $Z$ is a base case by \cref{lem:plateau-base,lem:directed-delta}; every other recursive child lies in $R-Z$.
\item {\bf Infinite plateau.} For $\mu=+\infty$, no $H_1$-path uses $Z$, because $d_1=+\infty$ on $Z$, and for $\mu=-\infty$, no $H_2$-path uses $Z$. Thus one boundary cut invokes a zero- or one-active case and leaves no two-path child on an infinite plateau.
\end{itemize}

Let $\Psi(R)$ be $n_R(m_R+n_R)$ for the row schedule and $n_R^\tau$ with $\tau=\omega+\varepsilon$ for the matrix schedule. By \cref{lem:directed-circuit,lem:extraction-step}, the expected work at a cut, including the child evaluations (at most $\Psi(L)+\Psi(U)\leq\Psi(R)$) and the regional scans, is $O(\Psi(R))$; for the matrix schedule this uses $m_R+n_R\leq2n_R^2$.

\emph{Contraction of the cost.} The recursive children have disjoint vertex sets and disjoint internal edge sets, and every child of a balanced cut has at most $3n_R/4$ vertices. Hence, at a balanced cut,
\begin{align*}
\sum_jn_{R_j}(m_{R_j}+n_{R_j})
&\leq\frac{3n_R}{4}\sum_j(m_{R_j}+n_{R_j})
\leq\frac34n_R(m_R+n_R),
\\
\sum_jn_{R_j}^\tau
&\leq\Bigl(\frac{3n_R}{4}\Bigr)^{\tau-1}\sum_jn_{R_j}
\leq\Bigl(\frac34\Bigr)^{\tau-1}n_R^\tau.
\end{align*}
The recursive children of a heavy-plateau phase lie outside $Z$ and contain fewer than $n_R/2$ vertices in total. Bounding the largest child by this total in the two estimates gives $\sum_j\Psi(R_j)\leq\Psi(R)/2$.

\emph{Total time and space.} A heavy-plateau phase performs at most two cuts, hence costs $O(\Psi(R))$. Plateau leaves cost $O(m+n)$ in total by \cref{lem:plateau-base}. Induction over balanced cuts and grouped heavy-plateau phases, using the contraction estimate, gives $O(\Psi(V))$ total expected work, since the potentials of the regions at successive levels of the grouped recursion decrease geometrically. The component preprocessing gives $n=O(m)$, so $\Psi(V)=O(mn)$ for the row schedule, and $\Psi(V)=n^{\omega+\varepsilon}$ for the matrix schedule. The one-time $O(n\log n)$ ordering is absorbed in both bounds.

Depth-first execution releases a parent circuit before constructing a child, and unvisited siblings have disjoint region lists. The largest live circuit has $O(\Psi(V))$ gates, and the gradient evaluations of \cref{lem:interpolated-extraction} store $O(1)$ field elements per gate, which gives the storage bounds.
\end{proof}

We now prove the formal version of \cref{thm:two-directed-informal}.

\begin{theorem}[Directed 2-DSP, formal version of \cref{thm:two-directed-informal}]
\label{thm:two-directed}
Let $G$ be a directed graph with $n$ vertices, $m$ edges, four distinct terminals, and real edge weights. Suppose every directed cycle has positive total weight. In the computational model of \cref{sec:prelim}, for every constant $c>0$ and every fixed $\varepsilon>0$, there is a randomized algorithm using expected
\begin{align*}
O\bigl(\min\{mn,n^{\omega+\varepsilon}\}\bigr)
\end{align*}
time and $O(\min\{mn,n^{\omega+\varepsilon}\})$ storage cells that never returns an invalid pair, returns \textnormal{\textsc{no}} on every infeasible instance, and reports two vertex-disjoint shortest paths with probability at least $1-(m+n)^{-c}$ on every feasible instance. Evaluating only the root circuit decides feasibility within the same time bound.
\end{theorem}
\begin{proof}
\label{proof:two-directed}
Compute $d_1$ and $d_2$ by \cref{lem:sssp} in expected $O(m+n)+m^{1+o(1)}=O(\min\{mn,n^{\omega+\varepsilon}\})$ time, since at most $n^2$ edges remain after parallel edges are merged, and use the row schedule if $mn\leq n^{\omega+\varepsilon}$ and the matrix schedule otherwise. At each strict cut, \cref{lem:cut-factorization,lem:directed-delta,lem:extraction-step} preserve nonzero subinstance evaluations; disjoint terminal ranges and single-value plateaus are base cases. The reporting recursion costs $O(\min\{mn,n^{\omega+\varepsilon}\})$ by \cref{lem:directed-cost}, and \cref{lem:probability} gives the one-sided error guarantee.
\end{proof}

\appendix
\section{Regional instances of the cancellation identities}\crefalias{section}{appendix}\label{app:regional}

This appendix completes the proof of \cref{lem:regional-circuits}. The polynomial of two disjoint shortest paths is computed in \cite{avw24} for the whole graph, with two terminal pairs $(s_i,t_i)$ and the graphs $G_i$ whose $s_i$--$t_i$ paths are exactly the shortest ones. We show that the same identities compute the regional polynomial $F_R$ of every region produced by the algorithms of \cref{sec:two-linear}, with the same circuit size.

Fix such a region $R$ with terminal pairs $(a_1,b_1)$ and $(a_2,b_2)$. By the first part of the proof of \cref{lem:regional-circuits}, we may assume that both paths contain an edge and that the boundary vertices required to be disjoint are distinct. We apply the identities of \cite{avw24} with $s_i$, $t_i$, and $G_i$ replaced by $a_i$, $b_i$, and $H_i[R]$. Every object in these identities is then defined from $H_1[R]$ and $H_2[R]$ alone:
\begin{itemize}
\item the path sums $L_i(v)$ of the $a_i$--$v$ paths and $R_i(v)$ of the $v$--$b_i$ paths in $H_i[R]$;
\item for each vertex $v$, the set $V_{\mathrm{out}}(v)$ of vertices $w$ such that $(v,w)$ is an arc of both $H_1[R]$ and $H_2[R]$, and the set $V_{\mathrm{mix}}(v)$ of vertices $u$ such that $(u,v)$ is an arc of $H_1[R]$ and $(v,u)$ is an arc of $H_2[R]$;
\item the source linkages $D(v)$, the target linkages $T(v)$, and the polynomials $H(v)$ of \cite[Lemmas~22, 36, and~38]{avw24}, which are sums of products of these path sums and edge variables.
\end{itemize}
In particular, no distance to a regional terminal is needed: a vertex on no $a_i$--$b_i$ path of $H_i[R]$ has $L_i(v)R_i(v)=0$ and contributes nothing, and the only distances used are the global labels $d_1,d_2$ that define $H_1$ and $H_2$.

\begin{lemma}[Regional transfer]
\label{lem:regional-transfer}
With these substitutions, the identities of \cite[Lemmas~22 and~23]{avw24} compute $F_R$ for every region $R$ of a DAG, and together with the identities of \cite[Lemmas~30, 36, 38, and~39]{avw24} they compute $F_R$ for every region $R$ of an undirected graph. Evaluating them uses $O(m_R+n_R)$ additions and multiplications over the field, and the circuit is constructed in $O(m_R+n_R)$ time.
\end{lemma}
\begin{proof}
Apart from counting identities that hold for any two directed graphs, such as the inclusion--exclusion over first edges behind \cite[Lemmas~21 and~36]{avw24}, the proofs in \cite[Section~4]{avw24} use the graphs $G_i$ only through four properties. We first verify each of them for $H_i[R]$, and then account for the cost.

The first property is that every $s_i$--$t_i$ path of $G_i$ is a shortest path of $G$, and that a walk of $G_i$ from $s_i$ to $t_i$ is such a path. For $H_i[R]$, every arc is tight for $d_i$, so an $a_i$--$b_i$ path of $H_i[R]$ has length $d_i(b_i)-d_i(a_i)$, which equals $\dist_G(a_i,b_i)$ by the triangle inequality, as in the proof of \cref{lem:regional-closure}. Hence it is a shortest path of $G$, and so is each of its subpaths. A walk of $H_i[R]$ repeats no vertex, because $H_i$ is acyclic. This also holds for DAGs with arbitrary real weights, since a DAG has no cycle at all.

The second property is that $G_i$ is acyclic, which the path-sum recurrences use. The graph $H_i[R]$ is a subgraph of the acyclic graph $H_i$.

The third property, used only for undirected graphs, is \cite[Proposition~24]{avw24}: if three vertices appear in some order on a shortest path of $G$, then on every shortest path of $G$ through all three, the middle one lies between the other two. This is a statement about shortest paths of $G$, and by the first property it applies to all regional paths and their subpaths.

The fourth property is the exchange of subpaths. Let $P$ and $P'$ be an $a_i$--$b_i$ path of $H_i[R]$ and an $a_{i'}$--$b_{i'}$ path of $H_{i'}[R]$ that both visit vertices $u$ and $v$. If both visit $u$ before $v$, the proofs exchange $P[u,v]$ and $P'[u,v]$ \cite[Lemma~14]{avw24}. In undirected graphs, if $P$ visits $u$ before $v$ and $P'$ visits $v$ before $u$, they exchange $P[u,v]$ with the reversal of $P'[v,u]$, and $P'[v,u]$ with the reversal of $P[u,v]$ \cite[proof of Lemma~33]{avw24}. In either case the two exchanged pieces are shortest $u$--$v$ paths of $G$ by the first property, so they have the same length, and the exchanged walk $Q$ from $a_i$ to $b_i$ has length $\dist_G(a_i,b_i)$ and uses only vertices of $R$. Prefix $Q$ by a shortest $s_i$--$a_i$ path. The result is an $s_i$--$b_i$ walk of length $d_i(b_i)$. A repeated vertex would contradict acyclicity in a DAG and, in an undirected graph, would close a cycle of positive weight whose deletion gives a shorter walk. Hence the walk is a shortest path, each of its edges is tight for $d_i$, and $Q$ is an $a_i$--$b_i$ path of $H_i[R]$. Every exchanged physical edge already lies on $P$ or $P'$, and a reversed edge carries the same variable $x_{uv}=x_{vu}$, so the multiset of edge variables is unchanged.

Every step in the proofs of the cited lemmas is a decomposition of a pair of paths at a vertex or at a common edge, which needs the first property; a statement about the order of vertices along shortest paths, which needs the third; or an involution that exchanges subpaths, which needs the fourth. All pieces lie in $R$, so the boundary of $R$ is never crossed, and the identities hold for $F_R$.

It remains to bound the cost. The global labels $d_1,d_2$ and the tight graphs $H_1,H_2$ are computed once at the root, and each region receives its vertex and adjacency lists from its parent, so $H_i[R]$ is available in $O(m_R+n_R)$ time. A topological order of $H_i[R]$ is found by depth-first search in $O(m_R+n_R)$ time, without sorting distance values. The path sums $L_i$ and $R_i$ follow by forward and backward dynamic programming in this order. For each vertex $v$, one scan of its incident edges determines $V_{\mathrm{out}}(v)$ and $V_{\mathrm{mix}}(v)$. As in the proof of \cite[Theorem~1]{avw24}, $D(v)$ then takes $O(\deg(v))$ operations by \cite[Lemma~22]{avw24}, and in an undirected graph so do $T(v)$ and $H(v)$ by \cite[Lemmas~36 and~38]{avw24}. The sums of \cite[Lemmas~23, 30, and~39]{avw24} add $O(m_R+n_R)$ further operations. All operations are additions and multiplications; in characteristic two, subtraction is addition.
\end{proof}

\PaperPrintBibliography

\section*{AI methodology statement}

We used Claude Opus 5.5 (Anthropic) and GPT-6 Astra (OpenAI) to assist with literature review, the design and analysis of the algorithms, computational checks on small instances, and manuscript drafting. These tools materially affected all sections, in particular the algorithms and proofs of \cref{sec:extraction,sec:dag,sec:two-linear,sec:directed} and \cref{app:regional}. The authors verified the correctness and originality of all content, including references.

\PaperEndMatter

\end{document}